\documentclass[11pt]{article}
\usepackage{amsmath}
\usepackage{graphicx}
\usepackage{enumerate}
\usepackage{hyperref}
\hypersetup{breaklinks,colorlinks,allcolors=blue,pdfpagemode=UseNone}

\usepackage{authblk,longtable,booktabs,array,enumitem,bm,bbm,float,setspace}
\usepackage{natbib}
\usepackage{tikz}
\usetikzlibrary{positioning, arrows.meta, decorations.markings, decorations.pathreplacing, calc, matrix, fit}

\usepackage[poorman,noabbrev]{cleveref}

\newcommand{\blind}{1}

\usepackage[margin=1in]{geometry}

\DeclareMathOperator*{\varop}{Var}

\newcommand{\cor}[2]{\mathrm{Cor}\!\left(#1, #2 \right)}
\newcommand{\cov}[2]{\mathrm{Cov}\!\left(#1, #2 \right)}
\newcommand{\numbereqn}{\addtocounter{equation}{1}\tag{\theequation}}
\newcommand{\ind}[1]{\mathbbm{1}\!\left\{ #1 \right\}}
\newcommand{\vctr}[1]{\bm{#1}}
\newcommand{\mtrx}[1]{\bm{#1}}

\newcommand{\var}[1]{\varop\!\left( #1 \right)}
\newcommand{\Tpre}{T_\text{pre}}
\newcommand{\Tpost}{T_\text{post}}
\newcommand{\ATT}{\mathrm{ATT}}
\newcommand{\disj}[3]{\mathcal{D}_\text{#1}^{\mathcal{C}_{#2} \backslash \mathcal{C}_{#3}}}
\newcommand{\ovrlp}[4]{\mathcal{O}_{\mathrm{#1,#2}}^{\mathcal{C_{#3}}, \mathcal{C}_{#4}}}
\newcommand{\ndisj}[2]{N_{#1 \backslash #2}}
\newcommand{\novrlp}[2]{N_{#1 \cap #2}}

\newlist{assumec}{enumerate}{2}
\setlist[assumec,1]{label*=\textup{A\arabic*},leftmargin=*,resume=assumec}
\setlist[assumec,2]{label=\textup{(\alph*)},leftmargin=*,ref=\textup{I\arabic{assumeci}(\alph{assumecii})}}
\crefname{assumeci}{identifiability assumption}{identifiability assumptions}
\crefalias{assumecii}{assumeci}

\begin{document}

\onehalfspacing


\if1\blind
{
  \title{\bf Shared Control Individuals in Health Policy Evaluations with
  	Application to Medical Cannabis Laws}
  	\author[1,2]{Nicholas J. Seewald}
  	\author[3]{Emma E. McGinty}
  	\author[3]{Kayla Tormohlen}
  	\author[3]{Ian Schmid}
  	\author[1,4,5]{Elizabeth A. Stuart}
  	\affil[1]{Department of Health Policy and Management, Johns Hopkins Bloomberg School of Public Health}
  	\affil[2]{Department of Biostatistics, Epidemiology, and Informatics, University of Pennsylvania Perelman School of Medicine}
  	\affil[3]{Division of Health Policy and Economics, Weill Cornell Medicine}
  	\affil[4]{Department of Mental Health, Johns Hopkins Bloomberg School of Public Health}
  	\affil[5]{Department of Biostatistics, Johns Hopkins Bloomberg School of Public Health}
  \maketitle
} \fi

\if0\blind
{
  \bigskip
  \bigskip
  \bigskip
  \begin{center}
    {\LARGE\bf Shared Control Individuals in Health Policy Evaluations with
    	Application to Medical Cannabis Laws}
\end{center}
  \medskip
} \fi

\bigskip
\begin{abstract}
Health policy researchers often have questions about the effects of a
policy implemented at some cluster-level unit, e.g., states, counties,
hospitals, etc. on individual-level outcomes collected over multiple
time periods. Stacked difference-in-differences is an increasingly
popular way to estimate these effects. This approach involves estimating
treatment effects for each policy-implementing unit, then, if
scientifically appropriate, aggregating them to an average effect
estimate. However, when individual-level data are available and
non-implementing units are used as comparators for multiple
policy-implementing units, data from untreated individuals may be used
across multiple analyses, thereby inducing correlation between effect
estimates. Existing methods do not quantify or account for this sharing
of controls. Here, we describe a stacked difference-in-differences study
investigating the effects of state medical cannabis laws on treatment
for chronic pain management that motivated this work, discuss a
framework for estimating and managing this correlation due to shared
control individuals, and show how accounting for it affects the
substantive results.
\end{abstract}

\noindent%
{\it Keywords:}  Difference in differences, causal inference, correlated data, insurance claims data, multilevel data
\vfill

\newpage
\section{Introduction} \label{sec:intro}

In 2021, the Centers for Disease Control and Prevention estimated that
opioid-related overdoses claimed over 80,000 lives in the United States \citep{cdcOverdoseDeaths2021}. The cannabis industry and advocates have
argued that state medical cannabis laws could provide a partial solution
to this crisis by introducing an alternative treatment for chronic
non-cancer pain, an important driver of opioid prescribing in the U.S.
\citep{cannabisindustryCombatingOpioidEpidemic}. Treatment guidelines for
chronic non-cancer pain, which is commonly related to low back pain,
fibromyalgia, chronic headaches (including migraine), arthritis, and
neuropathic pain, have deemphasized prescription opioids as first-line
treatments in recent years. However, these guidelines do not recommend
cannabis; rather, they emphasize treatments like non-opioid prescription
analgesic medications or procedures (e.g., physical therapy)
\citep{dowellCDCClinicalPractice2022}. Despite this, individuals with
chronic non-cancer pain are eligible to use medical cannabis for pain
management under all existing state medical cannabis laws in the U.S.
\citep{ncslStateCannabisPolicy}.

While recent survey evidence shows that chronic non-cancer pain patients
report substituting cannabis for prescription opioids
\citep{bicketUseCannabisOther2023}, empirical policy evaluations have
found mixed results on the effects of these laws on opioid prescribing
\citep{shahImpactMedicalMarijuana2019, rajiAssociationCannabisLaws2019, bradfordMedicalMarijuanaLaws2016, bradfordAssociationUSState2018, wenAssociationMedicalAdultUse2018, liangMedicalCannabisLegalization2018, powellMedicalMarijuanaLaws2018, bachhuberMedicalCannabisLaws2014}. However, these empirical studies
have largely used general-population and/or repeated cross-sectional
samples, which means they consist mostly of individuals without chronic
non-cancer pain diagnoses (who are likely unaffected by the law)
and/or are unable to follow individuals over time to observe changes in
pain treatment. In this work, we describe and reanalyze individual-level
data from a study estimating the effect of state medical cannabis laws
on opioid prescribing and non-opioid chronic pain treatment receipt for
individuals with chronic non-cancer pain
\citep{mcgintyEffectsStateMedical2023}.

There are a variety of statistical methods that, under assumptions,
enable inference for the causal effect of a policy on outcomes of
interest relative to a well-defined comparison condition. Broadly, the
class of methods that use changes over time across a set of treated and
comparison units to estimate effects are referred to as comparative
interrupted time series or difference-in-differences (DiD). DiD is a
popular approach for estimating policy effects that compares the change
over time in an outcome between treated and comparison groups,
essentially using the trends in the comparison groups as a proxy for how
outcomes would have evolved in the treated group in the absence of the
policy change \citep{wingDesigningDifferenceDifference2018}. The
inclusion of a comparison group allows investigators to control for
underlying secular trends that might affect both groups, making DiD a
stronger design for causal inference than ``uncontrolled'' approaches
that do not include a comparison group and simply compare trends over
time in a set of treated unit(s) \citep{stuartUsingPropensityScores2014}.

Traditionally, the two-way fixed effects (TWFE) model was widely used to
estimate the average treatment effect among the treated (ATT) in DiD
studies for policy evaluation. In its simplest form, TWFE DiD regresses
an outcome on fixed effects for unit and time and an indicator for
whether a unit has been treated by that time. Many recent advances in
DiD methods aim to solve problems with TWFE, specifically that it can
yield a biased estimate of the treatment effect on average over
policy-implementing units when those units implement the policy at
different times (``staggered adoption'') and the effect of the policy is
time-varying
\citep{goodman-baconDifferenceindifferencesVariationTreatment2021}. One
way to circumvent problems with TWFE under staggered adoption is to use
DiD to estimate separate treatment effects for each policy-implementing
unit, then combine those effect estimates for an overall estimate. This
approach is commonly known as ``stacking'' and is related to the idea of
serial ``trial emulation'' \citep{ben-michaelTrialEmulationApproach2021}:
effectively, stacking involves emulating a target trial for each treated
unit, then pooling effect estimates.

In general, stacking proceeds by identifying, for each treated unit, a
``time 0'' at which the policy is implemented. Then, a suitable pool of
comparison units is identified, and their time 0 is defined to be the
same as the treated unit's. These comparators are identified in a
principled way; for example, all units that do not implement the policy
of interest in a 7-year window around the treated unit's time 0. When
control units are chosen in such a way, it is likely that some will be
used as comparators for multiple treated units. In the remainder of the
article, we often refer to the cluster-level units that may have
implemented the policy of interest as ``states'', though they could also
be, e.g., counties, nations, hospitals, etc.

When individual-level data is used in stacked DiD, and when the pool of
comparison states is not distinct for each treated state, it is likely
that individuals in comparison states may contribute to effect
estimation for multiple treated states. These ``shared control
individuals'' meet eligibility criteria for multiple DiDs, and therefore
induce correlation between those effect estimates. In the medical
cannabis laws study, eligibility criteria were a qualifying diagnosis in
a treated state's pre-law period and continuous presence in the health
insurance claims database from which the data were collected; we provide
more details in \Cref{sec:cannabisStudy}. When aggregating per-state
effect estimates, as is often the goal, correlation induced by shared
control individuals must be accounted for in order to produce correct
inference. This problem may arise in any policy evaluation that uses
individual-level data with at least partial sharing of control
individuals across time and trial emulations. Our methodological
contribution is a procedure to, when individual-level data is available,
estimate the correlation between treatment effect estimates that is induced by
shared control individuals and to account for it when estimating the
variance of the aggregated effect estimate.

We start by describing the quantitative portion of the mixed-methods
study that motivates this work and that investigates the effects of
state medical cannabis laws on opioid and non-opioid prescribing for
individuals with chronic non-cancer pain
\citep{mcgintyProtocolMixedmethodsStudy2021, mcgintyEffectsStateMedical2023}. We then provide a brief review of
DiD in \Cref{sec:diffInDiff}, introduce an approach that estimates and
adjusts for the correlation across stacked DiD estimates in
\Cref{sec:correlation}, and reanalyze data from the medical cannabis laws
study in \Cref{sec:sims}.

\section{Motivation: State Medical Cannabis Laws and Opioid Prescribing}\label{sec:cannabisStudy}

The motivating example for this work is a study designed to estimate the effect of state medical cannabis laws on opioid and guideline-concordant
non-opioid prescribing for chronic non-cancer pain treatment among
commercially-insured U.S. adults
\citep{mcgintyProtocolMixedmethodsStudy2021, mcgintyEffectsStateMedical2023}.
The study identified a set of 12 ``treated'' states that enacted a
medical cannabis law between 2012 and 2019 and did not also enact a
recreational cannabis law within 4 years pre- or 3 years post-cannabis
law implementation (CT, MN, NY, NH, FL, MD, PA, OK, OH, ND, AK, LA) and
17 control states (AL, GA, ID, IN, IA, KS, KY, MS, NE, NC, SC, SD, TN,
TX, VA, WI, WY) that did not enact medical or recreational cannabis laws
over the same period. The primary scientific question asked about the
effect of implementing a medical cannabis law on chronic pain treatment
outcomes, relative to what would have happened in the absence of a law,
on average among the states that implemented such a law.

Outcomes of interest were measures of opioid and guideline-concordant
non-opioid prescribing and chronic pain procedures among individuals
with chronic non-cancer pain. Importantly, the study did not use a
general-population sample, as access to medical cannabis is restricted
by state laws to only individuals with a qualifying diagnosis
\citep{ncslStateCannabisPolicy}; therefore, individuals without chronic
non-cancer pain were excluded from the sample. Opioid-related outcomes
studied include receipt of any opioid prescription in a given month, the
number of opioid prescriptions among individuals who received at least
one, the total morphine milligram equivalents (MME) per day for those
prescriptions, the number of days' supply, receipt of an opioid
prescription with more than 7 days' supply, and receipt of more than 50
MME per day. The last two outcomes are indicators of high-risk opioid
prescribing that increases risk of overdose
\citep{dowellCDCClinicalPractice2022,
dowellCDCGuidelinePrescribing2016}. Guideline-concordant non-opioid
outcomes included receipt of any non-opioid analgesic prescription, the
number of such prescriptions among those who received at least one, and
receipt and number of treatment(s) via procedures (e.g., surgeries)
\citep{mcgintyEffectsStateMedical2023}.

This study used de-identified administrative claims data from the Optum
Labs Data Warehouse (OLDW), which includes medical and pharmacy claims,
laboratory results, and enrollment records for commercial and Medicare
Advantage enrollees. The database contains longitudinal health
information for over 200 million individuals, representing a mixture of
ages and geographic regions across the United States
\citep{optumlabs2022}. The study sample included individuals who reside
in a treated or control state, who were continuously enrolled in a
commercial or Medicare Advantage plan that provides claims data to OLDW,
and who meet eligibility criteria discussed below. Analytic data sets
were constructed at the patient-month level. Monthly data allowed for the capture of guideline-concordant prescribing for chronic non-cancer pain,
which is often given 30 days at a time.

\begin{figure}[ht]
	\centering
	\input{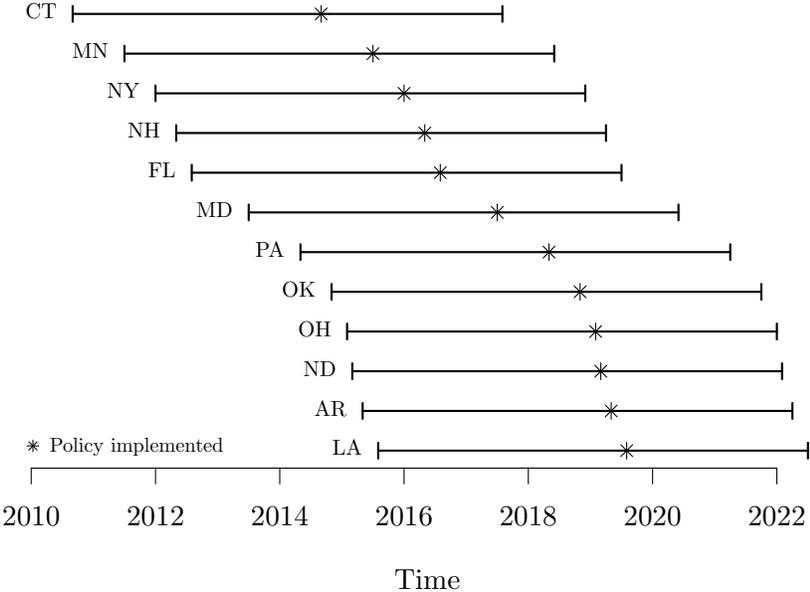}
	\caption{Diagram of study periods for each of the 12 treated states in the medical cannabis study. Stars (*) represent the date on which the state's medical cannabis policy went into effect.}
	\label{fig:implementation_dates}
\end{figure}

Each medical cannabis state's law implementation date was defined as the
first day of the month in which the state's first medical cannabis
dispensary opened. All 12 states had unique implementation dates;
therefore, using a traditional TWFE DiD approach may have led to a biased
estimate of the average treatment effect among the treated (ATT). In
this study, stacked DiD was used to solve this problem. For each treated
state, a unique 7-year study period was constructed, centered around that state's medical cannabis law implementation date: data was collected for 4 years
pre- and 3 years post-implementation. The study periods for all 12
treated states are depicted visually in \Cref{fig:implementation_dates}.

The key idea of stacking is to anchor time for the treated and
comparison states, given unique study periods for each treated state,
relative to the treated state's implementation date (i.e., set policy
implementation as time 0), estimate a treatment effect for each treated
state, then aggregate those estimates across all treated states. We
describe the specific estimands in more detail in \Cref{sec:diffInDiff}.
Each state-specific effect estimate was based on a treated state-specific dataset that we refer to as a \emph{cohort}. A treated state's cohort consisted of monthly measurements from individuals in the treated state of interest or one of the 17 control states who had at least two insurance claims, on
different days, with a primary ICD-9/ICD-10 diagnosis code related to
one or more of low back pain, fibromyalgia, chronic headaches (including
migraine), arthritis (including rheumatoid and osteoarthritis), or
neuropathic pain --- conditions commonly leading to chronic non-cancer
pain -- in the 4 years prior to law implementation \emph{and} who were
continuously enrolled in insurance and therefore present in the OLDW
data for the entire 7-year study period. Since each cohort contained
information from only one treated state, with clear time anchoring at
the time of the policy start date for the treated state, TWFE
DiD could be used to estimate the treatment effect. 

We highlight two important considerations that went into this stacked
design. First, using a common pool of ``never-treated'' control states avoided
difficulties in interpretation and comparison of effect estimates that
may arise when the pool of control individuals changes from treated
state to treated state, as would be the case if using control states
that are ``not yet treated''. Next, individuals were required to be
continuously present in the data over a treated state's 7-year study
period to contribute to that state's analysis. This is a key difference
between this application and those that motivate other recent DiD
advances, which typically assume group-panel data (e.g., annual homicide
rates for an entire state)
\citep{rothWhatTrendingDifferenceindifferences2023}. In this context, and
in many others, this data structure is not available: this study used
insurance claims data that individuals may enter and exit over time as
they enroll or disenroll from an insurance plan. Requiring individuals
be continuously enrolled over the study period minimized bias and
challenges in interpretation due to changing composition (of either the
treated or control group) over time; this is analogous to active efforts
to retain participants over time in a randomized trial.

Imposing a continuous enrollment requirement came at the cost of
generalizability: individuals enrolled in a commercial health insurance
plan for 7 years are different from those who may be enrolled for only
1-2 years at a time. However, in this case it is highly unlikely that
selection into the sample (i.e., whether an individual is continuously
enrolled in commercial health insurance over a particular 7-year study
period) was related to a state's implementation of a medical cannabis
policy, so concerns about bias due to selection on post-treatment
characteristics are minimal.

\begin{figure}
	\centering
	\includegraphics{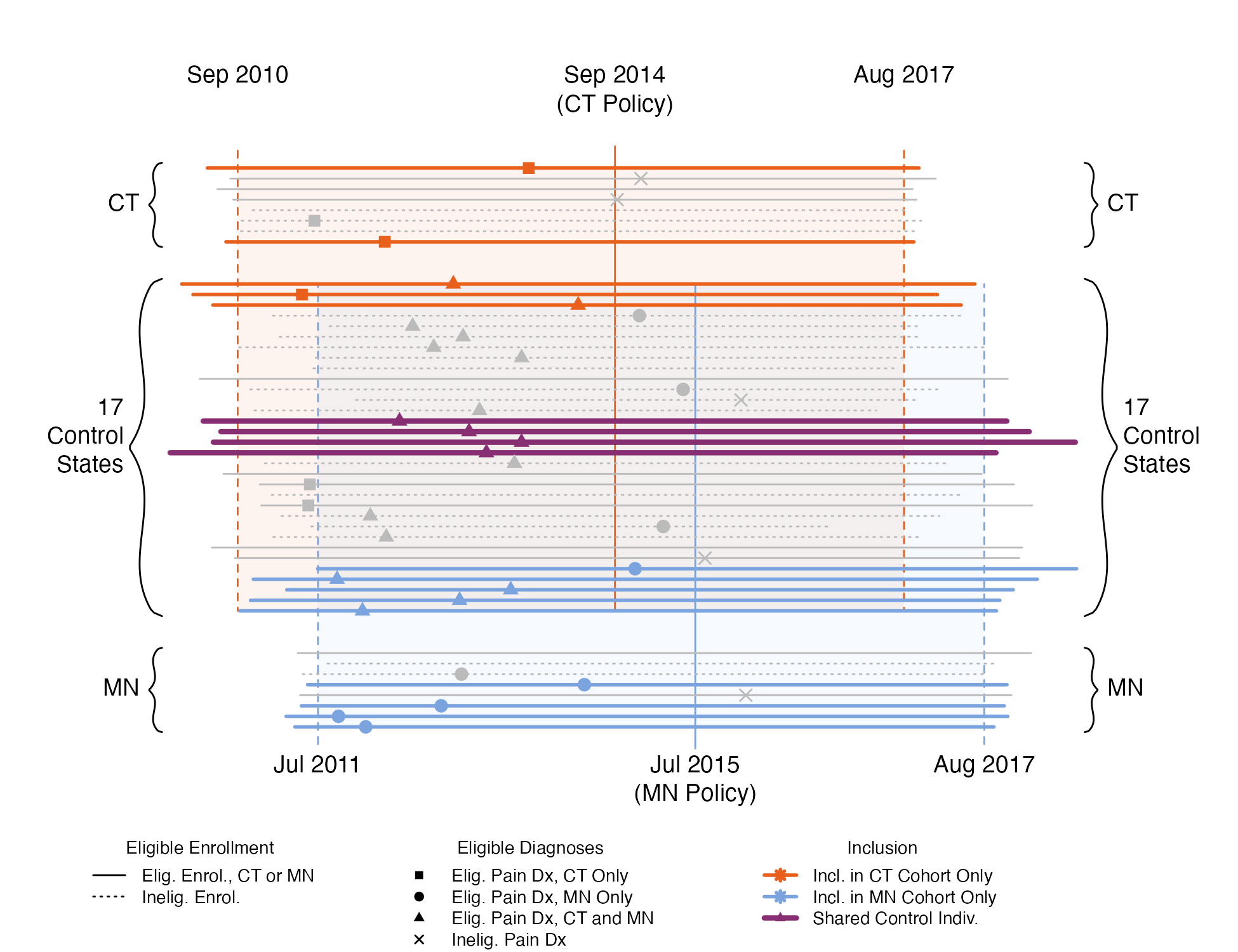}
	\caption{Cohort construction for Connecticut and Minnesota in the
		medical cannabis laws study. Horizontal lines represent times over which an individual is continuously present in the data: a solid line indicates continuous presence over either CT or MN's full study period (and thus eligibility for inclusion); a dotted line otherwise. The top and bottom groups of lines are individuals in CT and MN, respectively; the middle group individuals in one of the 17 control states. Markers on the lines indicate the time of a qualifying chronic pain diagnosis:  squares indicate a diagnosis time allowing inclusion in CT's cohort;  circles, only MN's cohort;  triangles, both; $\times$s, neither. Individuals in treated states are eligible for only their states' cohort. Individuals included in a cohort have thick, colored timelines, the color of which indicates the cohort. Very thick purple lines for control individuals indicate inclusion in both cohorts.
		Ineligible individuals have grey timelines.}
	\label{fig:cohort-building}
\end{figure}

These design considerations, combined with the use of individual-level
data, led to (partial) sharing of individuals in control states between
cohorts. To see this, consider \Cref{fig:cohort-building} and the cohorts
constructed for CT and MN's analyses. CT implemented its medical
cannabis law in September 2014; therefore, its distinct 7-year study
period covered September 2010 through August 2017. MN implemented its law
in July 2015; its 7-year study period ran from July 2011 through June
2018. Individuals living in CT or MN were included in the sample if they
were continuously enrolled for their state's respective 7-year study
period and had an eligible chronic pain diagnosis in their state's
pre-law period. Individuals living in one of the 17 control states who
were continuously enrolled from September 2010 through August 2017 and
who have a qualifying chronic non-cancer pain diagnosis between
September 2010 and August 2014 (CT's pre-law period) were included in
CT's cohort. The control individuals in MN's cohort include those living
in one of the 17 control states who were continuously enrolled from July
2011 through June 2018 with a qualifying diagnosis between July 2011 and
July 2015. Of the control individuals in either cohort, those who had a
qualifying diagnosis between July 2011 and August 2014 (while both CT
and MN are in their pre-law periods) and who were continuously enrolled
from September 2010 through June 2018 were shared between both CT and
MN's analyses. Across the entire study, 84.3\% of control individuals
contributed to two or more of the 12 cohorts; 4.2\% contributed to all
12.

\section{Difference-in-Differences Estimation}\label{sec:diffInDiff}

As discussed previously, most of the literature on methods for DiD
assume group-panel data, i.e., repeated measures over time at the state
level \citep{rothWhatTrendingDifferenceindifferences2023}. The situation,
and solution, we describe here is relevant in the wide-range of analyses
that use individual-level data on particular populations of interest,
aggregated up to some unit/time level (e.g., state-months).

The target estimand in DiD is the average treatment effect among the
treated (ATT), the expected difference in potential outcomes under
treatment and control conditioned on being treated. There are a number of
forms of the ATT that may be of interest, including at a particular time
in the post-treatment period or averaged across time and/or states, with
a variety of state weights possible (e.g., states weighted equally
or by population size). In the medical cannabis laws study, primary
scientific interest is in the ATT averaged across states and months over
the 3-year period following law implementation. We define this as 
\begin{equation}
	\label{eq:ATT}
	\ATT = \mathrm{E}\!\left[ \bar{Y}_{\left\{t\geq t_*\right\}}(1) - \bar{Y}_{\left\{t\geq t_*\right\}}(0) \mid A = 1 \right],
\end{equation}
where $Y(a)$ is the potential outcome under treatment
status $a\in\left\{0,1\right\}$, where $a = 1$ indicates treatment
and $0$ otherwise, and $A$ is the random variable indicating whether
or not a state is ever treated; the overbar and subscript
$\left\{t\geq t_*\right\}$ indicates averaging over the post-treatment
period (such that $t_*$ is the policy implementation date; see below
for more detail). Note that this is not a state-specific quantity: the
expectation is over all treated states. We make three assumptions to
identify the $\ATT$ from observed data:

\begin{assumec}
	\item \emph{No anticipation.} Potential outcomes prior to treatment are
	unaffected by treatment, i.e., $Y_{t}(1) = Y_{t}(0) = Y_t$ for $t$
	in the pre-treatment period ($t< t_*$). \label{assume:no-anticipation}
	
	\item \emph{Consistency.} Observed outcomes are equal to the corresponding
	potential outcomes under observed post-period treatment status, i.e.,
	$Y_{t} = AY_t(1) + (1-A)Y_t(0)$ for $t \geq t^*$.
	\label{assume:consistency}
	
	\item \emph{Parallel counterfactual trends.} In the absence of treatment,
	the mean outcome evolution in the treated group would be ``parallel'' to
	the mean outcome evolution in the control group. To identify the ATT as
	defined in \Cref{eq:ATT}, this requires that
	$\mathrm{E}\!\left[ \bar{Y}_{\left\{t\geq t^*\right\}}(0) - \bar{Y}_{\left\{t< t^*\right\}}(0) \mid A = 1 \right] = \mathrm{E}\!\left[ \bar{Y}_{\left\{t\geq t^*\right\}}(0) - \bar{Y}_{\left\{t < t^*\right\}}(0) \mid A = 0 \right]$.
	The specific form of this assumption varies with the choice of estimand
	(estimating the ATT at a particular time, for example, requires a
	different form), and might condition on covariates incorporated
	into an estimator. \label{assume:parallel}
\end{assumec}

For more details on the assumptions, see for example
\citet{zeldowConfoundingRegressionAdjustment2021}. Under \cref{assume:consistency,assume:no-anticipation,assume:parallel} and continuing the scenario in which there is one treated state (or all treated states implement the policy simultaneously), we can consistently estimate the ATT using a simple plug-in estimator: 
\begin{equation}
	\label{eq:diffInDiffEst}
	\widehat{\ATT} = 
	\left(\bar{Y}_{\text{tx,} \{t\geq t_*\}} -
	\bar{Y}_{\text{tx,} \{t< t_*\}}\right) -
	\left(\bar{Y}_{\text{ctrl,} \{t\geq t_*\}} -
	\bar{Y}_{\text{ctrl,} \{t< t_*\}}\right),
\end{equation}
where, e.g., \(\bar{Y}_{\text{ctrl,} \{t\geq t_*\}}\) is the average observed outcome
over the post-law period and over all control states
\citep{lechnerEstimationCausalEffects2011}. Alternatively, one could fit
a TWFE regression model, given by 
\begin{equation}
	\label{eq:TWFE}
	Y_{\gamma it} = \beta_{0,\gamma} + \beta_{1,t} + \beta_2 A_{\gamma t} +
	\epsilon_{\gamma it},
\end{equation}
where $Y_{\gamma it}$ is the observed outcome for
individual $i$ in state $\gamma$ at time $t$, $A_{\gamma t}$ is
a treatment indicator for whether state $\gamma$ has implemented the
policy (1) or not (0) at or before time $t$ and
$\epsilon_{\gamma it}$ is random mean-zero error. Because we are
considering a setting with a single treated state and are weighting each
post-treatment time point equally, the ordinary least-squares (OLS) estimate
of $\beta_2$ is equivalent to the simple estimator of $\ATT$
in \Cref{eq:diffInDiffEst}.

As discussed above, \citet{goodman-baconDifferenceindifferencesVariationTreatment2021} showed that the TWFE estimate $\hat{\beta}_2$ of the treatment effect can be
severely biased in settings with multiple treated states that implement
the treatment at different times, particularly when the true treatment
effect is time-varying. A common way to avoid such bias in a setting
with multiple treated units when the estimand is an overall ATT averaged
across treated units is to use an approach called \emph{stacking}.

\subsection{Stacked Difference-in-Differences}\label{sec:stacking}

Stacked DiD is an estimation strategy for the ATT in which we obtain
cohort-specific effect estimates and then pool them, i.e., combine them
by taking a (weighted) average. In contrast to a traditional TWFE
approach, stacking involves the construction of a series of cohorts, one
for each set of states that implemented the policy of interest at the
same point in time (at the time level of analysis; e.g., in the same
month for state-month level data), with a DiD model run separately for
each cohort and then the results averaged together. In
\Cref{sec:cannabisStudy} and \Cref{fig:cohort-building}, we briefly described
the construction of state-specific cohorts and how control individuals
may be shared across them. Here, we establish notation and formally
define shared control individuals in that context.

Consider a set of $S$ states $\Xi = \Xi_\text{tx} \cup \Xi_\text{ctrl}$ where $\Xi_\text{tx}$ is the set of \(S_\text{tx}\) treated states, $\Xi_\text{ctrl}$ the collection of $S_\text{ctrl}$ control states, and $\Xi_\text{tx} \cap \Xi_\text{ctrl} = \emptyset$. In the medical cannabis laws study, $S_\text{tx} = 12$ and $S_\text{ctrl} = 17$. For simplicity of exposition, we proceed as if all treated states implement the policy of interest at different times (i.e., there are $S_\text{tx}$ cohorts in the stacked DiD), though all results hold if some states implement simultaneously. For a state $\gamma \in \Xi_\text{tx}$, define its study period $\mathcal{T}_\gamma = \left\{t_{1\gamma}, \ldots, t_{*\gamma}, \ldots, t_{T\gamma}\right\}$ as the set of consecutive measurement occasions at which data is collected, with $t_{*\gamma}$ the first measurement after treatment. We assume the study periods $\mathcal{T}_\gamma$ ($\gamma\in\Xi_\text{tx}$) are all of length $T$ with $\Tpre$ measurements before and $\Tpost$ after treatment. This last condition is a design choice made in the medical cannabis laws study: because scientific interest is in an ATT on average over the post-law period, it was important to maintain identical study period durations.

Every state $\zeta\in\Xi$ is a collection of individuals; say
$i\in \zeta$ if individual $i$ lives in state $\zeta$. Individual
$i$ is continuously present in the data over $U_i$ consecutive
measurement occasions
$\mathcal{U}_i = \left\{u_{1i}, \ldots u_{U_i i}\right\}$. For a pair of states
$(\gamma, \zeta) \in \Xi_\text{tx} \times \Xi$, define
\[
	\mathcal{I}_\gamma(\zeta) = \left\{i\in \zeta : i \text{ meets inclusion criteria for analysis of treated state } \gamma \right\}
\]
as the collection of $N_\gamma(\zeta)$ individuals in state
$\zeta$ that contribute to the analysis for treated state $\gamma$.
If we impose a continuous presence requirement, the inclusion criteria for individual $i$ include $\mathcal{T}_\gamma \subset \mathcal{U}_i$. Additionally, by design, $\mathcal{I}_\gamma(\zeta) = \emptyset$ for
$\zeta \in \Xi_\mathrm{tx}/\left\{\gamma\right\}$. In the context of the medical cannabis laws study, $\mathcal{I}_\text{CT}(\text{AL})$ is the set of all individuals in the data who live in Alabama (a control state), had a chronic non-cancer pain diagnosis in the four years prior to Connecticut's medical cannabis policy implementation, and were continuously enrolled in commercial health insurance over Connecticut'sstudy period. The analytic sample for treated state $\gamma$ is  
\[
	\mathcal{I}_\gamma = \bigcup_{\zeta\in\Xi} \mathcal{I}_\gamma(\zeta) = \bigcup_{\zeta\in (\{\gamma\} \cup \Xi_\text{ctrl})}
	\mathcal{I}_{\gamma}(\zeta),
\] 
the union over $\gamma$ and control states of individuals in each state who contribute to analysis for state $\gamma$.

We define a treated state $\gamma$'s \emph{cohort}
$\mathcal{C}_\gamma$ as the collection of person-time that contributes
to estimation of the treatment effect for $\gamma$: 
\begin{equation}
	\label{eq:cohortDefn}
	\mathcal{C}_\gamma = \mathcal{I}_\gamma \times \mathcal{T}_\gamma = \left\{(i,t) : i\in \mathcal{I}_\gamma, t \in
	\mathcal{T}_\gamma\right\}.
\end{equation}
Using data from a treated unit's cohort, we can estimate  $\ATT_\gamma$, the average treatment effect in treated state $\gamma$, using the simple plug-in DiD estimator from \Cref{eq:diffInDiffEst}:
\begin{equation}
	\label{eq:stateATT}
	\begin{aligned}
		\widehat{\ATT}_\gamma =& 
		\frac{1}{N_{\gamma}(\gamma)} \sum_{i\in \mathcal{I}_\gamma(\gamma)}
		\left(\frac{1}{\Tpost} \sum_{\left\{t\geq t_{*\gamma}\right\}}  Y_{\gamma it} -
		\frac{1}{\Tpre} \sum_{\left\{t<t_{*\gamma}\right\}}  Y_{\gamma it} \right)
		\\
		\qquad & - \frac{1}{\sum_{\zeta\in\Xi_{\text{ctrl}}} N_\gamma(\zeta)}
		\sum_{\zeta\in\Xi_\text{ctrl}} \sum_{i\in\mathcal{I}_\gamma(\zeta)}
		\left(\frac{1}{\Tpost} \sum_{\left\{t\geq t_{*\gamma}\right\}} Y_{\zeta it}-
		\frac{1}{\Tpre} \sum_{\left\{t< t_{*\gamma}\right\}} Y_{\zeta it}\right).
	\end{aligned}
\end{equation}
As before, this estimator is equivalent to $\beta_2$ in a TWFE regression as specfied by \Cref{eq:TWFE} because there is only a single treated state. We proceed with a focus on this particular estimator for $\ATT_\gamma$, though the correlation results are assumed to extend and can be applied to other estimators as well.

The final step of a stacked DiD analysis is to pool the
$\widehat{\ATT}_\gamma$s. In a setting without shared control
individuals (i.e., when all effect estimates are uncorrelated), a
natural way to aggregate might be to use an inverse-variance weighted
average, which places more weight on more precise estimates and is the
minimum-variance aggregation strategy for uncorrelated estimates
\citep[ch. 4]{hartungStatisticalMetaanalysisApplications2008}: 
\begin{equation}
	\label{eq:ivwAgg}
	\widehat{\ATT} = \frac{1}{\sum_{\gamma\in\Xi_\text{tx}} 1/v_{\gamma\gamma}} \sum_{\gamma\in\Xi_\text{tx}} \frac{1}{v_{\gamma\gamma}} \widehat{\ATT}_\gamma,
\end{equation}
where $v_{\gamma\gamma} = \mathop{\mathrm{Var}}\!\left( \widehat{\ATT}_\gamma \right)$. In a setting with shared control individuals, and thus correlated estimates, we propose a using different weighted average arising from generalized least-squares that incorporates non-zero covariance between estimates. We discuss this in more detail in \Cref{sec:correlation}.

In practice, researchers who estimate the ATT using TWFE will typically
adjust the OLS estimate of the standard error of $\hat{\beta}_2$ in
\Cref{eq:TWFE} to account for clustering of individuals within states.
We do not do that here: in settings with one treated state and many
control states, \citet{rokickiInferenceDifferenceinDifferencesSmall2018} found that typical cluster adjustments lead to poor confidence interval
coverage and should not be used. Mathematically, our analyses without
cluster adjustment are equivalent to \citeauthor{rokickiInferenceDifferenceinDifferencesSmall2018}'s ``aggregated''
analysis strategy, which does not suffer from undercoverage problems. We defer our approach to estimating the variance of $\widehat{\ATT}$ to
\Cref{sec:correlation}.

\subsection{Shared Control Individuals in Stacked Difference-in-Differences}
\label{sec:shared-controls}

Consider two treated states $\gamma, \nu \in \Xi_\text{tx}$. An
individual in control unit $\zeta\in\Xi_\text{ctrl}$ is \emph{shared}
between cohorts $C_\gamma$ and $C_\nu$ if that individual is in the
intersection $\mathcal{I}_\gamma(\zeta) \cap \mathcal{I}_\nu(\zeta)$.
This sharing occurs as a consequence of inclusion criteria for each
cohort; the number of control individuals shared between two cohorts can
change based on continuous presence requirements or the amount of time
overlap between the cohorts' study periods $\mathcal{T}_\gamma$ and
$\mathcal{T}_\nu$.

For any pair of cohorts $\mathcal{C}_\gamma$ and $\mathcal{C}_\nu$,
we define the following quantities. Let $N_\gamma = \left|\bigcup_{\zeta\in\Xi} \mathcal{I}_\gamma(\zeta)\right|$ be the total number of individuals who contribute data to cohort $\mathcal{C}_\gamma$. For any $\zeta\in\Xi$, we decompose $N_\gamma(\zeta) = \left|\mathcal{I}_\gamma(\zeta)\right|$, the number
of individuals in state $\zeta$ that contribute to cohort $\mathcal{C}_\gamma$, into the sum $N_{\gamma/\nu}(\zeta) + N_{\gamma\cap \nu}(\zeta)$, where $N_{\gamma/\nu}(\zeta)$ is the number of individuals in state $\zeta$ who are included in $\mathcal{C}_\gamma$ but not $\mathcal{C}_\nu$ and $N_{\gamma\cap\nu}(\zeta)$ the number of individuals in state $\zeta$ who contribute to both $\mathcal{C}_\gamma$ and $\mathcal{C}_\nu$. Further define $N_{\gamma}^\text{ctrl} := \sum_{\zeta\in\Xi_\text{ctrl}} N_\gamma(\zeta) = N_\gamma - N_\gamma(\gamma)$. Finally, let $\Delta = \lvert t_{*\gamma}-t_{*\nu}\rvert$ be the number of measurement occasions between the implementation dates for treated states $\gamma$ and $\nu$.

\section{Correlation between Estimates due to Shared Control Individuals}\label{sec:correlation}

In order to understand the correlation between two state-specific
treatment effect estimates $\widehat{\ATT}_\gamma$ and
$\widehat{\ATT}_\nu$ induced by shared control individuals, we
make some simplifying assumptions about the dependence structure of data
within a state. Borrowing from the literature on multi-period
cluster-randomized trials, we consider three types of correlation:
within-person, within-period, and between-period
\citep{kaszaImpactNonuniformCorrelation2019}. First, within-person (or
intra-individual, longitudinal, or serial) correlation occurs when
repeated outcome measures are collected on the same individual over
time. Similarly, we expect individuals within the same unit to be
related to each other: this is between-person correlation, which has two
components: we expect observations from different individuals collected
at the same time within the same unit to be correlated (``within-period'' correlation), as well as observations from different individuals
in the same unit at different times (``between-period'' correlation). As is common in the policy evaluation methods literature, we
assume states are mutually independent; i.e., that
$\mathrm{Cor}\!\left(Y_{\gamma it}, Y_{\nu js} \right) = 0$ for all
$\gamma\neq\nu$.

For simplicity, we assume a block exchangeable correlation structure for
all states. For two individuals $i$, $j$ in the same state
$\gamma\in\Xi$ and timepoints $t$ and $s$, $\mathrm{Cor}\!\left(Y_{\gamma it}, Y_{\gamma is} \right) = \rho_\gamma$ (within-person correlation),
$\mathrm{Cor}\!\left(Y_{\gamma it}, Y_{\gamma jt} \right) = \phi_\gamma$ (within-period correlation), and $\mathrm{Cor}\!\left(Y_{\gamma it}, Y_{\gamma js} \right) = \psi_\gamma$ (between-period correlation). For a state $\gamma$, then, the correlation matrix for all observations is block-diagonal with
$\operatorname{Exch}_{T}(\rho_\gamma)$ correlation matrices on the diagonal and off-diagonal blocks $\psi_\gamma\bm{1}_T\bm{1}^\top_T + (\phi_\gamma - \psi_\gamma)I_T$, where $\operatorname{Exch}_{T}(\rho)$ is a $T\times T$ matrix with 1's on the diagonal and all off-diagonal elements are $\rho$, $\bm{1}_T$ is
a $T$-vector of 1's, and $I_T$ is the $T\times T$ identity matrix. This is depicted visually in \Cref{eq:sigma_s}. 
\begin{equation}
	\label{eq:sigma_s}
	\Sigma_\gamma := \mathop{\mathrm{Var}}\!\left( \bm{Y}_\gamma \right) = 
	\left(
		\begin{array}{@{}ccccccccccc@{}}
			1 & \rho_\gamma & \cdots & \rho_\gamma & & & & \phi_\gamma & \psi_\gamma & \cdots & \psi_\gamma \\
			\rho_\gamma & 1 & \cdots & \rho_\gamma & & & & \psi_\gamma & \phi_\gamma & \cdots & \psi_\gamma \\
			\vdots & \vdots & \ddots & \vdots & & \cdots & & \vdots & \vdots & \ddots & \vdots \\
			\rho_\gamma & \rho_\gamma & \cdots & 1 & & & & \psi_\gamma & \psi_\gamma & \cdots & \phi_\gamma \\ 
			& & \vdots & & & \ddots & & & & \vdots & \\ 
			\phi_\gamma & \psi_\gamma & \cdots & \psi_\gamma & & & & 1 & \rho_\gamma & \cdots & \rho_\gamma \\
			\psi_\gamma & \phi_\gamma & \cdots & \psi_\gamma & & & & \rho_\gamma & 1 & \cdots & \rho_\gamma \\
			\vdots & \vdots & \ddots & \vdots & & \cdots & & \vdots & \vdots & \ddots & \vdots \\ 
			\psi_\gamma & \psi_\gamma & \cdots & \phi_\gamma & & & & \rho_\gamma & \rho_\gamma & \cdots & 1
		\end{array}
	\right)
	\sigma^2_\gamma,
\end{equation}
with
$\sigma^2_\gamma = \mathop{\mathrm{Var}}\!\left( Y_{\gamma it} \right)$ for all $i$ and $t$.

Under this covariance structure, we can find an analytic form of both
the variance of single-unit DiD estimates $\widehat{ATT}_\gamma$ as in
\Cref{eq:stateATT} and the pairwise covariance between two such
estimates; all such derivations are given in \Cref{appendix:derivation}. For a given treated unit $\gamma\in\Xi_\text{tx}$, the
variance of $\widehat{\ATT}_\gamma$ is 
\begin{equation}
	\label{eq:varATThat}
	\var{\widehat{\ATT}_\gamma} =: v_{\gamma\gamma} = 
	\frac{T}{\Tpre \Tpost}
	\sum_{\zeta \in (\{\gamma\} \cup \Xi_\text{ctrl})}
	\frac{\sigma^2_\zeta}{\left(N_\gamma^{A_\zeta}\right)^2}
	\left[ N_\gamma(\zeta) (1 - \rho_\zeta) + N_\gamma(\zeta) (N_\gamma(\zeta) - 1) (\phi_\zeta - \psi_\zeta) \right],
\end{equation} 
where
$A_\zeta = \mathbbm{1}\!\left\{\zeta\in\Xi_\text{tx} \right\}$
is an indicator for whether state $\zeta$ was ever treated (as in
\Cref{eq:ATT}) such that $N_{\gamma}^{A_\zeta} = A_\zeta N_\gamma^{\text{tx}} + (1-A_\zeta) N_\gamma^\text{ctrl}$. That is, if $\zeta\in\Xi_\text{ctrl}$, then $N_\gamma^{A_\zeta} = N_\gamma^\text{ctrl}$ is the total number of
individuals in control states that contribute to $\widehat{ATT}_\gamma$. Similarly, if $\zeta\in\Xi_\text{tx}$ (i.e.,
$\zeta = \gamma)$, then $N_\gamma^{A_\zeta} = N_\gamma^\text{tx} = N_\gamma(\gamma)$.

For a treated state $\gamma\in\Xi_\text{tx}$, decompose the cohort
$\mathcal{C}_\gamma$ as 
\[
	\mathcal{C}_\gamma = \mathcal{C}_\gamma^\text{tx} \cup
	\mathcal{C}_\gamma^\text{ctrl} = \left(\mathcal{I}_\gamma(\gamma) \times
	\mathcal{T}_\gamma\right) \cup \left(\bigcup_{\zeta\in\Xi_\text{ctrl}}
	\mathcal{I}_\gamma(\zeta) \times \mathcal{T}_\gamma \right);
\]
that is, let $\mathcal{C}_\gamma^\text{tx}$ be the collection of person-times from the treated state $\gamma$ in cohort $\mathcal{C}_\gamma$ and similarly for
$\mathcal{C}_\gamma^\text{ctrl}$. Consider now any pair of treated states $\gamma,\nu\in\Xi_\text{tx}$. Under the assumption that states are independent,
$\mathrm{Cov}\!\left(\widehat{\ATT}_\gamma, \widehat{\ATT}_\nu \right)$ depends solely on information from control states, as they are the only
states to contribute to both estimates: 
\begin{equation}
	\label{eq:covATThat}
	\begin{aligned}
		\mathrm{Cov}\!\left(\widehat{\text{ATT}}_\gamma, \widehat{\text{ATT}}_\nu \right) &= 
	\mathrm{Cov}\!\left(\bar{Y}_{\mathcal{C}_\gamma^\text{ctrl},\left\{t\geq t_{*\gamma}\right\}}, \bar{Y}_{\mathcal{C}_\nu^\text{ctrl},\left\{t\geq t_{*\nu}\right\}} \right) + 
		\mathrm{Cov}\!\left(\bar{Y}_{\mathcal{C}_\gamma^\text{ctrl},\left\{t< t_{*\gamma}\right\}}, \bar{Y}_{\mathcal{C}_\nu^\text{ctrl},\left\{t< t_{*\nu}\right\}} \right) \\
		&\qquad - \mathrm{Cov}\!\left(\bar{Y}_{\mathcal{C}_\gamma^\text{ctrl},\left\{t\geq t_{*\gamma}\right\}}, \bar{Y}_{\mathcal{C}_\nu^\text{ctrl},\left\{t< t_{*\nu}\right\}} \right) -
		\mathrm{Cov}\!\left(\bar{Y}_{\mathcal{C}_\gamma^\text{ctrl},\left\{t< t_{*\gamma}\right\}}, \bar{Y}_{\mathcal{C}_\nu^\text{ctrl},\left\{t\geq t_{*\nu}\right\}} \right),
	\end{aligned}
\end{equation}
where, e.g.,
\[
	\bar{Y}_{\mathcal{C}_\gamma^\text{ctrl},\left\{t\geq t_{*\gamma}\right\}} = 
	\frac{1}{T_\text{post} N_\gamma^\text{ctrl}}
	\sum_{\zeta\in\Xi^\text{ctrl}} \sum_{i\in\mathcal{I}_\gamma(\zeta)}
	\sum_{\left\{t\geq t_{*\gamma}\right\}} Y_{\zeta it}
\]
is the average outcome over all post-treatment periods and all control states.

Again assuming the covariance structure in \Cref{eq:sigma_s}, then 
\begin{multline}
	\label{eq:didCov}
	\cov{\widehat{\text{ATT}}_\gamma}{\widehat{\text{ATT}}_\nu} = 
	\frac{f\left(\Tpre, \Tpost, \Delta\right)}
	{N^\mathrm{ctrl}_\gamma N^\mathrm{ctrl}_\nu} 
	\sum_{\zeta\in\Xi_\mathrm{ctrl}} \sigma^2_\zeta \left[ N_{\gamma}(\zeta)N_{\nu}(\zeta) \left(\phi_\zeta - \psi_\zeta\right) \right. 
	\\
	\left. +
	N_{\gamma\cap \nu}(\zeta)\left(1-\rho_\zeta - (\phi_\zeta - \psi_\zeta)\right)\right],
\end{multline}
where 
\begin{equation}
	\label{eq:timeFactor}
	\begin{aligned}
		f\left(\Tpre, \Tpost, \Delta\right) = 
		\frac{1}{\Tpre^2 \Tpost^2} \cdot 
		\left(\Tpre^2 \max\left(\Tpost - \Delta, 0\right)
		+ \Tpost^2 \max\left(\Tpre - \Delta, 0\right) \right.
		\\
		\left. - \Tpre\Tpost \min\left(\Tpre,
		\Tpost, \Delta, \max\left(\Tpre + \Tpost -
		\Delta, 0\right)\right)\right).
	\end{aligned}
\end{equation}
See \Cref{appendix:derivation} for a full derivation.

The summand in \Cref{eq:didCov} is strictly positive under the
reasonable assumption that for any $\zeta \in \Xi$,
$\rho_\zeta \geq \phi_\zeta \geq \psi_\zeta$, i.e., that within-person
correlation is higher than between-person correlation and that
observations among different individuals at the same time are more
correlated than observations at different times. However, the subtracted
component of the ``time factor'' $f(\cdot)$ (\Cref{eq:timeFactor}) suggests that this multiplier on the covariance may not be everywhere-positive.
Indeed, $f(\cdot)$ has two zeros in $\Delta$ at $\Delta^* := (\Tpre^2 \Tpost + \Tpre \Tpost^2) / (\Tpre^2 + \Tpre \Tpost + \Tpost^2)$ and $\Delta^\dagger := \Tpre + \Tpost$. Note that $\Delta^* < \Delta^\dagger$ for positive $\Tpre$ and $\Tpost$. For $\Delta \in \left[\Delta^*, \Delta^\dagger\right]$,
$f(\Tpre, \Tpost, \Delta) \leq 0$, for $\Delta < \Delta^*$, $f(\Tpre, \Tpost, \Delta) > 0$, and $f(\Tpre, \Tpost, \Delta) = 0$ for $\Delta > \Delta^\dagger$. The factor $f\left(\Tpre, \Tpost, \Delta\right)$ thus controls the sign of the covariance in \Cref{eq:covATThat}, depending on the amount of time overlap between cohorts' study periods. When there is no time overlap, the covariance is exactly zero: no observations are shared between cohorts' DiD analyses.

\begin{figure}
	\centering
\begin{tikzpicture}[x=1pt,y=1pt]
\definecolor{fillColor}{RGB}{255,255,255}
\path[use as bounding box,fill=fillColor,fill opacity=0.00] (0,0) rectangle (361.35,216.81);
\begin{scope}
\path[clip] (  0.00,  0.00) rectangle (361.35,216.81);
\definecolor{drawColor}{RGB}{0,0,0}

\path[draw=drawColor,line width= 0.4pt,line join=round,line cap=round] ( 57.14, 61.20) -- (325.52, 61.20);

\path[draw=drawColor,line width= 0.4pt,line join=round,line cap=round] ( 57.14, 61.20) -- ( 57.14, 55.20);

\path[draw=drawColor,line width= 0.4pt,line join=round,line cap=round] (110.82, 61.20) -- (110.82, 55.20);

\path[draw=drawColor,line width= 0.4pt,line join=round,line cap=round] (164.50, 61.20) -- (164.50, 55.20);

\path[draw=drawColor,line width= 0.4pt,line join=round,line cap=round] (218.17, 61.20) -- (218.17, 55.20);

\path[draw=drawColor,line width= 0.4pt,line join=round,line cap=round] (271.85, 61.20) -- (271.85, 55.20);

\path[draw=drawColor,line width= 0.4pt,line join=round,line cap=round] (325.52, 61.20) -- (325.52, 55.20);

\node[text=drawColor,anchor=base,inner sep=0pt, outer sep=0pt, scale=  1.00] at ( 57.14, 39.60) {0};

\node[text=drawColor,anchor=base,inner sep=0pt, outer sep=0pt, scale=  1.00] at (110.82, 39.60) {20};

\node[text=drawColor,anchor=base,inner sep=0pt, outer sep=0pt, scale=  1.00] at (164.50, 39.60) {40};

\node[text=drawColor,anchor=base,inner sep=0pt, outer sep=0pt, scale=  1.00] at (218.17, 39.60) {60};

\node[text=drawColor,anchor=base,inner sep=0pt, outer sep=0pt, scale=  1.00] at (271.85, 39.60) {80};

\node[text=drawColor,anchor=base,inner sep=0pt, outer sep=0pt, scale=  1.00] at (325.52, 39.60) {100};

\path[draw=drawColor,line width= 0.4pt,line join=round,line cap=round] ( 49.20, 67.52) -- ( 49.20,174.60);

\path[draw=drawColor,line width= 0.4pt,line join=round,line cap=round] ( 49.20, 67.52) -- ( 43.20, 67.52);

\path[draw=drawColor,line width= 0.4pt,line join=round,line cap=round] ( 49.20, 85.36) -- ( 43.20, 85.36);

\path[draw=drawColor,line width= 0.4pt,line join=round,line cap=round] ( 49.20,103.21) -- ( 43.20,103.21);

\path[draw=drawColor,line width= 0.4pt,line join=round,line cap=round] ( 49.20,121.06) -- ( 43.20,121.06);

\path[draw=drawColor,line width= 0.4pt,line join=round,line cap=round] ( 49.20,138.90) -- ( 43.20,138.90);

\path[draw=drawColor,line width= 0.4pt,line join=round,line cap=round] ( 49.20,156.75) -- ( 43.20,156.75);

\path[draw=drawColor,line width= 0.4pt,line join=round,line cap=round] ( 49.20,174.60) -- ( 43.20,174.60);

\node[text=drawColor,rotate= 90.00,anchor=base,inner sep=0pt, outer sep=0pt, scale=  1.00] at ( 34.80, 67.52) {-0.02};

\node[text=drawColor,rotate= 90.00,anchor=base,inner sep=0pt, outer sep=0pt, scale=  1.00] at ( 34.80,103.21) {0.00};

\node[text=drawColor,rotate= 90.00,anchor=base,inner sep=0pt, outer sep=0pt, scale=  1.00] at ( 34.80,138.90) {0.02};

\node[text=drawColor,rotate= 90.00,anchor=base,inner sep=0pt, outer sep=0pt, scale=  1.00] at ( 34.80,174.60) {0.04};

\path[draw=drawColor,line width= 0.4pt,line join=round,line cap=round] ( 49.20, 61.20) --
	(336.15, 61.20) --
	(336.15,191.61) --
	( 49.20,191.61) --
	cycle;
\end{scope}
\begin{scope}
\path[clip] (  0.00,  0.00) rectangle (361.35,216.81);
\definecolor{drawColor}{RGB}{0,0,0}

\node[text=drawColor,anchor=base,inner sep=0pt, outer sep=0pt, scale=  1.00] at (192.68, 15.60) {$\Delta$};

\node[text=drawColor,rotate= 90.00,anchor=base,inner sep=0pt, outer sep=0pt, scale=  1.00] at ( 10.80,126.41) {$f(T_\text{pre}, T_\text{post}, \Delta)$};
\end{scope}
\begin{scope}
\path[clip] ( 49.20, 61.20) rectangle (336.15,191.61);
\definecolor{drawColor}{RGB}{211,211,211}

\path[draw=drawColor,line width= 0.4pt,line join=round,line cap=round] (130.26, 61.20) -- (130.26,191.61);

\path[draw=drawColor,line width= 0.4pt,line join=round,line cap=round] (282.58, 61.20) -- (282.58,191.61);
\definecolor{drawColor}{RGB}{0,0,0}

\path[draw=drawColor,line width= 0.4pt,line join=round,line cap=round] ( 49.20,103.21) -- (336.15,103.21);

\path[draw=drawColor,line width= 0.8pt,line join=round,line cap=round] ( 59.83,186.78) --
	( 62.51,183.60) --
	( 65.20,180.41) --
	( 67.88,177.23) --
	( 70.56,174.04) --
	( 73.25,170.86) --
	( 75.93,167.67) --
	( 78.61,164.49) --
	( 81.30,161.30) --
	( 83.98,158.12) --
	( 86.67,154.94) --
	( 89.35,151.75) --
	( 92.03,148.57) --
	( 94.72,145.38) --
	( 97.40,142.20) --
	(100.08,139.01) --
	(102.77,135.83) --
	(105.45,132.64) --
	(108.14,129.46) --
	(110.82,126.28) --
	(113.50,123.09) --
	(116.19,119.91) --
	(118.87,116.72) --
	(121.55,113.54) --
	(124.24,110.35) --
	(126.92,107.17) --
	(129.61,103.98) --
	(132.29,100.80) --
	(134.97, 97.62) --
	(137.66, 94.43) --
	(140.34, 91.25) --
	(143.03, 88.06) --
	(145.71, 84.88) --
	(148.39, 81.69) --
	(151.08, 78.51) --
	(153.76, 75.33) --
	(156.44, 74.55) --
	(159.13, 73.78) --
	(161.81, 73.00) --
	(164.50, 72.23) --
	(167.18, 71.45) --
	(169.86, 70.68) --
	(172.55, 69.90) --
	(175.23, 69.13) --
	(177.91, 68.35) --
	(180.60, 67.58) --
	(183.28, 66.80) --
	(185.97, 66.03) --
	(188.65, 67.06) --
	(191.33, 68.10) --
	(194.02, 69.13) --
	(196.70, 70.16) --
	(199.38, 71.19) --
	(202.07, 72.23) --
	(204.75, 73.26) --
	(207.44, 74.29) --
	(210.12, 75.33) --
	(212.80, 76.36) --
	(215.49, 77.39) --
	(218.17, 78.42) --
	(220.85, 79.46) --
	(223.54, 80.49) --
	(226.22, 81.52) --
	(228.91, 82.55) --
	(231.59, 83.59) --
	(234.27, 84.62) --
	(236.96, 85.65) --
	(239.64, 86.69) --
	(242.32, 87.72) --
	(245.01, 88.75) --
	(247.69, 89.78) --
	(250.38, 90.82) --
	(253.06, 91.85) --
	(255.74, 92.88) --
	(258.43, 93.92) --
	(261.11, 94.95) --
	(263.80, 95.98) --
	(266.48, 97.01) --
	(269.16, 98.05) --
	(271.85, 99.08) --
	(274.53,100.11) --
	(277.21,101.14) --
	(279.90,102.18) --
	(282.58,103.21) --
	(285.27,103.21) --
	(287.95,103.21) --
	(290.63,103.21) --
	(293.32,103.21) --
	(296.00,103.21) --
	(298.68,103.21) --
	(301.37,103.21) --
	(304.05,103.21) --
	(306.74,103.21) --
	(309.42,103.21) --
	(312.10,103.21) --
	(314.79,103.21) --
	(317.47,103.21) --
	(320.15,103.21) --
	(322.84,103.21) --
	(325.52,103.21);

\path[] (266.83, 87.63) rectangle (298.33, 66.03);

\node[text=drawColor,anchor=base west,inner sep=0pt, outer sep=0pt, scale=  0.90] at (283.03, 73.76) {$\Delta^\dagger$};

\path[] (114.32, 87.63) rectangle (146.20, 66.03);

\node[text=drawColor,anchor=base west,inner sep=0pt, outer sep=0pt, scale=  0.90] at (130.52, 73.76) {$\Delta^*$};
\end{scope}
\end{tikzpicture}
	\caption{The ``time factor'' in \Cref{eq:covATThat} for $\Tpre = 48$ and $\Tpost = 36$ as in the medical cannabis laws study.}
	\label{fig:timeFactor}
\end{figure}
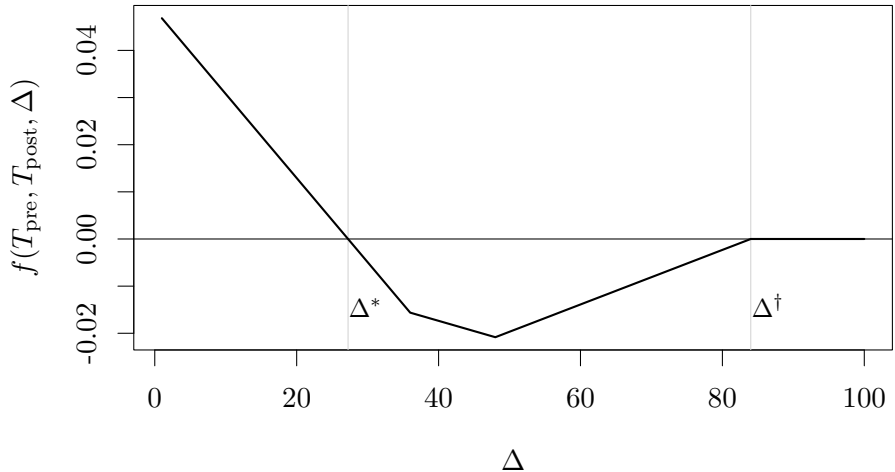

In the context of the medical cannabis laws study, where $\Tpre = 48$ and $\Tpost = 36$, $\Delta^* = 27.24$ and $\Delta^\dagger = 84$. $f(48, 36, \Delta)$ is nonpositive for $\Delta \in \left\{28, \ldots, 84\right\}$ (see \Cref{fig:timeFactor}). Therefore, the DiD estimate for Connecticut, for example, will have positive covariance with the estimates from Minnesota, New York, New Hampshire, and Florida ($\Delta = 10, 16, 20, 23$, respectively) and negative covariance with
the estimates from Maryland, Pennsylvania, Oklahoma, Ohio, North Dakota, Arkansas, and Louisiana ($\Delta = 34, 44, 50, 53, 54, 56, 59$, respectively).

This sign change may be rather surprising: one might expect that
allowing information to contribute to multiple effect estimates would
induce strictly positive correlation between those estimates. However,
this is not always the case in a DiD setting due to the idiosyncratic
nature of the DiD estimator and the use of some time periods as
pre-policy in one cohort and post-policy in another. These
``cross-period'' timepoints induce the ``cross-period'' covariances that
are subtracted in \Cref{eq:covATThat}. As $\Delta$ increases from 0,
the number of measurement occasions in which cohorts $C_\gamma$ and $C_\nu$
are \emph{both} in their pre-treatment periods (and thus both in their
post-treatment periods as well) decreases. When $\Delta > \Delta^*$,
this is no longer true: the number of cross-period measurement occasions
exceeds those in both cohorts' pre- or post-treatment periods. The
number of measurement occasions in a period is a proxy for the amount of
information that period contributes to a treatment effect estimate. When
the bulk of information shared between two estimates comes from periods
in which the treated states have different treatment statuses, this
increases the subtracted components of \Cref{eq:covATThat}.

As an example, consider Maryland and Oklahoma, for which
$\Delta_\text{MD,OK} = 16$ months. Maryland implemented its medical
cannabis law in July 2017; Oklahoma in November 2018. Both states'
cohorts are in their pre-law periods for 32 months, from November 2014
through June 2017, and both are in their post-law periods for 20 months,
from November 2018 through June 2020. By contrast, there are only 16
months of cross-period time, from July 2017 through October 2018, while
Maryland's law was in place and Oklahoma's was not. Thus, the DiD
analyses for Maryland and Oklahoma use data from 52 measurement
occasions in the same way (i.e., as either pre- or post-law), and data
from only 16 measurement occasions are used differently. We would
therefore expect positive covariance between the two DiD estimates in
this case, and that is the case: $f(48, 36, 16) = 0.02 > 0$. By
contrast, for New York and Oklahoma, $\Delta_\text{NY,OK} = 34$. Now,
there are only 16 months in which both states are either pre-law (14
months) or post-law (2 months), and 34 cross-period months. Because more
information is being used in different ways by the two analyses, we now
expect negative covariance between the estimates, and indeed
$f(48, 36, 34) = -0.01 < 0$.

Finally, for completeness, we define the correlation between
$\widehat{\ATT}_\gamma$ and $\widehat{\ATT}_\nu$ as
\begin{equation}
	\label{eq:didCor}
	\cor{\widehat{\ATT}_\gamma}{\widehat{\ATT}_\nu} =
	\frac{\cov{\widehat{\ATT}_\gamma}{\widehat{\ATT}_\nu}}%
	{\var{\widehat{\ATT}_\gamma}^{1/2}
		\var{\widehat{\ATT}_\gamma}^{1/2}},
\end{equation}
which we estimate with a simple plug-in estimator.

\subsection{Aggregation of Correlated Effect Estimates} \label{sec:aggregation}

In the medical cannabis laws study, the estimand of interest was the ATT
on average over all 12 treated states. As mentioned above, one strategy for aggregating uncorrelated unit-specific estimates $\{\widehat{\ATT}_\gamma\}$ is inverse-variance weighting (IVW). Define the $S_\text{tx} \times S_\text{tx}$ diagonal matrix $\mtrx{V} = \operatorname{diag}\left(\left\{v_{\gamma\gamma}\right\}_{\gamma\in\Xi_\text{tx}}\right)$ that has as its diagonal entries the variances of the $\{\widehat{\ATT}_\gamma\}$ as defined in \Cref{eq:varATThat}. Then the IVW estimate of the overall $\ATT$, as in \Cref{eq:ivwAgg}, is 
\begin{equation}
	\label{eq:ivw-wtd-est}
	\widehat{\ATT}_\text{ivw} := \left(\vctr{1}_{S_{\text{tx}}}^\top \mtrx{V}^{-1} \vctr{1}_{S_{\text{tx}}}\right)^{-1} \vctr{1}_{S_{\text{tx}}} \mtrx{V}^{-1} \widehat{\vctr{\ATT}}_{\Xi_\text{tx}} = \frac{1}{\sum_{\gamma\in\Xi_\text{tx}} v_{\gamma\gamma}^{-1}} \sum_{\gamma\in\Xi_\text{tx}} v_{\gamma\gamma}^{-1} \widehat{\ATT}_\gamma,
\end{equation}
where $\widehat{\vctr{\ATT}}_{\Xi_\text{tx}}$ is the $S_\text{tx}$-vector of treated state-specific $\widehat{\ATT}$s. This estimator has variance 
\begin{equation}
	\label{eq:ivwAggVarNaive}
	\operatorname{Var}\!\left(\widehat{\ATT}_\text{ivw}\right) = \left(\vctr{1}_{S_\text{tx}}^\top \mtrx{V}^{-1} \vctr{1}_{S_\text{tx}}\right)^{-1} = 
	\frac{1}{\sum_{\gamma\in\Xi_\text{tx}} 1/v_{\gamma\gamma}}.
\end{equation}

The above IVW procedure assumes that the $\{\widehat{\ATT}_\gamma\}$ are uncorrelated. Now define the $S_\text{tx} \times S_\text{tx}$ matrix $\mtrx{W} = \var{\widehat{\bm{\ATT}}_{\Xi_\text{tx}}}$ with entries $w_{\gamma\nu} = \cov{\widehat{\ATT}_\gamma}{\widehat{\ATT}_\nu}$.

We propose an aggregation strategy that accounts for correlated effect estimates by substituting $\mtrx{W}$ for $\mtrx{V}$ in \Cref{eq:ivw-wtd-est,eq:ivwAggVarNaive}. This is equivalent to estimating the overall $\ATT$ by fitting an intercept-only model with generalized least squares (GLS), following the approach of \citet{linMetaAnalysisGenomewideAssociation2009}. Our ``GLS-aggregated'' estimate $\widehat{\ATT}_\text{gls}$ of the overall $\ATT$ is the weighted average
\begin{equation}
	\label{eq:gls-wtd-est}
	\widehat{\ATT}_\text{gls} = \left(\vctr{1}_{S_{\text{tx}}}^\top \mtrx{W}^{-1} \vctr{1}_{S_{\text{tx}}}\right)^{-1} \vctr{1}_{S_{\text{tx}}} \mtrx{W}^{-1} \widehat{\vctr{\ATT}}_{\Xi_\text{tx}}
\end{equation}
where $\widehat{\vctr{\ATT}}_{\Xi_\text{tx}}$ is the vector of treated state-specific $\widehat{\ATT}$s. This estimator explicitly adjusts for the between-estimate correlation and has variance
\begin{equation}
	\label{eq:gls-wtd-var}
	\var{\widehat{\ATT}_\text{gls}} = \left(\vctr{1}_{S_\text{tx}}^\top \mtrx{W}^{-1} \vctr{1}_{S_\text{tx}}\right)^{-1}.
\end{equation}
Note that, if $\mtrx{W}$ is diagonal (i.e., $\mtrx{W} = \mtrx{V}$), \Cref{eq:gls-wtd-est,eq:ivw-wtd-est} are equivalent.

To illustrate the substantive difference between variances \labelcref{eq:ivwAggVarNaive,eq:gls-wtd-var}, consider a simplified setting with just two treated units. If the between-estimate correlation is $\kappa = w_{12}/\sqrt{w_{11}w_{22}}$, then $\var{\widehat{\ATT}_\text{gls}} = \frac{w_{11} + w_{22} - 2\kappa\sqrt{w_{11}w_{22}}}{\left(w_{11} + w_{22}\right)\left(1-\kappa^2\right)}$ and $\var{\widehat{\ATT}_\text{ivw}} = \frac{w_{11} + w_{22}}{w_{11}w_{22}}$. When $0 \leq \kappa \leq \frac{2\sqrt{w_{11}w_{22}}}{w_{11} + w_{22}}$, the GLS-adjusted estimator \labelcref{eq:gls-wtd-est} has \textit{smaller} variance (i.e., is more efficient) than the na\"ive estimator in \cref{eq:ivw-wtd-est}. Otherwise, accounting for between-estimate correlation leads to larger standard errors in the two-cohort setting.

\section{Simulations and Reanalysis of Medical Cannabis Laws Study} \label{sec:sims}

We designed a simulation study to investigate the properties of \Cref{eq:didCor}. We simulate outcomes $Y_{\gamma it}$ from a modified version of model (2) of \citet{kaszaImpactNonuniformCorrelation2019} such that 
\begin{equation}
	\label{eq:genModel}
	Y_{\gamma it} = \beta_0 +
	\beta_1(t) + \beta_2 A_{\gamma t} + b_{\gamma i} + c_{\gamma t} +
	\epsilon_{\gamma it},
\end{equation}
where $b_{\gamma i} \sim N(0, \sigma^2_{b\gamma})$ is an individual-specific
random intercept for person $i$ in state $\gamma$, $c_{\gamma t}$
is the \(t\)-th element of the \(T\) vector of state-time random effects
$\bm{c}_{\gamma} \sim N_T(\bm{0}, \sigma^2_{c\gamma} \bm{R}_\gamma)$
where $\bm{R}_\gamma$ has $(s,t)$th element
$\mathrm{Cor}\!\left(c_{\gamma s}, c_{\gamma t} \right) = \psi_\gamma/ \phi_\gamma$, and $\epsilon_{\gamma it} \sim N(0, \sigma^2_{e\gamma})$ is random error. To achieve the correlation structure in \Cref{eq:sigma_s}, we set $\sigma^2_{b\gamma} = \frac{\rho_\gamma - \psi_\gamma}{(1-\rho_\gamma) - (\phi_\gamma - \psi_\gamma)}\cdot \sigma^2_{e\gamma}$ and $\sigma^2_{c\gamma} = \frac{\phi_\gamma}{(1-\rho_\gamma) - (\phi_\gamma - \psi_\gamma)}\cdot \sigma^2_{e\gamma}$; $\sigma^2_{e\gamma}$ is allowed to vary. Note that we do not model effect heterogeneity across states: such heterogeneity would likely make aggregating effect estimates inappropriate (see \Cref{sec:discussion}).

We consider two treated states (i.e., two cohorts) under a range of
scenarios, varying the number of control states, $T_\text{pre}$,
$T_\text{post}$, $\Delta$, within- and between-person correlations,
and the percent of individuals in each control state shared between cohorts (assumed constant). For
simplicity, we set $\rho_\gamma = \rho$, $\phi_\gamma = \phi$ and
$\psi_\gamma = \psi$ for all states $\gamma$ in each simulation
setting. We only consider settings in which the generated correlation
structure matches \Cref{eq:sigma_s}. The correlation estimate and
variance correction developed here explicitly assume that structure and
do not apply to other, non-exchangeable structures. When the
covariance structure is correctly specified, we expect an unbiased
estimate of the correlation between cohort-specific DiD estimates. We
also expect to see nominal coverage of confidence intervals for the
inverse-variance weighted $\widehat{\ATT}_\text{ivw}$ when we use the
correlation-corrected variance formula in \Cref{eq:gls-wtd-var}, with deviations
from nominal coverage increasing as magnitude of the between-cohort correlation
becomes larger. Reported correlations are averages of 100 estimates, each of which is computed from 100 pairs of simulated $\widehat{ATT}$s.

\begin{table}
\centering
\begingroup\scriptsize
\begin{tabular}{lllllll|llllllll}
  \toprule & & & & & & & \multicolumn{2}{c}{$\cor{\widehat{\ATT}_\gamma}{\widehat{\ATT}_\nu}$} & \multicolumn{3}{c}{$\widehat{\mathrm{ATT}}_\text{ivw}$} & \multicolumn{3}{c}{$\widehat{\mathrm{ATT}}_\text{gls}$} \\ \cmidrule{8-15} 
$T_\text{pre}$ & $T_\text{post}$ & $\Delta$ & \% Shared & $\rho$ & $\phi$ & $\psi$ & True Cor. & Est. Bias & Bias & SE & 95\% Covg. & Bias & SE & 95\% Covg. \\ 
  \midrule
1 & 1 & 1 & 0.25 & 0.10 & 0.06 & 0.02 & -0.108 & -0.00 & -0.00 & 0.27 & 0.962 & -0.00 & 0.26 & 0.950 \\ 
   &  &  &  & 0.60 & 0.40 & 0.20 & -0.124 & 0.00 & 0.01 & 1.16 & 0.961 & 0.01 & 1.09 & 0.944 \\ 
   &  &  & 0.75 & 0.10 & 0.06 & 0.02 & -0.119 & 0.01 & 0.00 & 0.27 & 0.965 & 0.00 & 0.26 & 0.950 \\ 
   &  &  &  & 0.60 & 0.40 & 0.20 & -0.125 & 0.01 & -0.01 & 1.16 & 0.965 & -0.01 & 1.09 & 0.950 \\ 
   &  & 2 & 0.25 & 0.10 & 0.06 & 0.02 & 0.000 & 0.01 & -0.00 & 0.27 & 0.952 & -0.00 & 0.27 & 0.952 \\ 
   &  &  &  & 0.60 & 0.40 & 0.20 & 0.000 & -0.01 & 0.00 & 1.16 & 0.953 & 0.00 & 1.16 & 0.953 \\ 
   &  &  & 0.75 & 0.10 & 0.06 & 0.02 & 0.000 & -0.00 & -0.00 & 0.27 & 0.951 & -0.00 & 0.27 & 0.951 \\ 
   &  &  &  & 0.60 & 0.40 & 0.20 & 0.000 & 0.01 & -0.01 & 1.16 & 0.949 & -0.01 & 1.16 & 0.949 \\ 
  5 & 5 & 3 & 0.25 & 0.10 & 0.06 & 0.02 & 0.022 & 0.00 & 0.00 & 0.12 & 0.944 & 0.00 & 0.12 & 0.947 \\ 
   &  &  &  & 0.60 & 0.40 & 0.20 & 0.025 & -0.01 & -0.00 & 0.52 & 0.948 & -0.00 & 0.53 & 0.951 \\ 
   &  &  & 0.75 & 0.10 & 0.06 & 0.02 & 0.024 & 0.01 & -0.00 & 0.12 & 0.946 & -0.00 & 0.12 & 0.950 \\ 
   &  &  &  & 0.60 & 0.40 & 0.20 & 0.025 & -0.00 & -0.00 & 0.52 & 0.949 & -0.00 & 0.53 & 0.952 \\ 
   &  & 6 & 0.25 & 0.10 & 0.06 & 0.02 & -0.087 & 0.01 & 0.00 & 0.12 & 0.958 & 0.00 & 0.12 & 0.951 \\ 
   &  &  &  & 0.60 & 0.40 & 0.20 & -0.099 & 0.01 & 0.00 & 0.52 & 0.958 & 0.00 & 0.49 & 0.943 \\ 
   &  &  & 0.75 & 0.10 & 0.06 & 0.02 & -0.096 & -0.00 & -0.00 & 0.12 & 0.960 & -0.00 & 0.12 & 0.951 \\ 
   &  &  &  & 0.60 & 0.40 & 0.20 & -0.100 & 0.00 & 0.00 & 0.52 & 0.961 & 0.00 & 0.49 & 0.949 \\ 
  7 & 3 & 3 & 0.25 & 0.10 & 0.06 & 0.02 & -0.028 & 0.00 & -0.00 & 0.13 & 0.955 & -0.00 & 0.13 & 0.951 \\ 
   &  &  &  & 0.60 & 0.40 & 0.20 & -0.032 & 0.02 & 0.00 & 0.57 & 0.953 & 0.00 & 0.56 & 0.948 \\ 
   &  &  & 0.75 & 0.10 & 0.06 & 0.02 & -0.031 & -0.01 & -0.00 & 0.13 & 0.957 & -0.00 & 0.13 & 0.955 \\ 
   &  &  &  & 0.60 & 0.40 & 0.20 & -0.032 & -0.01 & -0.00 & 0.57 & 0.955 & -0.00 & 0.56 & 0.950 \\ 
   &  & 6 & 0.25 & 0.10 & 0.06 & 0.02 & -0.056 & 0.02 & 0.00 & 0.13 & 0.955 & 0.00 & 0.13 & 0.948 \\ 
   &  &  &  & 0.60 & 0.40 & 0.20 & -0.064 & 0.00 & -0.00 & 0.57 & 0.956 & -0.00 & 0.55 & 0.948 \\ 
   &  &  & 0.75 & 0.10 & 0.06 & 0.02 & -0.061 & 0.00 & -0.00 & 0.13 & 0.955 & -0.00 & 0.13 & 0.948 \\ 
   &  &  &  & 0.60 & 0.40 & 0.20 & -0.064 & -0.00 & 0.01 & 0.57 & 0.959 & 0.01 & 0.55 & 0.952 \\ 
   \bottomrule
\end{tabular}
\endgroup
\caption{Simulated between-estimate correlations along with standard error and 95\% confidence interval coverage for aggregated estimates of $\widehat{\ATT}$ for a variety of generative model parameters, 100 individuals per state, and 3 control states. Reported correlations are averages of 100 estimates generated from 100 simulations.} 
\label{tab:sims3}
\end{table}

\begin{figure}
	\centering
	\input{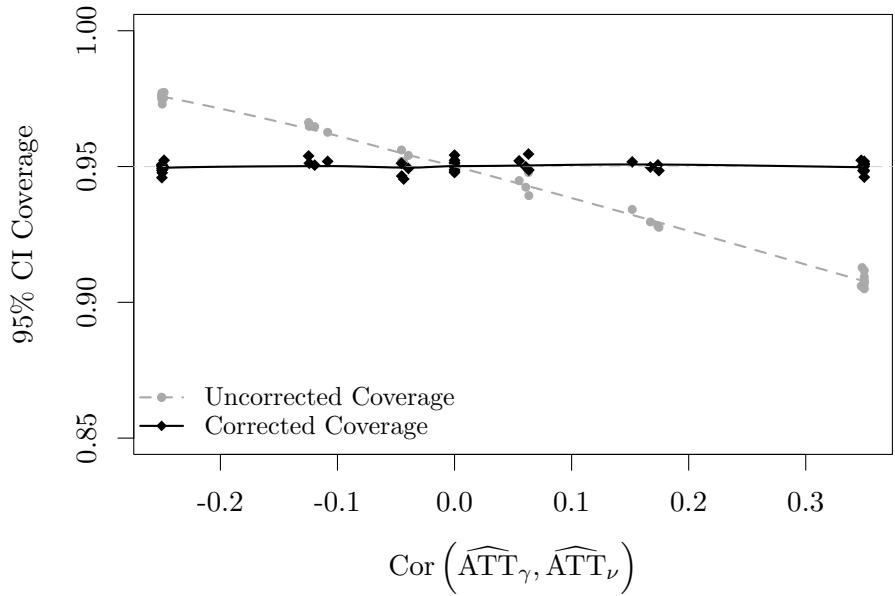}
	\caption{95\% confidence interval coverage when applying and not applying the correlation correction as a function of between-estimate 			correlation. Coverage is estimated based on 10,000 simulations of 			cohorts with 10 total measurements, 5 pre-treatment and 5 			post-treatment, with 1 control unit. Lines are loess smoothers with bandwidth 3/4.}
	\label{fig:covg}
\end{figure}

Our hypotheses are confirmed in \Cref{tab:sims3}, which presents results for settings with 3 control states (see \Cref{appendix:sims} for 10 controls). Over a variety of scenarios, we see very small bias in the estimated correlation. Of primary interest, however, are the corrected standard errors of the inverse variance weighted average ATT over both simulated treated states. Both $\widehat{\ATT}_\text{ivw}$ and $\widehat{\ATT}_\text{gls}$ are unbiased, but have different standard errors, and therefore coverage (\Cref{fig:covg}). As the magnitude of the correlation increases, uncorrected (IVW-based) coverage suffers. When correlations are negative, the uncorrected standard errors of $\widehat{\ATT}_\text{ivw}$ are too large; when positive, too small. The correlation-corrected intervals around $\widehat{\ATT}_\text{gls}$, though, achieve nominal coverage across scenarios. 

We turn now to the medical cannabis laws study. In order to estimate the
between-ATT correlations due to shared control individuals, we need
estimates of the within- and between-person correlations
$\rho_\gamma$, $\phi_\gamma$ and $\psi_\gamma$ for each state
$\gamma\in\Xi$. While the formulas in \Cref{eq:didCov,eq:varATThat}
allow for state-specific correlation estimates, obtaining such estimates
can be challenging, particularly in larger datasets: the medical
cannabis laws study's analytic sample consists of 583,820 individuals
measured for at least 84 months (treated and disjoint control individuals contribute exactly 84; shared control individuals contribute strictly more than 84). As such, we follow the literature for cluster-randomized trials
\citep{ouyangEstimatingIntraclusterCorrelation2023, kaszaImpactNonuniformCorrelation2019}, and estimate these
correlations averaged over all states using a mixed-effects modeling
approach, specifically by fitting the model 
\begin{equation}
	\label{eq:mixedModel}
	Y_{\gamma i t} = \beta_0 + \beta_{1,t} + \beta_2 A_{\gamma t} + b_{i} +
	b_{\gamma} + b_{\gamma t} + \epsilon_{\gamma it},
\end{equation}
where $A_{\gamma t}$ is a treatment indicator  in \Cref{eq:TWFE}, $b_{i}$ is a person-specific random effect, $b_{\gamma}$ a state-specific random effect, and $b_{\gamma t}$ a state-time-specific random effect as in \Cref{eq:genModel}. Correlation estimates are functions of estimates of the variances of the random effects; see Table 1 of \citet{ouyangEstimatingIntraclusterCorrelation2023} for details.

For this reanalysis, we focus on three outcomes: the proportion of
chronic non-cancer pain patients receiving any opioid prescription in a
given month; the proportion receiving any non-cannabis, non-opioid prescription analgesic in a given month; and the proportion receiving any procedure
for chronic pain in a given month. These are state-time outcomes
aggregated up from individual-level data, and are reported in percentage
points. The block exchangeable correlation structure in
\Cref{eq:sigma_s} is scientifically reasonable here, as we expect
within- and between-person relationships in the outcomes of interest to
be relatively stable over time in the population of interest
(commercially-insured U.S. adults with a chronic non-cancer pain
diagnosis). \Cref{tbl:rho} contains estimates of $\rho$, $\phi$, and $\psi$
on average across all states for each outcome, averaged across 100
resamples of 0.1\% of the 583,820 individuals in the full sample such
that each resample is balanced in the number of individuals per state.
Note that our assumption that $\rho > \phi > \psi$ appears valid based
on these estimates. In \Cref{appendix:medical-cannabis-law-study} we present
per-cohort sample sizes as well as counts of disjoint and shared control
individuals per cohort pair.

\begin{table}
	\centering
	\begin{tabular}{lrrr}
		\toprule
		& $\hat{\rho}$ & $\hat{\phi}$ & $\hat{\psi}$ \\
		\midrule
		Any Opioid Rx & 0.463 & 0.024 & 0.023 \\
		Any Non-Opioid Rx & 0.318 & 0.014 & 0.013 \\
		Any Procedure & 0.181 & 0.006 & 0.004 \\ 
		\bottomrule
	\end{tabular}
	\caption{Estimated within- and between-person correlations in the medical cannabis laws study.}
 	\label{tbl:rho}
\end{table}

\begin{figure}
	\centering
\begin{tikzpicture}[x=1pt,y=1pt]
\definecolor{fillColor}{RGB}{255,255,255}
\path[use as bounding box,fill=fillColor,fill opacity=0.00] (0,0) rectangle (433.62,216.81);
\begin{scope}
\path[clip] ( 49.20, 25.20) rectangle (408.42,191.61);
\definecolor{fillColor}{RGB}{211,211,211}

\path[fill=fillColor] ( 73.59, 97.34) --
	(162.29, 97.34) --
	(162.29,137.81) --
	( 73.59,137.81) --
	cycle;
\definecolor{drawColor}{RGB}{0,0,0}

\path[draw=drawColor,line width= 1.2pt,line join=round] ( 73.59,116.84) -- (162.29,116.84);

\path[draw=drawColor,line width= 0.4pt,dash pattern=on 4pt off 4pt ,line join=round,line cap=round] (117.94, 72.76) -- (117.94, 97.34);

\path[draw=drawColor,line width= 0.4pt,dash pattern=on 4pt off 4pt ,line join=round,line cap=round] (117.94,188.97) -- (117.94,137.81);

\path[draw=drawColor,line width= 0.4pt,line join=round,line cap=round] ( 95.77, 72.76) -- (140.11, 72.76);

\path[draw=drawColor,line width= 0.4pt,line join=round,line cap=round] ( 95.77,188.97) -- (140.11,188.97);

\path[draw=drawColor,line width= 0.4pt,line join=round,line cap=round] ( 73.59, 97.34) --
	(162.29, 97.34) --
	(162.29,137.81) --
	( 73.59,137.81) --
	cycle;

\path[fill=fillColor] (184.46, 99.25) --
	(273.16, 99.25) --
	(273.16,134.72) --
	(184.46,134.72) --
	cycle;

\path[draw=drawColor,line width= 1.2pt,line join=round] (184.46,115.44) -- (273.16,115.44);

\path[draw=drawColor,line width= 0.4pt,dash pattern=on 4pt off 4pt ,line join=round,line cap=round] (228.81, 77.41) -- (228.81, 99.25);

\path[draw=drawColor,line width= 0.4pt,dash pattern=on 4pt off 4pt ,line join=round,line cap=round] (228.81,175.36) -- (228.81,134.72);

\path[draw=drawColor,line width= 0.4pt,line join=round,line cap=round] (206.64, 77.41) -- (250.98, 77.41);

\path[draw=drawColor,line width= 0.4pt,line join=round,line cap=round] (206.64,175.36) -- (250.98,175.36);

\path[draw=drawColor,line width= 0.4pt,line join=round,line cap=round] (184.46, 99.25) --
	(273.16, 99.25) --
	(273.16,134.72) --
	(184.46,134.72) --
	cycle;

\path[fill=fillColor] (295.33, 96.72) --
	(384.03, 96.72) --
	(384.03,135.54) --
	(295.33,135.54) --
	cycle;

\path[draw=drawColor,line width= 1.2pt,line join=round] (295.33,116.46) -- (384.03,116.46);

\path[draw=drawColor,line width= 0.4pt,dash pattern=on 4pt off 4pt ,line join=round,line cap=round] (339.68, 91.17) -- (339.68, 96.72);

\path[draw=drawColor,line width= 0.4pt,dash pattern=on 4pt off 4pt ,line join=round,line cap=round] (339.68,150.24) -- (339.68,135.54);

\path[draw=drawColor,line width= 0.4pt,line join=round,line cap=round] (317.51, 91.17) -- (361.85, 91.17);

\path[draw=drawColor,line width= 0.4pt,line join=round,line cap=round] (317.51,150.24) -- (361.85,150.24);

\path[draw=drawColor,line width= 0.4pt,line join=round,line cap=round] (295.33, 96.72) --
	(384.03, 96.72) --
	(384.03,135.54) --
	(295.33,135.54) --
	cycle;
\end{scope}
\begin{scope}
\path[clip] (  0.00,  0.00) rectangle (433.62,216.81);
\definecolor{drawColor}{RGB}{0,0,0}

\path[draw=drawColor,line width= 0.4pt,line join=round,line cap=round] (117.94, 25.20) -- (339.68, 25.20);

\path[draw=drawColor,line width= 0.4pt,line join=round,line cap=round] (117.94, 25.20) -- (117.94, 19.20);

\path[draw=drawColor,line width= 0.4pt,line join=round,line cap=round] (228.81, 25.20) -- (228.81, 19.20);

\path[draw=drawColor,line width= 0.4pt,line join=round,line cap=round] (339.68, 25.20) -- (339.68, 19.20);

\node[text=drawColor,anchor=base,inner sep=0pt, outer sep=0pt, scale=  1.00] at (117.94,  3.60) {Any Opioid Rx};

\node[text=drawColor,anchor=base,inner sep=0pt, outer sep=0pt, scale=  1.00] at (228.81,  3.60) {Any Non-Opioid Rx};

\node[text=drawColor,anchor=base,inner sep=0pt, outer sep=0pt, scale=  1.00] at (339.68,  3.60) {Any Procedure};

\path[draw=drawColor,line width= 0.4pt,line join=round,line cap=round] ( 49.20, 44.20) -- ( 49.20,172.61);

\path[draw=drawColor,line width= 0.4pt,line join=round,line cap=round] ( 49.20, 44.20) -- ( 43.20, 44.20);

\path[draw=drawColor,line width= 0.4pt,line join=round,line cap=round] ( 49.20, 65.60) -- ( 43.20, 65.60);

\path[draw=drawColor,line width= 0.4pt,line join=round,line cap=round] ( 49.20, 87.00) -- ( 43.20, 87.00);

\path[draw=drawColor,line width= 0.4pt,line join=round,line cap=round] ( 49.20,108.41) -- ( 43.20,108.41);

\path[draw=drawColor,line width= 0.4pt,line join=round,line cap=round] ( 49.20,129.81) -- ( 43.20,129.81);

\path[draw=drawColor,line width= 0.4pt,line join=round,line cap=round] ( 49.20,151.21) -- ( 43.20,151.21);

\path[draw=drawColor,line width= 0.4pt,line join=round,line cap=round] ( 49.20,172.61) -- ( 43.20,172.61);

\node[text=drawColor,rotate= 90.00,anchor=base,inner sep=0pt, outer sep=0pt, scale=  1.00] at ( 34.80, 44.20) {-0.15};

\node[text=drawColor,rotate= 90.00,anchor=base,inner sep=0pt, outer sep=0pt, scale=  1.00] at ( 34.80, 87.00) {-0.05};

\node[text=drawColor,rotate= 90.00,anchor=base,inner sep=0pt, outer sep=0pt, scale=  1.00] at ( 34.80,129.81) {0.05};

\node[text=drawColor,rotate= 90.00,anchor=base,inner sep=0pt, outer sep=0pt, scale=  1.00] at ( 34.80,172.61) {0.15};
\end{scope}
\begin{scope}
\path[clip] (  0.00,  0.00) rectangle (433.62,216.81);
\definecolor{drawColor}{RGB}{0,0,0}

\path[draw=drawColor,line width= 0.4pt,line join=round,line cap=round] ( 49.20, 25.20) --
	(408.42, 25.20) --
	(408.42,191.61) --
	( 49.20,191.61) --
	cycle;
\end{scope}
\begin{scope}
\path[clip] (  0.00,  0.00) rectangle (433.62,216.81);
\definecolor{drawColor}{RGB}{0,0,0}

\node[text=drawColor,rotate= 90.00,anchor=base,inner sep=0pt, outer sep=0pt, scale=  1.00] at ( 10.80,108.41) {$\cor{\widehat{\mathrm{ATT}}_\gamma}{\widehat{\mathrm{ATT}}_\nu}$};
\end{scope}
\end{tikzpicture}
	\caption{Boxplots of computed correlations between state-specific ATT
		estimates for each outcome in the medical cannabis laws study; $(\gamma,\nu)\in\Xi_\text{tx} \times \Xi_\text{tx}$ where $\Xi_\text{tx} = \{\text{CT}, \ldots, \text{LA}\}$.}
	\label{fig:corBoxplots}
\end{figure}
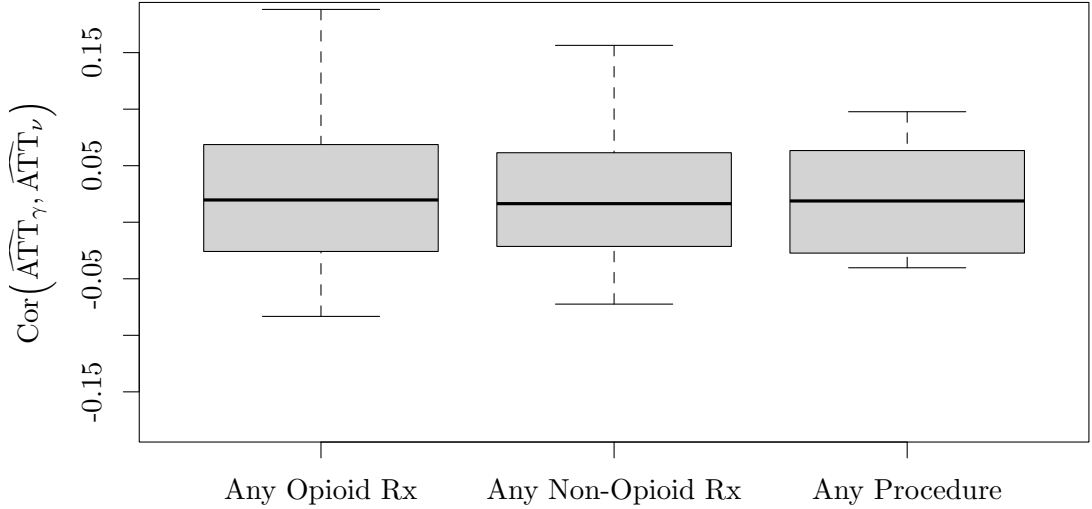

Using the estimates in \Cref{tbl:rho}, we can compute pairwise correlations
between state-specific $\widehat{\ATT}$s, which are summarized
for each outcome in \Cref{fig:corBoxplots}. We note that the
correlations are not centered at zero. In the medical cannabis laws
study, $T_\text{pre} = 48$ and $T_\text{post} = 36$, correlations
are positive for $\Delta < \Delta^* = 27.2$ months, and the median
spacing between cohorts is 21.5 months. Also note that some of the
correlations are quite sizeable: values for the outcome indicating
receipt of any opioid prescription in a given month range from -0.095 to
0.185.

\begin{figure}
	\centering
\begin{tikzpicture}[x=1pt,y=1pt]
\definecolor{fillColor}{RGB}{255,255,255}
\path[use as bounding box,fill=fillColor,fill opacity=0.00] (0,0) rectangle (361.35,252.94);
\begin{scope}
\path[clip] ( 61.20, 61.20) rectangle (348.15,239.75);
\definecolor{drawColor}{RGB}{169,169,169}
\definecolor{fillColor}{RGB}{169,169,169}

\path[draw=drawColor,line width= 0.4pt,line join=round,line cap=round,fill=fillColor] ( 69.83,152.45) rectangle ( 73.82,156.44);
\definecolor{drawColor}{RGB}{0,0,0}
\definecolor{fillColor}{RGB}{0,0,0}

\path[draw=drawColor,line width= 0.4pt,line join=round,line cap=round,fill=fillColor] (109.78,150.85) --
	(112.60,153.67) --
	(109.78,156.49) --
	(106.96,153.67) --
	cycle;
\definecolor{drawColor}{RGB}{169,169,169}
\definecolor{fillColor}{RGB}{169,169,169}

\path[draw=drawColor,line width= 0.4pt,line join=round,line cap=round,fill=fillColor] (183.70,152.73) rectangle (187.69,156.72);
\definecolor{drawColor}{RGB}{0,0,0}
\definecolor{fillColor}{RGB}{0,0,0}

\path[draw=drawColor,line width= 0.4pt,line join=round,line cap=round,fill=fillColor] (223.65,150.82) --
	(226.47,153.64) --
	(223.65,156.46) --
	(220.83,153.64) --
	cycle;
\definecolor{drawColor}{RGB}{169,169,169}
\definecolor{fillColor}{RGB}{169,169,169}

\path[draw=drawColor,line width= 0.4pt,line join=round,line cap=round,fill=fillColor] (297.57,134.63) rectangle (301.56,138.61);
\definecolor{drawColor}{RGB}{0,0,0}
\definecolor{fillColor}{RGB}{0,0,0}

\path[draw=drawColor,line width= 0.4pt,line join=round,line cap=round,fill=fillColor] (337.52,133.87) --
	(340.34,136.69) --
	(337.52,139.51) --
	(334.70,136.69) --
	cycle;
\end{scope}
\begin{scope}
\path[clip] (  0.00,  0.00) rectangle (361.35,252.94);
\definecolor{drawColor}{RGB}{0,0,0}

\path[draw=drawColor,line width= 0.4pt,line join=round,line cap=round] ( 61.20, 67.81) -- ( 61.20,233.13);

\path[draw=drawColor,line width= 0.4pt,line join=round,line cap=round] ( 61.20, 67.81) -- ( 55.20, 67.81);

\path[draw=drawColor,line width= 0.4pt,line join=round,line cap=round] ( 61.20,109.14) -- ( 55.20,109.14);

\path[draw=drawColor,line width= 0.4pt,line join=round,line cap=round] ( 61.20,150.47) -- ( 55.20,150.47);

\path[draw=drawColor,line width= 0.4pt,line join=round,line cap=round] ( 61.20,191.80) -- ( 55.20,191.80);

\path[draw=drawColor,line width= 0.4pt,line join=round,line cap=round] ( 61.20,233.13) -- ( 55.20,233.13);

\node[text=drawColor,rotate= 90.00,anchor=base,inner sep=0pt, outer sep=0pt, scale=  1.00] at ( 46.80, 67.81) {-1.0};

\node[text=drawColor,rotate= 90.00,anchor=base,inner sep=0pt, outer sep=0pt, scale=  1.00] at ( 46.80,109.14) {-0.5};

\node[text=drawColor,rotate= 90.00,anchor=base,inner sep=0pt, outer sep=0pt, scale=  1.00] at ( 46.80,150.47) {0.0};

\node[text=drawColor,rotate= 90.00,anchor=base,inner sep=0pt, outer sep=0pt, scale=  1.00] at ( 46.80,191.80) {0.5};

\node[text=drawColor,rotate= 90.00,anchor=base,inner sep=0pt, outer sep=0pt, scale=  1.00] at ( 46.80,233.13) {1.0};

\path[draw=drawColor,line width= 0.4pt,line join=round,line cap=round] ( 61.20, 61.20) --
	(348.15, 61.20) --
	(348.15,239.75) --
	( 61.20,239.75) --
	cycle;
\end{scope}
\begin{scope}
\path[clip] (  0.00,  0.00) rectangle (361.35,252.94);
\definecolor{drawColor}{RGB}{0,0,0}

\node[text=drawColor,anchor=base,inner sep=0pt, outer sep=0pt, scale=  1.00] at (204.67, 15.60) {Type of Chronic Non-Cancer Pain Treatment};

\node[text=drawColor,rotate= 90.00,anchor=base,inner sep=0pt, outer sep=0pt, scale=  1.00] at ( 10.80,150.47) {Percentage Point Difference in the Proportion of};

\node[text=drawColor,rotate= 90.00,anchor=base,inner sep=0pt, outer sep=0pt, scale=  1.00] at ( 22.80,150.47) {Patients Recieving Treatment Attributable to the Law};
\end{scope}
\begin{scope}
\path[clip] ( 61.20, 61.20) rectangle (348.15,239.75);
\definecolor{drawColor}{RGB}{169,169,169}

\path[draw=drawColor,line width= 0.8pt,line join=round,line cap=round] ( 71.83,140.77) -- ( 71.83,168.12);

\path[draw=drawColor,line width= 0.8pt,line join=round,line cap=round] (185.70,140.07) -- (185.70,169.37);

\path[draw=drawColor,line width= 0.8pt,line join=round,line cap=round] (299.57,116.10) -- (299.57,157.14);
\definecolor{drawColor}{RGB}{0,0,0}

\path[draw=drawColor,line width= 0.8pt,line join=round,line cap=round] (109.78,138.59) -- (109.78,168.75);

\path[draw=drawColor,line width= 0.8pt,line join=round,line cap=round] (223.65,137.34) -- (223.65,169.94);

\path[draw=drawColor,line width= 0.8pt,line join=round,line cap=round] (337.52,114.13) -- (337.52,159.26);
\end{scope}
\begin{scope}
\path[clip] (  0.00,  0.00) rectangle (361.35,252.94);
\definecolor{drawColor}{RGB}{0,0,0}

\node[text=drawColor,anchor=base,inner sep=0pt, outer sep=0pt, scale=  1.00] at ( 90.81, 48.78) {Any Opioid Rx};

\node[text=drawColor,anchor=base,inner sep=0pt, outer sep=0pt, scale=  1.00] at (204.67, 48.78) {Any Non-Opioid Rx};

\node[text=drawColor,anchor=base,inner sep=0pt, outer sep=0pt, scale=  1.00] at (318.54, 48.78) {Any Procedure};
\end{scope}
\begin{scope}
\path[clip] ( 61.20, 61.20) rectangle (348.15,239.75);
\definecolor{drawColor}{RGB}{0,0,0}

\path[draw=drawColor,line width= 0.4pt,line join=round,line cap=round] (272.54,239.75) rectangle (348.15,203.75);
\definecolor{fillColor}{RGB}{169,169,169}

\path[fill=fillColor] (279.54,225.75) rectangle (283.53,229.74);
\definecolor{fillColor}{RGB}{0,0,0}

\path[fill=fillColor] (281.54,212.93) --
	(284.36,215.75) --
	(281.54,218.56) --
	(278.72,215.75) --
	cycle;

\node[text=drawColor,anchor=base west,inner sep=0pt, outer sep=0pt, scale=  1.00] at (290.54,224.33) {Uncorrected};

\node[text=drawColor,anchor=base west,inner sep=0pt, outer sep=0pt, scale=  1.00] at (290.54,212.33) {Corrected};
\end{scope}
\end{tikzpicture}
	\caption{Percentage point differences in proportions of patients
	receiving chronic non-cancer pain treatment attributable to medical 		cannabis laws on average over the first 3 years of law implementation, 		with and without the correction for correlation due to shared controls. Any opioid Rx: McGinty et al. reported an estimated effect of 0.05 percentage points (95\% CI, -0.12 to 0.21 percentage points); our correlation-corrected estimate is 0.04 percentage points (95\% CI, -0.14 to 0.22 percentage points). Any non-opioid Rx: McGinty et al., 0.05 percentage points (95\% CI, -0.13 to 0.23 percentage points); corrected, 0.04 percentage points (95\% CI -0.16 to 0.24 percentage points). Any procedure: McGinty et al., -0.17 percentage points (95\% CI -0.42 to 0.08 percentage points); corrected, -0.17 percentage points (95\% CI -0.44, 0.11 percentage points).}
	\label{fig:cannabis}
\end{figure}
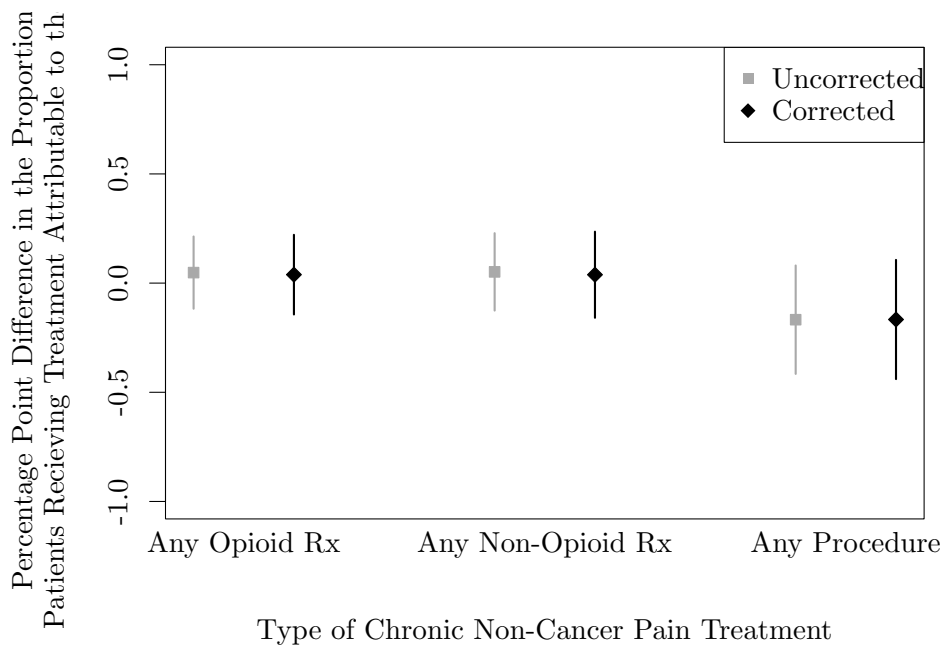

In our reanalysis, updated point estimates are not meaningfully different from those reported by \citet{mcgintyEffectsStateMedical2023}, but the standard errors that account for correlation induced by control individuals shared over the 12 stacked DiD analyses are larger (and thus confidence intervals are wider); inference remains the same. After accounting for correlation due to shared control individuals, we estimate an average difference of 0.04 percentage points (95\% CI -0.14 to 0.22 percentage points), 0.04 percentage points (CI -0.16 to 0.24 percentage points), and -0.17 percentage points (CI -0.44, 0.11 percentage points) in the proportion of individuals receiving any opioid prescription, any nonopioid prescription, and any chronic pain proceudre, respectively, attributable to state medical cannabis laws in a given month during the first three years of law implementation. These effects are small, and the confidence intervals remain narrow after accounting for between-estimate correlation, ruling out meaningful
effects of medical cannabis laws on chronic non-cancer pain treatment in
either direction. These results are both quantitatively and qualitatively similar to the unadjusted results in \citet{mcgintyEffectsStateMedical2023}; our correlation-corrected confidence intervals are 0.03, 0.04, and 0.05 percentage points (about 10\%) wider than the uncorrected intervals (see \Cref{fig:cannabis}, which recreates Figure 1 of \citet{mcgintyEffectsStateMedical2023}). An important note is that the original analysis uses the augmented synthetic control method to obtain state-specific $\widehat{\ATT}_\gamma$s: this approach is conceptually similar to DiD with weights on the control units \citep{ben-michaelAugmentedSyntheticControl2021}. While our current results do not accommodate such  weights, we  do not believe the substantive findings would not meaningfully change by doing so.

\section{Discussion}\label{sec:discussion}

We have developed a method for correcting standard errors for average
ATTs in a stacked DiD study in the presence of shared control
individuals and when individual-level data is available. The reuse of
data from individuals in control states across multiple stacked analyses
induces meaningful correlation between estimates that can be quantified
and accounted for when pooling effect estimates. Assuming a
block-exchangeable within- and between-person correlation structure, the
sign of the correlation between two DiD effect estimates is entirely
determined by timing -- the duration of the units' study periods and how
much or little overlap there is in those study periods in calendar time.
Failure to account for this correlation can lead to over- or
under-estimation of the standard error of the pooled ATT estimate,
depending on the sign of the correlation.

We note that this method applies only in scenarios in which scientific
interest is in a pooled ATT estimate averaged across multiple treated
units. In the medical cannabis laws example, we found no evidence of effect
heterogeneity across policy-implementing states, and therefore
comfortably pooled those results. In the presence of meaningful
differences in effects, scientists may not want to focus on an average
ATT, and instead report unit-specific $\widehat{\ATT}_\gamma$s
alongside deeper qualitative analysis of the results.

Though we found no substantive difference in corrected and uncorrected standard errors in the medical cannabis laws study, we believe the correction could be quite impactful in other settings in which states implement policies much closer together or farther apart in time (i.e., when $\Delta$s are consistently small or large). As an example, if we pooled over just CT, MN, and NY, the width of the corrected confidence interval for the proportion of patients receiving any opioid prescription would more than triple compared to the uncorrected interval (0.33 percentage points versus 1.06 percentage points). In a setting with a true policy effect, this may be enough to change inference. Additionally, we may see more meaningful changes in the medical cannabis law confidence intervals under a different correlation structure.

An important limitation is the imposition of a particular correlation
structure, though its choice was motivated by similar work in the
cluster-randomized trials literature
\citep{kaszaImpactNonuniformCorrelation2019, ouyangAccountingComplexIntracluster2023, ouyangEstimatingIntraclusterCorrelation2023}. The formulae presented
above do not, for instance, allow for decaying within- and
between-person correlations over time. An alternative approach might be
the bootstrap; however, it is unclear how one should perform resampling
in this setting to preserve the data structure involving shared
controls. Bootstrapping the entire dataset for all states simultaneously was  computationally infeasible for the medical cannabis laws study; this difficulty will likely translate to most policy evaluations with individual-level data. Pairwise resampling for each combination of cohorts would be more computationally feasible and adapting a cluster bootstrap to this setting may be promising, though future work is needed. We therefore trade an assumption on correlation structure for interpretability --- in that we developed a closed form of the between-estimate correlation --- and speed.

Future work will accommodate additional correlation structures,
including those with autoregressive decay over time, and incorporate
synthetic control weights to more accurately reflect realistic
correlation structures in data used for health policy evaluation.
Additionally, software to compute the between-estimate correlation is
available at \href{https://github.com/nickseewald/didsharedctrls}{https://github.com/nickseewald/didsharedctrls}.

\bibliography{shared-controls}

\appendix
\singlespacing
\newgeometry{margin=.5in}

\section{Derivation of DiD Correlation Formula}
\label{appendix:derivation}

We continue with the notation established in the main text. Consider two treated states $\gamma, \nu \in \Xi_\text{tx}$, each of which have study periods of length $T = \Tpre + \Tpost$ which begin $\Delta = \left|t_{*\gamma} - t_{*\nu}\right|$ periods apart. Recall the plug-in DiD estimator for treated state $\gamma$:
\begin{equation}
	\label{eq:did-estimator}
	\widehat{\mathrm{ATT}}_\gamma = 
	\left(\bar{Y}_{\gamma, \{t\geq t_{*\gamma}\}} -
	\bar{Y}_{\gamma, \{t< t_{*\gamma}\}}\right) -
	\left(\bar{Y}_{\text{ctrl,} \{t\geq t_{*\gamma}\}} -
	\bar{Y}_{\text{ctrl,} \{t< t_{*\gamma}\}}\right),
\end{equation}
where, for example,
\begin{equation}
	\bar{Y}_{\text{ctrl,} \{t\geq t_*\}} = \frac{1}{N_\gamma^\text{ctrl} \Tpost} \sum_{\zeta \in \Xi_\text{ctrl}} \sum_{t=t_{*\gamma}}^{t_{T\gamma}}  \sum_{i=1}^{N_\gamma(\zeta)}  Y_{\zeta it}
\end{equation}
is the mean outcome over all control individuals in cohort $\mathcal{C}_\gamma$ at all post-treatment measurement occasions.

For simplicity, we assume a block exchangeable correlation structure for all states. For two individuals $i$, $j$ in the same state $\gamma\in\Xi$ and timepoints $t$ and $s$, $\cor{Y_{\gamma it}}{Y_{\gamma is}} = \rho_\gamma$ (within-person correlation), $\cor{Y_{\gamma it}}{Y_{\gamma jt}} = \phi_\gamma$ (within-period correlation), and $\cor{Y_{\gamma it}}{Y_{\gamma js}} = \psi_\gamma$ (between-period correlation). For a state $\gamma$, then, the correlation matrix for all observations is block-diagonal with $\operatorname{Exch}_{T}(\rho_\gamma)$ correlation matrices on the diagonal and off-diagonal blocks $\psi_\gamma\bm{1}_T\bm{1}^\top_T + (\phi_\gamma - \psi_xi)I_T$, where $\operatorname{Exch}_{T}(\rho)$ is a $T\times T$ matrix with 1's on the diagonal and all off-diagonal elements are $\rho$, $\bm{1}_T$ is a $T$-vector of 1's, and $I_T$ is the $T\times T$ identity matrix. This is depicted visually in \cref{eq:sigma_s}. 

\subsection{Derivation of DiD Variance} \label{appendix:did-var-derivation}

We start by deriving an expression for the variance of the DiD estimator of $ATT_\gamma$ in \cref{eq:did-estimator}. We have
\begin{multline}
	\label{eq:did-var}
	\var{\widehat{\mathrm{ATT}}_\gamma} = 
	\var{\bar{Y}_{\gamma, \{t\geq t_{*\gamma}\}}} + \var{\bar{Y}_{\gamma, \{t< t_{*\gamma}\}}} + \var{\bar{Y}_{\text{ctrl},\{t\geq t_{*\gamma}\}}} + 
	\var{\bar{Y}_{\text{ctrl},\{t< t_{*\gamma}\}}} \\
	-2\cov{\bar{Y}_{\gamma, \{t\geq t_{*\gamma}\}}}{\bar{Y}_{\gamma, \{t< t_{*\gamma}\}}} - 
	2\cov{\bar{Y}_{\text{ctrl}, \{t\geq t_{*\gamma}\}}}{\bar{Y}_{\text{ctrl}, \{t< t_{*\gamma}\}}}.
\end{multline}
Note that there are no covarances between $\gamma$ and control states since, under the assumption that individuals living in different states are independent of one another, they are identically zero.

For treated state $\gamma$,
\begin{align*}
	\var{\bar{Y}_{\gamma, \{t\geq t_{*\gamma}\}}} =& \var{\frac{1}{N_\gamma(\gamma) \Tpost} \sum_{i=1}^{N_\gamma(\gamma)}\sum_{t=t_{*\gamma}}^{t_{T\gamma}} Y_{\gamma i t}} \\
	=& \left(N_\gamma(\gamma) \Tpost\right)^{-2} \left(\sum_{i}\sum_{t \geq t_{*\gamma}}\var{Y_{\gamma it}} + \sum_{i}\sum_{t\neq t' \geq t_{*\gamma}} \cov{ Y_{\gamma it}}{Y_{\gamma it'}}\right. \\
	&\phantom{\left(N_\gamma(\gamma) \Tpost\right)^{-2}} \quad  \left. + \sum_{i\neq j}\sum_{t \geq t_{*\gamma}}\cov{Y_{\gamma it}}{ Y_{\gamma jt}} +  
	\sum_{i\neq j}\sum_{t \neq t' \geq t_{*\gamma}} \cov{Y_{\gamma it}}{Y_{\gamma jt'}}\right) \\
	=& \left(N_\gamma(\gamma) \Tpost\right)^{-2} \left(N_\gamma(\gamma)\Tpost + N_\gamma(\gamma)\Tpost(\Tpost - 1)\rho_\gamma + \Tpost N_\gamma(\gamma)\left(N_\gamma(\gamma) - 1\right) \phi_\gamma + \right. \\
	&\phantom{\left(N_\gamma(\gamma) \Tpost\right)^{-2}} \quad
	\left.  + \Tpost (\Tpost - 1) N_\gamma(\gamma)\left(N_\gamma(\gamma) - 1\right) \psi_\gamma \right) \sigma^2_\gamma \\
	&= \left(N_\gamma(\gamma) \Tpost\right)^{-1} \left(1 + (\Tpost - 1) \rho_\gamma + \left(N_\gamma(\gamma) - 1\right)\left(\phi_\gamma + (\Tpost - 1)\psi_\gamma\right)\right) \sigma^2_\gamma. \numbereqn \label{eq:var-tx-post}
\end{align*}
Note that $\var{\bar{Y}_{\gamma, \{t< t_{*\gamma}\}}}$ is identical to the expression in \cref{eq:var-tx-post}, replacing $\Tpost$ with $\Tpre$.

Turning now to the control states, we have
\begin{align*}
	\var{\bar{Y}_{\text{ctrl},\{t\geq t_{*\gamma}\}}} =& \var{\frac{1}{N_\gamma^\text{ctrl} \Tpost} \sum_{\zeta \in \Xi_\text{ctrl}} \sum_{t=t_{*\gamma}}^{t_{T\gamma}} \sum_{i=1}^{N_\gamma(\zeta)}  Y_{\zeta it}} \numbereqn \label{eq:did-var-ctrl-step1} \\
	=& \left(N_\gamma^\text{ctrl} \Tpost\right)^{-2} \sum_{\zeta \in \Xi_\text{ctrl}} \left(\sum_{i}\sum_{t \geq t_{*\gamma}}\var{Y_{\gamma it}} + \sum_{i}\sum_{t\neq t' \geq t_{*\gamma}} \cov{ Y_{\gamma it}}{Y_{\gamma it'}}\right.  \\
	&\phantom{\left(N_\gamma^\text{ctrl} \Tpost\right)^{-2} \sum_{\zeta \in \Xi_\text{ctrl}}}  \left. + \sum_{i\neq j}\sum_{t \geq t_{*\gamma}}\cov{Y_{\gamma it}}{ Y_{\gamma jt}} +  
	\sum_{i\neq j}\sum_{t \neq t' \geq t_{*\gamma}} \cov{Y_{\gamma it}}{Y_{\gamma jt'}}\right) \numbereqn \label{eq:did-var-ctrl-step2} \\
	=& \left(N_\gamma^\text{ctrl} \Tpost\right)^{-2} \sum_{\zeta \in \Xi_\text{ctrl}} \left(N_\gamma(\zeta) \Tpost + N_\gamma(\zeta) \Tpost (\Tpost - 1) \rho_\zeta + N_\gamma(\zeta) \left(N_\gamma(\zeta) - 1\right) \Tpost \phi_\zeta \right. \\
	&\phantom{\left(N_\gamma^\text{ctrl} \Tpost\right)^{-2} \sum_{\zeta \in \Xi_\text{ctrl}}} \quad  \left. +  N_\gamma(\zeta) \left(N_\gamma(\zeta) - 1\right)\Tpost (\Tpost - 1) \psi_\zeta\right) \sigma^2_\zeta\\
	=& \left(N_\gamma^\text{ctrl} \Tpost\right)^{-2} \sum_{\zeta\in\Xi_\text{ctrl}} N_\gamma(\zeta) \Tpost \left(1 +\left(\Tpost-1\right)\rho_\zeta + \left(N_\gamma(\zeta) - 1\right)\left(\phi_\zeta + \left(\Tpost-1\right) \psi_\zeta\right)\right) \sigma^2_\zeta. \numbereqn \label{eq:var-ctrl-post}
\end{align*}
Note that \cref{eq:did-var-ctrl-step2} follows from \cref{eq:did-var-ctrl-step1} under the assumption that individuals living in different states are independent of one another. As with the treated states, the pre-treatment analogue of \cref{eq:var-ctrl-post} is identical, replacing $\Tpost$ with $\Tpre$. 

We now derive expressions for the covariance terms in \cref{eq:did-var} under the covariance structure on the data given in \cref{eq:sigma_s}. Note that for simplicity we use $(t_{*\gamma}-1)$ as shorthand to refer to the measurement occasion just prior to treatment initiation / policy implementation; we do not require that measurements be one time unit apart. We have
\begin{align*}
	\cov{\bar{Y}_{\gamma, \{t\geq t_{*\gamma}\}}}{\bar{Y}_{\gamma, \{t< t_{*\gamma}\}}} &= \cov{\frac{1}{N_\gamma(\gamma)\Tpost} \sum_{t=t_{*\gamma}}^{t_{T\gamma}} \sum_{i=1}^{N_{\gamma}(\gamma)} Y_{\gamma it}}{\frac{1}{N_\gamma(\gamma)\Tpre} \sum_{t=t_{1\gamma}}^{t_{*\gamma}-1} \sum_{i=1}^{N_{\gamma}(\gamma)} Y_{\gamma it}} \\
	&= \frac{1}{N_\gamma(\gamma)^2\Tpre\Tpost} \sum_{t=t_{1\gamma}}^{t_{*\gamma}-1} \sum_{t'=t_{*\gamma}}^{t_{T\gamma}} \left(\sum_{i=1}^{N_\gamma(\gamma)} \cov{Y_{\gamma it}}{Y_{\gamma it'}} + \sum_{i\neq j} \cov{Y_{\gamma it}}{Y_{\gamma jt'}}\right) \\
	&= \frac{1}{N_\gamma(\gamma)^2\Tpre\Tpost} \sum_{t=t_{1\gamma}}^{t_{*\gamma}-1} \sum_{t'=t_{*\gamma}}^{t_{T\gamma}} \left(N_\gamma(\gamma) \rho_\gamma + N_\gamma(\gamma) \left(N_\gamma(\gamma) - 1\right) \psi_\gamma\right)\sigma^2_\gamma \\
	&= \frac{1}{N_\gamma(\gamma)} \left(\rho_\gamma + \left(N_\gamma(\gamma) - 1\right) \psi_\gamma \right) \sigma^2_\gamma. \numbereqn \label{eq:did-var-tx-covar}
\end{align*}
Similarly, for the control states, we have
\begin{align}
	\cov{\bar{Y}_{\text{ctrl}, \{t\geq t_{*\gamma}\}}}{\bar{Y}_{\text{ctrl}, \{t< t_{*\gamma}\}}} &= \cov{\frac{1}{N_\gamma^\text{ctrl} \Tpost} \sum_{\zeta \in \Xi_\text{ctrl}} \sum_{t=t_{*\gamma}}^{t_{T\gamma}}  \sum_{i=1}^{N_\gamma(\zeta)}  Y_{\zeta it}}{\frac{1}{N_\gamma^\text{ctrl} \Tpost} \sum_{\zeta \in \Xi_\text{ctrl}} \sum_{t=t_{1\gamma}}^{t_{*\gamma}-1}  \sum_{i=1}^{N_\gamma(\zeta)}  Y_{\zeta it}} \label{eq:did-var-ctrl-covar-step1} \\
	&= \frac{1}{\left(N_\gamma^\text{ctrl}\right)^2 \Tpre\Tpost} \sum_{\zeta\in\Xi_\text{ctrl}} \sum_{t=t_{1\gamma}}^{t_{*\gamma}-1} \sum_{t'=t_{*\gamma}}^{t_{T\gamma}} \left(\sum_{i=1}^{N_\gamma(\zeta)} \cov{Y_{\zeta it}}{Y_{\zeta it'}} + \sum_{i\neq j} \cov{Y_{\zeta it}}{Y_{\zeta jt'}}\right) \label{eq:did-var-ctrl-covar-step2} \\
	&= \frac{1}{\left(N_\gamma^\text{ctrl}\right)^2 \Tpre\Tpost} \sum_{\zeta\in\Xi_\text{ctrl}} \sum_{t=t_{1\gamma}}^{t_{*\gamma}-1} \sum_{t'=t_{*\gamma}}^{t_{T\gamma}} \left(N_\gamma(\zeta) \rho_\zeta + N_\gamma(\zeta)\left(N_\gamma(\zeta) - 1\right) \psi_\zeta \right) \sigma^2_\zeta \nonumber \\
	&= \frac{1}{\left(N_\gamma^\text{ctrl}\right)^2} \sum_{\zeta\in\Xi_\text{ctrl}} \left(N_\gamma(\zeta) \rho_\zeta + N_\gamma(\zeta)\left(N_\gamma(\zeta) - 1\right) \psi_\zeta \right) \sigma^2_\zeta, \label{eq:did-var-ctrl-covar}
\end{align}
with \cref{eq:did-var-ctrl-covar-step2} following from \cref{eq:did-var-ctrl-covar-step1} under independence of states.

Now, using \cref{eq:var-tx-post,eq:did-var-tx-covar} (and the pre-treatment analogue of \cref{eq:var-tx-post}), we have 
\begin{align*}
	&\var{\bar{Y}_{\gamma, \{t\geq t_{*\gamma}\}}} + \var{\bar{Y}_{\gamma, \{t< t_{*\gamma}\}}} 
	-2\cov{\bar{Y}_{\gamma, \{t\geq t_{*\gamma}\}}}{\bar{Y}_{\gamma, \{t< t_{*\gamma}\}}} \\
	&{}\quad = \frac{ \sigma^2_\gamma}{\left(N_\gamma(\gamma) \Tpre \Tpost\right)^2}
	\left[N_\gamma(\gamma)\left(\Tpost^2 \Tpre + \Tpost^2 \Tpre^2 \rho_\gamma - \Tpost^2 \Tpre\rho_\gamma + \Tpost\Tpre^2 \right.\right. \\
	&{}\quad \phantom{= \frac{ \sigma^2_\gamma}{\left(N_\gamma(\gamma) \Tpre \Tpost\right)^2} \big[ N_\gamma(\gamma) } \quad
	\left.\left. + \Tpost^2\Tpre^2\rho_\gamma - \Tpost\Tpre^2\rho_\gamma - 2\Tpost^2\Tpre^2\rho_\gamma \right) \right. \\
	&{}\quad \phantom{= \frac{ \sigma^2_\gamma}{\left(N_\gamma(\gamma) \Tpre \Tpost\right)^2} \big[ } \quad 
	N_\gamma(\gamma)\left(N_\gamma(\gamma) - 1\right) \left(\Tpost^2\Tpre\phi_\gamma + \Tpost\Tpre^2\phi_\gamma + \Tpost^2\Tpre^2\psi_\gamma - \Tpost^2\Tpre\psi_\gamma \right.\\
	&{}\quad \phantom{= \frac{ \sigma^2_\gamma}{\left(N_\gamma(\gamma) \Tpre \Tpost\right)^2} \big[ N_\gamma(\gamma)\left(N_\gamma(\gamma) - 1\right) \big( } \quad \left.\left. + \Tpost^2\Tpre^2 \psi_\gamma - \Tpost\Tpre^2\psi_\gamma - 2\Tpre^2\Tpost^2\psi_\gamma\right)\right] \\
	&{}\quad = \frac{ \sigma^2_\gamma}{\left(N_\gamma(\gamma) \Tpre \Tpost\right)^2} \left(N_\gamma(\gamma)\Tpost\Tpre \left(\Tpre + \Tpost\right)\right) \left(\left(1-\rho\right) + \left(N_\gamma(\gamma) - 1\right) \left(\phi_\gamma - \psi_\gamma\right)\right) \\
	&{}\quad = \frac{\Tpre + \Tpost}{\left(N_\gamma(\gamma)\right)^2\Tpre\Tpost} \left(N_\gamma(\gamma)\left(1-\rho\right) + N_\gamma(\gamma)\left(N_\gamma(\gamma) - 1\right) \left(\phi_\gamma - \psi_\gamma\right)\right) \sigma^2_\gamma. \numbereqn \label{eq:did-var-tx-component-sum}
\end{align*}
Proceeding similarly with \cref{eq:var-ctrl-post,eq:did-var-ctrl-covar} (and the pre-treatment analogue of \cref{eq:var-ctrl-post}), we have
\begin{multline}
	\label{eq:did-var-ctrl-component-sum}
	\var{\bar{Y}_{\text{ctrl},\{t\geq t_{*\gamma}\}}} + 
	\var{\bar{Y}_{\text{ctrl},\{t< t_{*\gamma}\}}} -  2\cov{\bar{Y}_{\text{ctrl}, \{t\geq t_{*\gamma}\}}}{\bar{Y}_{\text{ctrl}, \{t< t_{*\gamma}\}}} \\
	= \frac{\Tpre + \Tpost}{\left(N_\gamma^\text{ctrl}\right)^2 \Tpre\Tpost} \sum_{\zeta \in \Xi_\text{ctrl}} \left(N_\gamma(\zeta) \left(1 - \rho_\zeta\right) + N_\gamma(\zeta) \left(N_\gamma(\zeta) - 1\right) \left(\phi_\zeta - \psi_\zeta\right)\right)\sigma^2_\zeta.
\end{multline}
Finally, summing \cref{eq:did-var-tx-component-sum,eq:did-var-ctrl-component-sum}, we have
\begin{equation}
	\boxed{
		\var{\widehat{\mathrm{ATT}}_\gamma} = \frac{\Tpre + \Tpost}{\Tpre\Tpost} \sum_{\zeta\in (\{\gamma\} \cup \Xi_\text{ctrl})} \frac{\sigma^2_\zeta}{\left(N_\gamma^{A_\gamma(\zeta)}\right)^2} \left[N_\gamma(\zeta) \left(1-\rho_\zeta\right) + N_\gamma(\zeta)\left(N_\gamma(\zeta) -1\right)\left(\phi_\zeta - \psi_\zeta\right)\right],}
\end{equation}
where $A_\zeta = \ind{\zeta\in\Xi_\text{tx}}$ is an indicator for whether state \(\zeta\) was ever treated such that $N_{\gamma}^{A_\zeta} = A_\zeta N_\gamma(\gamma) + (1-A_\zeta) N_\gamma^\text{ctrl}$. That is, if $\zeta\in\Xi_\text{ctrl}$, then	$N_\gamma^{A_\zeta} = N_\gamma^\text{ctrl}$ is the total number of 	individuals in control states that contribute to $\widehat{ATT}_\gamma$. Similarly, if \(\zeta\in\Xi_\text{tx}\) (i.e., $\zeta = \gamma)$, then 	$N_\gamma^{A_\zeta} = N_\gamma^\text{tx} = N_\gamma(\gamma)$.

\subsection{Derivation of Covariance between DiD Estimates} \label{appendix:did-covar-derivation}

Consider two cohorts, $\mathcal{C}_\gamma$ and $\mathcal{C}_\nu$, separated by $\Delta$ measurement occasions. By linearity of covariance and independence of individuals from different states, 
\begin{multline}
	\label{eq:covar-expansion-means}
	\cov{\widehat{\mathrm{ATT}}_\gamma}{\widehat{\mathrm{ATT}}_\nu} = 
	\cov{\bar{Y}_{\text{ctrl},\{t\geq t_{*\gamma}\}}}{\bar{Y}_{\text{ctrl},\{t\geq t_{*\nu}\}}} + \cov{\bar{Y}_{\text{ctrl},\{t< t_{*\gamma}\}}}{\bar{Y}_{\text{ctrl},\{t< t_{*\nu}\}}} \\ 
	- \cov{\bar{Y}_{\text{ctrl},\{t\geq t_{*\gamma}\}}}{\bar{Y}_{\text{ctrl},\{t < t_{*\nu}\}}} - \cov{\bar{Y}_{\text{ctrl},\{t< t_{*\gamma}\}}}{\bar{Y}_{\text{ctrl},\{t\geq t_{*\nu}\}}}.
\end{multline}
Notice that the covariance only involves data from control individuals: since individuals in treated states $\gamma$ and $\nu$ are independent of individuals in every other state, terms involving $\bar{Y}_{\gamma,\{t\}}$ and $\bar{Y}_{\nu,\{t\}}$ drop out of the covariance.

\begin{figure}
	\resizebox{\textwidth}{!}{\input{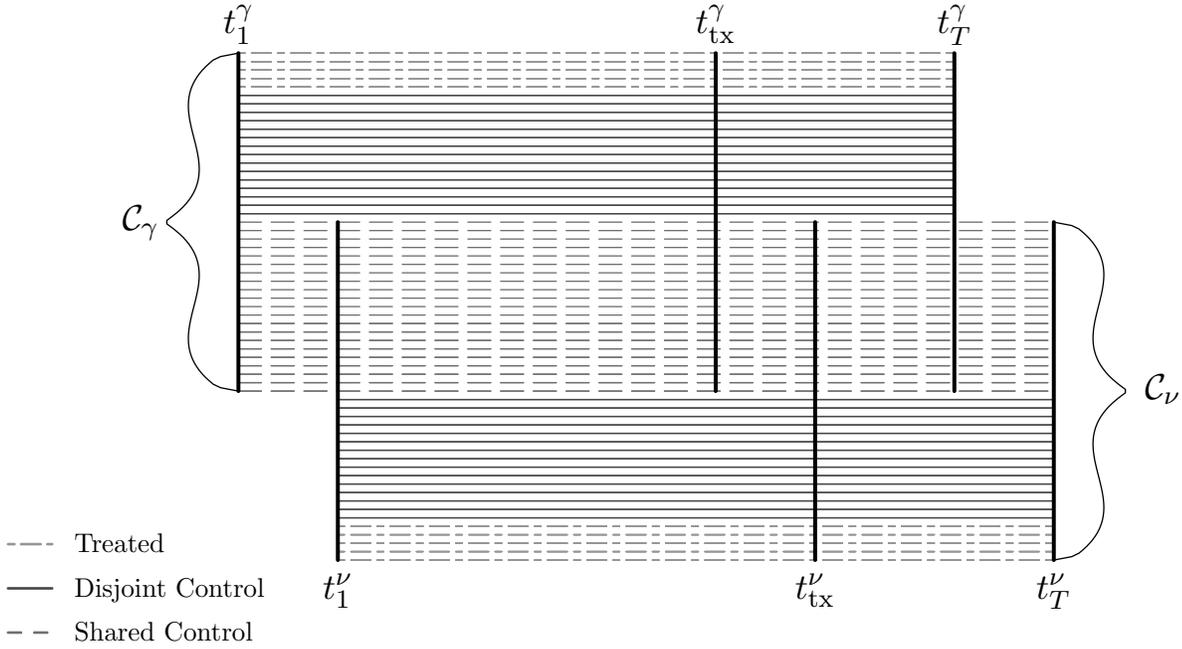}}
	\caption{A schematic depiction of two cohorts' overlapping study periods and shared control individuals. Horizontal lines represent individual ``timelines'' over which data are collected from that individual. Dot-dashed lines represent individuals in a treated state, solid lines represent individuals in a control state who contribute to exactly one of $\mathcal{C}_{\gamma}$ and $\mathcal{C}_{\nu}$ (``disjoint'' control individuals), and dashed lines shared control individuals.}
	\label{fig:two-cohorts}
\end{figure}

Consider \cref{fig:two-cohorts}, which depicts overlap between cohorts $\mathcal{C}_\gamma$ and $\mathcal{C}_\nu$ when $\Delta < \Tpost < \Tpre$. In the figure, dot-dashed lines represent study timelines for treated individuals in either state $\gamma$ or state $\nu$. Solid lines represent timelines for ``disjoint'' control individuals in some control state $\zeta \in \Xi_\text{ctrl}$ who contribute to \textit{either} the cohort for state $\gamma$ or for state $\nu$ but not both (i.e., individuals $i \in (\mathcal{I}_\gamma / \mathcal{I}_\gamma(\gamma)) \sqcup (\mathcal{I}_\nu / \mathcal{I}_\nu(\nu)))$), and dashed lines represent ``shared control'' individuals that contribute to cohorts for both $\gamma$ and $\nu$ (i.e., individuals $i\in (\mathcal{I}_\gamma \cap \mathcal{I}_\nu)$. Similar diagrams can be created for other configurations of $\Tpre$, $\Tpost$, and $\Delta$. 

We begin by dividing the person-time contributed by control individuals to each cohort into a number of disjoint ``windows''. Definitions of the windows are given in \cref{tab:windows}. Data from disjoint control individuals fall into one of 4 possible windows: pre- or post-treatment time in either $\mathcal{C}_\gamma$ or $\mathcal{C}_\nu$. As an example, we denote the set of \textit{disjoint} control person-time in $\mathcal{C}_\gamma$'s pre-treatment period as $\mathcal{D}_\text{pre}^{\mathcal{C}_\gamma/\mathcal{C}_\nu}$; the superscript is meant to invoke the set difference between the two cohorts.

\begingroup
\renewcommand{\arraystretch}{2}
\begin{table}
	\small
	\centering
	\caption{Descriptions of the 11 possible disjoint subdivisions (``windows'') of person-time contributed by control individuals in state $\zeta\in\Xi_\text{ctrl}$ to either cohort $\mathcal{C}_\gamma$ or $\mathcal{C}_\nu$, where $\gamma,\nu\in\Xi_\text{tx}$ and $t_{1\gamma} - t_{1\nu} = \Delta > 0$. Note that $(x)_+ := \max(x, 0)$.}
	\label{tab:windows}
	\begin{tabular}{lp{3.7in}c}
		\toprule
		Window & Definition & Time Duration \\
		\midrule 
		$\mathcal{D}^{\mathcal{C}_\gamma\backslash \mathcal{C}_\nu}_\text{pre}(\zeta)$ & 
		\raggedright Control person-time in state $\zeta$ contributing \textit{only} to $\mathcal{C}_\gamma$ in its pre-treatment period. \newline
		$\displaystyle \left\{(i, t) : i\in \mathcal{I}_{\mathcal{C}_\gamma}(\zeta) \backslash \mathcal{I}_{\mathcal{C}_\nu}(\zeta), t < t_{*\gamma}\right\}$ &
		$\Tpre$ \\
		$\mathcal{D}^{\mathcal{C}_\nu\backslash \mathcal{C}_\gamma}_\text{pre}(\zeta)$ &
		\raggedright Control person-time in state $\zeta$ contributing \textit{only} to $\mathcal{C}_\nu$ in its pre-treatment period.  \newline
		$\displaystyle \left\{(i, t) : i\in \mathcal{I}_{\mathcal{C}_\nu}(\zeta) \backslash \mathcal{I}_{\mathcal{C}_\gamma}(\zeta), t < t_{*\nu}\right\}$ &
		$\Tpre$ \\
		$\mathcal{D}^{\mathcal{C}_\gamma\backslash \mathcal{C}_\nu}_\text{post}(\zeta)$ &
		\raggedright Control person-time in state $\zeta$ contributing \textit{only} to $\mathcal{C}_\gamma$ in its post-treatment period. \newline
		$\displaystyle \left\{(i, t) : i\in \mathcal{I}_{\mathcal{C}_\gamma}(\zeta) \backslash \mathcal{I}_{\mathcal{C}_\nu}(\zeta), t \geq t_{*\gamma}\right\}$ &
		$\Tpost$ \\
		$\mathcal{D}^{\mathcal{C}_\nu\backslash \mathcal{C}_\gamma}_\text{post}(\zeta)$ &
		\raggedright Control person-time in state $\zeta$ contributing \textit{only} to $\mathcal{C}_\nu$ in its post-treatment period. \newline
		$\displaystyle \left\{(i, t) : i\in \mathcal{I}_{\mathcal{C}_\nu}(\zeta) \backslash \mathcal{I}_{\mathcal{C}_\gamma}(\zeta), t \geq t_{*\nu}\right\}$ &
		$\Tpost$ \\ \midrule
		$\mathcal{O}^{\mathcal{C}_\gamma,\mathcal{C}_\nu}_{\text{pre},\cdot}(\zeta)$ &
		\raggedright Person-time from shared controls in state $\zeta$ contributing to $\mathcal{C}_\gamma$ in its pre-treatment period while $\mathcal{C}_\nu$'s study period has not started. \newline
		$\displaystyle \left\{(i,t) : i\in \mathcal{I}_{\mathcal{C}_\gamma}(\zeta) \cap \mathcal{I}_{\mathcal{C}_\nu}(\zeta), t < t_{1\nu}\right\}$ &
		$\min\left(\Tpre, \Delta\right)$ \\
		$\mathcal{O}^{\mathcal{C}_\gamma,\mathcal{C}_\nu}_{\text{pre},\text{pre}}(\zeta)$ &
		\raggedright Person-time from shared controls in state $\zeta$ contributing to both $\mathcal{C}_\gamma$ and $\mathcal{C}_\nu$ in their pre-treatment periods.
		$\displaystyle \left\{(i,t) : i\in \mathcal{I}_{\mathcal{C}_\gamma}(\zeta) \cap \mathcal{I}_{\mathcal{C}_\nu}(\zeta), t_{1\nu} \leq t < t_{*\gamma}\right\}$ &
		$\left(\Tpre-\Delta\right)_+$ \\
		$\mathcal{O}^{\mathcal{C}_\gamma,\mathcal{C}_\nu}_{\text{post},\cdot}(\zeta)$ &
		\raggedright Person-time from shared controls in state $\zeta$ contributing to $\mathcal{C}_\gamma$ in its post-treatment period while $\mathcal{C}_\nu$'s study period has not started. \newline
		$\displaystyle \left\{(i,t) : i\in \mathcal{I}_{\mathcal{C}_\gamma}(\zeta) \cap \mathcal{I}_{\mathcal{C}_\nu}(\zeta), t_{*\gamma} \leq t < t_{1\nu}\right\}$ &
		$\min\left(\left(\Delta - \Tpre\right)_+, \Tpost\right)$ \\
		$\mathcal{O}^{\mathcal{C}_\gamma,\mathcal{C}_\nu}_{\text{post},\text{pre}}(\zeta)$ &
		\raggedright Person-time from shared controls in state $\zeta$ contributing to $\mathcal{C}_\gamma$ in its post-treatment period and $\mathcal{C}_\nu$ in its pre-treatment period. \newline
		$\displaystyle \left\{(i,t) : i\in \mathcal{I}_{\mathcal{C}_\gamma}(\zeta) \cap \mathcal{I}_{\mathcal{C}_\nu}(\zeta), \max\left(t_{*\gamma}, t_{1\nu}\right) \leq t < \min\left(t_{T\gamma}, t_{*\nu}\right)\right\}$ &
		$\begin{aligned}
			\min\bigg( & \Tpre, \Tpost, \Delta, \\[-.7em] & \left. \left(\Tpre + \Tpost - \Delta\right)_+\right)
		\end{aligned}$ \\
		$\mathcal{O}^{\mathcal{C}_\gamma,\mathcal{C}_\nu}_{\text{post},\text{post}}(\zeta)$ &
		\raggedright Person-time from shared controls in state $\zeta$ contributing to both $\mathcal{C}_\gamma$ and $\mathcal{C}_\nu$ in their post-treatment periods. \newline
		$\displaystyle \left\{(i,t) : i\in \mathcal{I}_{\mathcal{C}_\gamma}(\zeta) \cap \mathcal{I}_{\mathcal{C}_\nu}(\zeta), t_{*_\nu} \leq t \leq t_{*\gamma}\right\}$ &
		$\left(\Tpost - \Delta\right)_+$ \\
		$\mathcal{O}^{\mathcal{C}_\gamma,\mathcal{C}_\nu}_{\cdot,\text{pre}}(\zeta)$ &
		\raggedright Person-time from shared controls in state $\zeta$ contributing to $\mathcal{C}_\nu$ in its pre-treatment period after $\mathcal{C}_\gamma$'s study period has ended. \newline
		$\displaystyle \left\{(i,t) : i\in \mathcal{I}_{\mathcal{C}_\gamma}(\zeta) \cap \mathcal{I}_{\mathcal{C}_\nu}(\zeta), t_{T\gamma} < t < t_{*\nu}\right\}$ &
		$\left(\Delta - \Tpost\right)_+$ \\
		$\mathcal{O}^{\mathcal{C}_\gamma,\mathcal{C}_\nu}_{\cdot,\text{post}}(\zeta)$ &
		\raggedright Person-time from shared controls in state $\zeta$ contributing to $\mathcal{C}_\nu$ in its post-treatment period after $\mathcal{C}_\gamma$'s study period has ended. \newline
		$\displaystyle \left\{(i,t) : i\in \mathcal{I}_{\mathcal{C}_\gamma}(\zeta) \cap \mathcal{I}_{\mathcal{C}_\nu}(\zeta), \max\left(t_{T\gamma} + 1, t_{*\nu}\right) \leq t \leq t_{T\nu}\right\}$ &
		$\min\left(\Tpost, \Delta\right)$ \\
		\bottomrule
	\end{tabular}
\end{table}
\endgroup

\begin{figure}
	\resizebox{\textwidth}{!}{\input{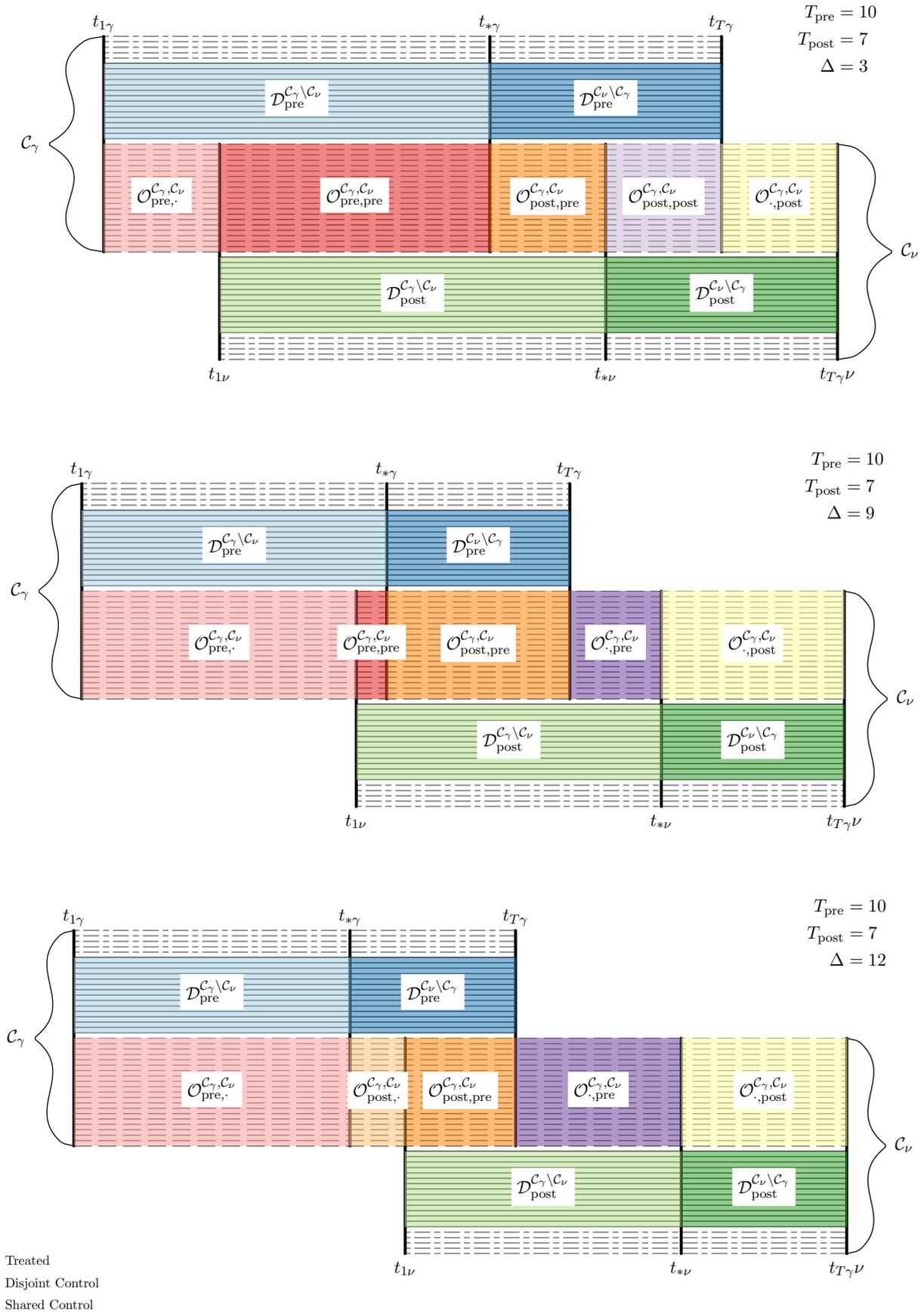}}
	\caption{Schematic depictions of overlapping cohorts with varying $\Delta$, highlighted to show person-time windows defined in \cref{tab:windows}. The value of $\Delta$ relative to $\Tpre$ and $\Tpost$ determines which windows are present in the shared control person-time.}
	\label{fig:two-cohorts-labeled-col-stack}
\end{figure}

The person-time shared between cohorts $\mathcal{C}_\gamma$ and $\mathcal{C}_\nu$ can be divided into either 4 or 5 mutually disjoint ``windows'' depending on the size of $\Delta$ relative to $\Tpre$ and $\Tpost$. These windows are based on the cohort's treatment status at a given time. There are 7 possible windows, of which at most 5 can be present in a particular configuration of $\Tpre$, $\Tpost$, and $\Delta$ (this is easily seen by inspecting the ``Time Duration'' column of \cref{tab:windows}). As an example, we denote the set of \textit{overlapping} control person-time when $\mathcal{C}_\gamma$ is in its post-treatment period and $\mathcal{C}_\nu$ is in its pre-treatment period as $\mathcal{O}^{\mathcal{C}_\gamma, \mathcal{C}_\nu}_{\text{post,pre}}$. In this notation, we use $\cdot$ to denote a period in which one of the cohorts' study periods has not started or has ended. \Cref{fig:two-cohorts-labeled-col-stack} shows three configurations of $\Tpre$, $\Tpost$, and $\Delta$ along with the person-time windows induced by each configuration.

Following the notation in \cref{tab:windows}, we write each mean in \cref{eq:covar-expansion-means} as an average over the person-time that contributes to that mean. For example, in the top diagram in \cref{fig:two-cohorts-labeled-col-stack}, with $\Delta = 3$, the pre-treatment mean outcome in the control individuals in cohort $\mathcal{C}_\gamma$, $\bar{Y}_{\text{ctrl},\{t<t_{*\gamma}\}}$, is an average over $\cup_{\zeta\in\Xi_\text{ctrl}} \mathcal{D}_\text{pre}^{\mathcal{C}_\gamma \backslash \mathcal{C}_\nu}(\zeta)$, $\cup_{\zeta\in\Xi_\text{ctrl}} \mathcal{O}_{\text{pre},\cdot}^{\mathcal{C}_\gamma, \mathcal{C}_\nu}(\zeta)$, and $\cup_{\zeta\in\Xi_\text{ctrl}}\mathcal{O}_{\text{pre},\text{pre}}^{\mathcal{C}_\gamma, \mathcal{C}_\nu}(\zeta)$. More specifically, we can write
\begin{equation}
	\label{eq:mean-as-sum-over-windows}
	\bar{Y}_{\text{ctrl},\{t<t_{*\gamma}\}} = 
	\frac{1}{\Tpre N_\gamma^\text{ctrl}}\sum_{\zeta\in\Xi_\text{ctrl}} \left[
	\sum_{(i,t)\in \mathcal{D}_\text{pre}^{\mathcal{C}_\gamma \backslash \mathcal{C}_\nu}(\zeta)} Y_{\zeta it} + \sum_{(i,t) \in \mathcal{O}_{\text{pre},\cdot}^{\mathcal{C}_\gamma, \mathcal{C}_\nu}(\zeta)} Y_{\zeta it} + \sum_{\mathcal{O}_{\text{pre},\text{pre}}^{\mathcal{C}_\gamma, \mathcal{C}_\nu}(\zeta)} Y_{\zeta it}
	\right].
\end{equation}
Similar expansions of each mean in the difference-in-differences estimator in \cref{eq:did-estimator} can be written over each of their component person-time windows listed in \cref{tab:mean-windows}.

\begingroup
\renewcommand{\arraystretch}{2}
\begin{table}
	\centering
	\caption{Windows of person-time composing each pre- and post-treatment mean in \cref{eq:covar-expansion-means}.}
	\label{tab:mean-windows}
	\begin{tabular}{lp{2in}l}
		\toprule 
		Mean & Description & Component Windows \\ \midrule
		$\bar{Y}_{\text{ctrl},\{t\geq t_{*\gamma}\}}$ &
		\raggedright Mean over control individuals in $\mathcal{C}_\gamma$ in its post-treatment period. & 
		$\displaystyle \bigcup_{\zeta\in\Xi_\text{ctrl}} \left(\mathcal{D}_\text{post}^{\mathcal{C}_\gamma \backslash \mathcal{C}_\nu}(\zeta) \cup
		\mathcal{O}_{\text{post}, \cdot}^{\mathcal{C}_\gamma, \mathcal{C}_\nu}(\zeta) \cup \mathcal{O}_\text{post,pre}^{\mathcal{C}_\gamma, \mathcal{C}_\nu}(\zeta) \cup
		\mathcal{O}_\text{post,post}^{\mathcal{C}_\gamma, \mathcal{C}_\nu}(\zeta)\right)$ \\
		$\bar{Y}_{\text{ctrl},\{t\geq t_{*\nu}\}}$ &
		\raggedright Mean over control individuals in $\mathcal{C}_\nu$ in its post-treatment period. & 
		$\displaystyle \bigcup_{\zeta\in\Xi_\text{ctrl}} \left(\mathcal{D}_\text{post}^{\mathcal{C}_\nu \backslash \mathcal{C}_\gamma}(\zeta) \cup
		\mathcal{O}_\text{post,post}^{\mathcal{C}_\gamma, \mathcal{C}_\nu}(\zeta) \cup
		\mathcal{O}_{\cdot,\text{post}}^{\mathcal{C}_\gamma, \mathcal{C}_\nu}(\zeta)\right)$ \\
		$\bar{Y}_{\text{ctrl},\{t< t_{*\gamma}\}}$ &
		\raggedright Mean over control individuals in $\mathcal{C}_\gamma$ in its pre-treatment period. & 
		$\displaystyle \bigcup_{\zeta\in\Xi_\text{ctrl}} \left( \mathcal{D}_\text{pre}^{\mathcal{C}_\gamma \backslash \mathcal{C}_\nu}(\zeta) \cup 
		\mathcal{O}_{\text{pre},\cdot}^{\mathcal{C}_\gamma, \mathcal{C}_\nu}(\zeta) \cup
		\mathcal{O}_\text{pre,pre}^{\mathcal{C}_\gamma, \mathcal{C}_\nu}\right)$\\
		$\bar{Y}_{\text{ctrl},\{t< t_{*\nu}\}}$ & 
		\raggedright Mean over control individuals in $\mathcal{C}_\nu$ in its pre-treatment period. & 
		$\displaystyle \bigcup_{\zeta\in\Xi_\text{ctrl}} \left(
		\mathcal{D}_\text{pre}^{\mathcal{C}_\nu \backslash \mathcal{C}_\gamma}(\zeta) \cup
		\mathcal{O}_{\text{pre,pre}}^{\mathcal{C}_\gamma, \mathcal{C}_\nu}(\zeta) \cup 
		\mathcal{O}_{\text{post,pre}}^{\mathcal{C}_\gamma, \mathcal{C}_\nu}(\zeta) \cup
		\mathcal{O}_{\cdot\text{,pre}}^{\mathcal{C}_\gamma, \mathcal{C}_\nu}(\zeta)\right) $ \\
		\bottomrule
	\end{tabular}
\end{table}
\endgroup

Start by considering the first summand on the right-hand side of \cref{eq:covar-expansion-means}, the covariance between post-treatment means among control individuals in cohorts $\mathcal{C}_\gamma$ and $\mathcal{C}_\nu$. When we expand each of these means as in \cref{eq:mean-as-sum-over-windows} and use bilinearity of covariance, we can write this as a sum of 12 covariances:
\begingroup
\small
\allowdisplaybreaks
\begin{align*}
	&\cov{\bar{Y}_{\text{ctrl},\{t\geq t_{*\gamma}\}}}{\bar{Y}_{\text{ctrl},\{t\geq t_{*\nu}\}}} = \\
	&\quad
	\left(N_\gamma^{\text{ctrl}} N_{\nu}^\text{ctrl} \Tpost^2\right)^{-1}
	\left[
	\cov{\sum_{(i,t)\in\disj{post}{\gamma}{\nu}(\zeta)} Y_{\zeta it}}{\sum_{(i,t)\in\disj{post}{\nu}{\gamma}(\zeta)} Y_{\zeta it}}
	+ 
	\cov{\sum_{(i,t)\in\disj{post}{\gamma}{\nu}(\zeta)} Y_{\zeta it}}{\sum_{(i,t)\in\ovrlp{post}{post}{\gamma}{\nu}(\zeta)} Y_{\zeta it}}  \right. \\
	&\quad \phantom{\left(N_\gamma^{\text{ctrl}} N_{\nu}^\text{ctrl} \Tpost^2\right)^{-1}\bigg[} 
	+
	\cov{\sum_{(i,t)\in\disj{post}{\gamma}{\nu}(\zeta)} Y_{\zeta it}}{\sum_{(i,t)\in \ovrlp{\cdot}{post}{\gamma}{\nu}(\zeta)} Y_{\zeta it}}  
	+
	\cov{\sum_{(i,t)\in \ovrlp{post}{\cdot}{\gamma}{\nu}(\zeta)} Y_{\zeta it}}{\sum_{(i,t)\in \disj{post}{\nu}{\gamma}(\zeta)} Y_{\zeta it}}\\
	&\quad \phantom{\left(N_\gamma^{\text{ctrl}} N_{\nu}^\text{ctrl} \Tpost^2\right)^{-1}\bigg[}
	+ 
	\cov{\sum_{(i,t)\in \ovrlp{post}{pre}{\gamma}{\nu}(\zeta)} Y_{\zeta it}}{\sum_{(i,t)\in \disj{post}{\nu}{\gamma}(\zeta)} Y_{\zeta it}} 
	+
	\cov{\sum_{(i,t)\in \ovrlp{post}{post}{\gamma}{\nu}(\zeta)} Y_{\zeta it}}{\sum_{(i,t)\in \disj{post}{\nu}{\gamma}(\zeta)} Y_{\zeta it}} \\
	&\quad \phantom{\left(N_\gamma^{\text{ctrl}} N_{\nu}^\text{ctrl} \Tpost^2\right)^{-1}\bigg[}
	+ 
	\cov{\sum_{(i,t)\in \ovrlp{post}{\cdot}{\gamma}{\nu}(\zeta)} Y_{\zeta it}}{\sum_{(i,t)\in \ovrlp{post}{post}{\nu}{\gamma}(\zeta)} Y_{\zeta it}} 
	+
	\cov{\sum_{(i,t)\in \ovrlp{post}{\cdot}{\gamma}{\nu}(\zeta)} Y_{\zeta it}}{\sum_{(i,t)\in \ovrlp{\cdot}{post}{\nu}{\gamma}(\zeta)} Y_{\zeta it}} \\
	&\quad \phantom{\left(N_\gamma^{\text{ctrl}} N_{\nu}^\text{ctrl} \Tpost^2\right)^{-1}\bigg[}
	+ 
	\cov{\sum_{(i,t)\in \ovrlp{post}{pre}{\gamma}{\nu}(\zeta)} Y_{\zeta it}}{\sum_{(i,t)\in \ovrlp{post}{post}{\nu}{\gamma}(\zeta)} Y_{\zeta it}} 
	+
	\cov{\sum_{(i,t)\in \ovrlp{post}{pre}{\gamma}{\nu}(\zeta)} Y_{\zeta it}}{\sum_{(i,t)\in \ovrlp{\cdot}{post}{\nu}{\gamma}(\zeta)} Y_{\zeta it}} \\
	&\quad \phantom{\left(N_\gamma^{\text{ctrl}} N_{\nu}^\text{ctrl} \Tpost^2\right)^{-1}\bigg[} \left.
	+ 
	\var{\sum_{(i,t)\in \ovrlp{post}{post}{\gamma}{\nu}(\zeta)} Y_{\zeta it}}
	+
	\cov{\sum_{(i,t)\in \ovrlp{post}{post}{\gamma}{\nu}(\zeta)} Y_{\zeta it}}{\sum_{(i,t)\in \ovrlp{\cdot}{post}{\nu}{\gamma}(\zeta)} Y_{\zeta it}}
	\right]. \numbereqn \label{eq:post-post-covar-blowup}
\end{align*}
\endgroup

We now handle each of these covariances in turn. Under a block-exchangeable covariance structure for the data (\cref{eq:sigma_s}), deriving an expression for each of these component covariances becomes a simple counting problem. For each pair of person-time windows, we simply count the amount of person-time shared within the same individual, shared within the same time, and shared across time. Let $\ndisj{\gamma}{\nu}(\zeta)$ be the number of disjoint individuals in state $\zeta$ that contribute to $\mathcal{C}_\gamma$ but not $\mathcal{C}_\nu$, and $\novrlp{\gamma}{\nu}(\zeta)$ the number of control individuals in state $\zeta$ shared between $\mathcal{C}_\gamma$ and $\mathcal{C}_\nu$. Let $T_\mathrm{post,post} := (\Tpost - \Delta)_+$ be the number of measurement ocassions that contribute to the post-treatment periods of both $\mathcal{C}_\gamma$ and $\mathcal{C}_\nu$, and similarly for all other combinations of pre- and post-treatment periods in \cref{tab:windows}.

As in \cref{sec:did-var-derivation}, we can write each of the component covariances above into sums of covariances betweeen outcomes in the same person at the same time, the same person at different times, different people at the same times, and different people at different times. 
As an example, between person time in $\disj{post}{\gamma}{\nu}$ and person time in $\ovrlp{post}{post}{\gamma}{\nu}$, there are $\ndisj{\gamma}{\nu}\novrlp{\gamma}{\nu}$ combinations of distinct individuals (i.e., no within-individual combinations), each of which containing $\Tpost$ combinations of time. Of those time combinations, $T_\mathrm{post,post}$ are between the same time and $\Tpost^2-T_\mathrm{post,post}$ are between different times. Therefore, under \cref{eq:sigma_s},
\begin{equation}
	\cov{\sum_{(i,t)\in\disj{post}{\gamma}{\nu}(\zeta)} Y_{\zeta it}}{\sum_{(i,t)\in\ovrlp{post}{post}{\gamma}{\nu}(\zeta)} Y_{\zeta it}} = \ndisj{\gamma}{\nu}\novrlp{\gamma}{\nu} \left(T_\mathrm{post,post} \phi_\zeta + \left(\Tpost^2 - T_\mathrm{post,post}\right)\psi_\zeta\right) \sigma^2_\zeta.
\end{equation}
Using the same reasoning, we can find expressions for every covariance on the right-hand side of \cref{eq:post-post-covar-blowup}. We provide them in \cref{tab:post-post-blowup}. Similarly, we provide expressions for all component covariances of the remaining terms on the right-hand side of \cref{eq:covar-expansion-means} in \cref{tab:pre-pre-blowup,tab:post-pre-blowup,tab:pre-post-blowup}. 

\begingroup
\renewcommand{\arraystretch}{2.5}
\begin{table}
	\centering
	\caption{Closed-form expressions of component covariances on the right-hand side of \cref{eq:post-post-covar-blowup}, the covariance of post-treatment means over control individuals for cohorts $\mathcal{C}_\gamma$ and $\mathcal{C}_\nu$.}
	\label{tab:post-post-blowup}
	\begin{tabular}{ccl}
		\toprule
		$X_1$ & $X_2$ & $\displaystyle \cov{\sum_{(i,t)\in X_1} Y_{\zeta it}}{\sum_{(i,t)\in X_2} Y_{\zeta it}}$ \\
		\midrule
		$\disj{post}{\gamma}{\nu}(\zeta)$ & $\disj{post}{\nu}{\gamma}(\zeta)$ &
		$\ndisj{\gamma}{\nu}(\zeta) \ndisj{\nu}{\gamma}(\zeta) \left(T_\mathrm{post,post}\phi_\zeta + \left(\Tpost^2 - T_\mathrm{post,post}\right)\psi_\zeta\right) \sigma^2_\zeta$
		\\
		$\disj{post}{\gamma}{\nu}(\zeta)$ &
		$\ovrlp{post}{post}{\gamma}{\nu}(\zeta)$ &
		$\ndisj{\gamma}{\nu}(\zeta) \novrlp{\gamma}{\nu}(\zeta) T_\mathrm{post,post} \left(\phi_\zeta +  \left(\Tpost - 1\right) \psi_\zeta\right) \sigma^2_\zeta$
		\\
		$\disj{post}{\gamma}{\nu}(\zeta)$ &
		$\ovrlp{\cdot}{post}{\gamma}{\nu}(\zeta)$ &
		$\ndisj{\gamma}{\nu}(\zeta) \novrlp{\gamma}{\nu}(\zeta) \Tpost T_\mathrm{\cdot,post} \psi_\zeta \sigma^2_\zeta$ 
		\\
		$\ovrlp{post}{\cdot}{\gamma}{\nu}(\zeta)$ & 
		$\disj{post}{\nu}{\gamma}(\zeta)$ &
		$\ndisj{\gamma}{\nu}(\zeta) \novrlp{\gamma}{\nu}(\zeta) T_\mathrm{post,\cdot}\Tpost\psi_\zeta \sigma^2_\zeta$ 
		\\
		$\ovrlp{post}{pre}{\gamma}{\nu}(\zeta)$ &
		$\disj{post}{\nu}{\gamma}(\zeta)$ &
		$\ndisj{\gamma}{\nu}(\zeta) \novrlp{\gamma}{\nu}(\zeta) \Tpost T_\mathrm{post,pre}\psi_\zeta \sigma^2_\zeta$
		\\
		$\ovrlp{post}{post}{\gamma}{\nu}(\zeta)$ &
		$\disj{post}{\nu}{\gamma}(\zeta)$ &
		$\ndisj{\gamma}{\nu}(\zeta) \novrlp{\gamma}{\nu}(\zeta) \left(T_\mathrm{post,post} \phi_\zeta T_\mathrm{post,post}\left(\Tpost - 1\right)\psi_\zeta \right) \sigma^2_\zeta$
		\\
		$\ovrlp{post}{\cdot}{\gamma}{\nu}(\zeta)$ &
		$\ovrlp{post}{post}{\gamma}{\nu}(\zeta)$ &
		$\novrlp{\gamma}{\nu}(\zeta) T_\mathrm{post,\cdot} T_\mathrm{post,post} \left(\rho_\zeta + \left(\novrlp{\gamma}{\nu}(\zeta) - 1\right) \psi_\zeta\right) \sigma^2_\zeta$ 
		\\
		$\ovrlp{post}{\cdot}{\gamma}{\nu}(\zeta)$ &
		$\ovrlp{\cdot}{post}{\gamma}{\nu}(\zeta)$ &
		$\novrlp{\gamma}{\nu}(\zeta) T_\mathrm{post,\cdot} T_\mathrm{\cdot,post} \left(\rho_\zeta + \left(\novrlp{\gamma}{\nu}(\zeta) - 1\right) \psi_\zeta\right) \sigma^2_\zeta$ 
		\\
		$\ovrlp{post}{pre}{\gamma}{\nu}(\zeta)$ &
		$\ovrlp{post}{post}{\gamma}{\nu}(\zeta)$ &
		$\novrlp{\gamma}{\nu}(\zeta) T_\mathrm{post,pre} T_\mathrm{post,post} \left(\rho_\zeta + \left(\novrlp{\gamma}{\nu}(\zeta) - 1\right) \psi_\zeta\right) \sigma^2_\zeta$ 
		\\
		$\ovrlp{post}{pre}{\gamma}{\nu}(\zeta)$ &
		$\ovrlp{\cdot}{post}{\gamma}{\nu}(\zeta)$ &
		$\novrlp{\gamma}{\nu}(\zeta) T_\mathrm{post,pre} T_\mathrm{\cdot,post} \left(\rho_\zeta + \left(\novrlp{\gamma}{\nu}(\zeta) - 1\right) \psi_\zeta\right) \sigma^2_\zeta$ 
		\\
		$\ovrlp{post}{post}{\gamma}{\nu}(\zeta)$ &
		$\ovrlp{post}{post}{\gamma}{\nu}(\zeta)$ &
		$\begin{aligned}
			\novrlp{\gamma}{\nu}(\zeta) T_\mathrm{post,post} 		
			&\left( 1 + \left(T_\mathrm{post,post} - 1\right)\rho_\zeta \right. \\
			&\quad  + \left(\novrlp{\gamma}{\nu}(\zeta) - 1\right)  \left(\phi_\zeta \right. \left.\left. + \left(T_\mathrm{post,post} - 1\right) \psi_\zeta\right)\right) \sigma^2_\zeta 
		\end{aligned}$ 
		\\
		$\ovrlp{post}{post}{\gamma}{\nu}(\zeta)$ &
		$\ovrlp{\cdot}{post}{\gamma}{\nu}(\zeta)$ &
		$\novrlp{\gamma}{\nu}(\zeta) T_\mathrm{post,post} T_\mathrm{\cdot,post} \left(\rho_\zeta + \left(\novrlp{\gamma}{\nu}(\zeta) - 1\right) \psi_\zeta\right) \sigma^2_\zeta$ 
		\\
		\bottomrule
	\end{tabular}
\end{table}
\endgroup

\begingroup
\renewcommand{\arraystretch}{2.5}
\begin{table}
	\centering
	\caption{Closed-form expressions of component covariances of the covariance of pre-treatment means over control individuals for cohorts $\mathcal{C}_\gamma$ and $\mathcal{C}_\nu$.}
	\label{tab:pre-pre-blowup}
	\begin{tabular}{ccl}
		\toprule
		$X_1$ & $X_2$ & $\displaystyle \cov{\sum_{(i,t)\in X_1} Y_{\zeta it}}{\sum_{(i,t)\in X_2} Y_{\zeta it}}$ \\
		\midrule
		$\disj{pre}{\gamma}{\nu}(\zeta)$ &
		$\disj{pre}{\nu}{\gamma}(\zeta)$ &
		$\ndisj{\gamma}{\nu}(\zeta) \ndisj{\nu}{\gamma}(\zeta) \left(T_\mathrm{pre,pre}\phi_\zeta + \left(\Tpre^2 - T_\mathrm{pre,pre}\right)\psi_\zeta\right) \sigma^2_\zeta$
		\\
		$\disj{pre}{\gamma}{\nu}(\zeta)$ &
		$\ovrlp{pre}{pre}{\gamma}{\nu}(\zeta)$ &
		$\ndisj{\gamma}{\nu}(\zeta) \novrlp{\gamma}{\nu}(\zeta) T_\mathrm{pre,pre} \left(\phi_\zeta + \left(\Tpre - 1\right) \psi_\zeta\right) \sigma^2_\zeta$
		\\
		$\disj{pre}{\gamma}{\nu}(\zeta)$ &
		$\ovrlp{post}{pre}{\gamma}{\nu}(\zeta)$ &
		$\ndisj{\gamma}{\nu}(\zeta) \novrlp{\gamma}{\nu}(\zeta) \Tpre T_\mathrm{post,pre} \psi_\zeta \sigma^2_\zeta$ 
		\\
		$\disj{pre}{\gamma}{\nu}(\zeta)$ &
		$\ovrlp{\cdot}{pre}{\gamma}{\nu}(\zeta)$ & 
		$\ndisj{\gamma}{\nu}(\zeta) \novrlp{\gamma}{\nu}(\zeta) T_\mathrm{\dot,pre}\Tpre \psi_\zeta \sigma^2_\zeta$ 
		\\
		$\ovrlp{pre}{\cdot}{\gamma}{\nu}(\zeta)$ &
		$\disj{pre}{\nu}{\gamma}(\zeta)$ &
		$\ndisj{\nu}{\gamma}(\zeta) \novrlp{\gamma}{\nu}(\zeta) \Tpre T_\mathrm{pre,\cdot}\psi_\zeta \sigma^2_\zeta$
		\\
		$\ovrlp{pre}{pre}{\gamma}{\nu}(\zeta)$ &
		$\disj{pre}{\nu}{\gamma}(\zeta)$ &
		$\ndisj{\nu}{\gamma}(\zeta) \novrlp{\gamma}{\nu}(\zeta) \left(T_\mathrm{pre,pre} \phi_\zeta T_\mathrm{pre,pre}\left(\Tpre - 1\right)\psi_\zeta \right) \sigma^2_\zeta$
		\\
		$\ovrlp{pre}{\cdot}{\gamma}{\nu}(\zeta)$ &
		$\ovrlp{pre}{pre}{\gamma}{\nu}(\zeta)$ &
		$\novrlp{\gamma}{\nu}(\zeta) T_\mathrm{pre,\cdot} T_\mathrm{pre,pre} \left(\rho_\zeta + \left(\novrlp{\gamma}{\nu}(\zeta) - 1\right) \psi_\zeta\right) \sigma^2_\zeta$ 
		\\
		$\ovrlp{pre}{\cdot}{\gamma}{\nu}(\zeta)$ &
		$\ovrlp{post}{pre}{\gamma}{\nu}(\zeta)$ &
		$\novrlp{\gamma}{\nu}(\zeta) T_\mathrm{pre,\cdot} T_\mathrm{post,pre} \left(\rho_\zeta + \left(\novrlp{\gamma}{\nu}(\zeta) - 1\right) \psi_\zeta\right) \sigma^2_\zeta$ 
		\\
		$\ovrlp{pre}{\cdot}{\gamma}{\nu}(\zeta)$ &
		$\ovrlp{\cdot}{pre}{\gamma}{\nu}(\zeta)$ &
		$\novrlp{\gamma}{\nu}(\zeta) T_\mathrm{pre,\cdot} T_\mathrm{\cdot,pre}
		\left(\rho_\zeta + \left(\novrlp{\gamma}{\nu}(\zeta) - 1\right)  \psi_\zeta\right) \sigma^2_\zeta$ 
		\\
		$\ovrlp{pre}{pre}{\gamma}{\nu}(\zeta)$ &
		$\ovrlp{pre}{pre}{\gamma}{\nu}(\zeta)$ &
		$\begin{aligned}
			\novrlp{\gamma}{\nu}(\zeta) T_\mathrm{pre,pre} 		
			&\left( 1 + \left(T_\mathrm{pre,pre} - 1\right)\rho_\zeta \right. \\
			&\quad  + \left(\novrlp{\gamma}{\nu}(\zeta) - 1\right)  \left(\phi_\zeta \right. \left.\left. + \left(T_\mathrm{pre,pre} - 1\right) \psi_\zeta\right)\right) \sigma^2_\zeta 
		\end{aligned}$ 
		\\
		$\ovrlp{pre}{pre}{\gamma}{\nu}(\zeta)$ &
		$\ovrlp{post}{pre}{\gamma}{\nu}(\zeta)$ &
		$\novrlp{\gamma}{\nu}(\zeta) T_\mathrm{pre,pre} T_\mathrm{post,pre} \left(\rho_\zeta + \left(\novrlp{\gamma}{\nu}(\zeta) - 1\right) \psi_\zeta\right) \sigma^2_\zeta$ 
		\\
		$\ovrlp{pre}{pre}{\gamma}{\nu}(\zeta)$ &
		$\ovrlp{\cdot}{pre}{\gamma}{\nu}(\zeta)$ &
		$\novrlp{\gamma}{\nu}(\zeta) T_\mathrm{pre,pre} T_\mathrm{\cdot,pre} \left(\rho_\zeta + \left(\novrlp{\gamma}{\nu}(\zeta) - 1\right) \psi_\zeta\right) \sigma^2_\zeta$ 
		\\
		\bottomrule
	\end{tabular}
\end{table}
\endgroup

\begingroup
\renewcommand{\arraystretch}{2.5}
\begin{table}
	\centering
	\caption{Closed-form expressions of component covariances of the covariance between the post-treatment mean over control individuals for cohort $\mathcal{C}_\gamma$ and the pre-treatment mean over control individuals for cohort $\mathcal{C}_\nu$. Note that some covariances are identically zero because the two windows cannot both have non-zero duration; see \cref{tab:windows}.}
	\label{tab:post-pre-blowup}
	\begin{tabular}{ccl}
		\toprule
		$X_1$ & $X_2$ & $\displaystyle \cov{\sum_{(i,t)\in X_1} Y_{\zeta it}}{\sum_{(i,t)\in X_2} Y_{\zeta it}}$ \\
		\midrule
		$\disj{post}{\gamma}{\nu}(\zeta)$ &
		$\disj{pre}{\nu}{\gamma}(\zeta)$ &
		$\ndisj{\gamma}{\nu}(\zeta) \ndisj{\nu}{\gamma}(\zeta) \left(T_\mathrm{post,pre}\phi_\zeta + \left(\Tpre\Tpost - T_\mathrm{post,pre}\right)\psi_\zeta\right) \sigma^2_\zeta$
		\\
		$\disj{post}{\gamma}{\nu}(\zeta)$ &
		$\ovrlp{\cdot}{pre}{\gamma}{\nu}(\zeta)$ &
		$\ndisj{\gamma}{\nu}(\zeta) \novrlp{\gamma}{\nu}(\zeta) \Tpost T_\mathrm{\cdot,pre} \left(\phi_\zeta + \left(\Tpre - 1\right) \psi_\zeta\right) \sigma^2_\zeta$
		\\
		$\disj{post}{\gamma}{\nu}(\zeta)$ &
		$\ovrlp{post}{pre}{\gamma}{\nu}(\zeta)$ &
		$\ndisj{\gamma}{\nu}(\zeta) \novrlp{\gamma}{\nu}(\zeta) T_\mathrm{post,pre} \left(\phi_\zeta + (\Tpost - 1) \psi_\zeta \right) \sigma^2_\zeta$ 
		\\
		$\disj{post}{\gamma}{\nu}(\zeta)$ &
		$\ovrlp{pre}{pre}{\gamma}{\nu}(\zeta)$ & 
		$\ndisj{\gamma}{\nu}(\zeta) \novrlp{\gamma}{\nu}(\zeta) \Tpost T_\mathrm{pre,pre}\psi_\zeta \sigma^2_\zeta$ 
		\\
		$\ovrlp{post}{\cdot}{\gamma}{\nu}(\zeta)$ &
		$\disj{pre}{\nu}{\gamma}(\zeta)$ &
		$\ndisj{\nu}{\gamma}(\zeta) \novrlp{\gamma}{\nu}(\zeta) \Tpre T_\mathrm{post,\cdot}\psi_\zeta \sigma^2_\zeta$
		\\
		$\ovrlp{post}{pre}{\gamma}{\nu}(\zeta)$ &
		$\disj{pre}{\nu}{\gamma}(\zeta)$ &
		$\ndisj{\nu}{\gamma}(\zeta) \novrlp{\gamma}{\nu}(\zeta) T_\mathrm{post, pre} \left(\phi_\zeta + \left(\Tpre - 1\right) \psi_\zeta\right) \sigma^2_\zeta$ 
		\\
		$\ovrlp{post}{post}{\gamma}{\nu}(\zeta)$ &
		$\disj{pre}{\nu}{\gamma}(\zeta)$ &
		$\ndisj{\nu}{\gamma}(\zeta) \novrlp{\gamma}{\nu}(\zeta) \left(T_\mathrm{post,post} \Tpre \psi_\zeta \right) \sigma^2_\zeta$
		\\
		$\ovrlp{post}{\cdot}{\gamma}{\nu}(\zeta)$ &
		$\ovrlp{pre}{pre}{\gamma}{\nu}(\zeta)$ &
		$\novrlp{\gamma}{\nu}(\zeta) T_\mathrm{post,\cdot} T_\mathrm{pre,pre} \left(\rho_\zeta + \left(\novrlp{\gamma}{\nu}(\zeta) - 1\right) \psi_\zeta\right) \sigma^2_\zeta \equiv 0$  
		\\
		$\ovrlp{post}{\cdot}{\gamma}{\nu}(\zeta)$ &
		$\ovrlp{post}{pre}{\gamma}{\nu}(\zeta)$ &
		$\novrlp{\gamma}{\nu}(\zeta) T_\mathrm{post,\cdot} T_\mathrm{post,pre} \left(\rho_\zeta + \left(\novrlp{\gamma}{\nu}(\zeta) - 1\right) \psi_\zeta\right) \sigma^2_\zeta$ 
		\\
		$\ovrlp{post}{\cdot}{\gamma}{\nu}(\zeta)$ &
		$\ovrlp{\cdot}{pre}{\gamma}{\nu}(\zeta)$ &
		$\novrlp{\gamma}{\nu}(\zeta) T_\mathrm{post,\cdot} T_\mathrm{\cdot,pre} \left(\rho_\zeta + \left(\novrlp{\gamma}{\nu}(\zeta) - 1\right) \psi_\zeta\right) \sigma^2_\zeta$ 
		
		\\
		$\ovrlp{post}{pre}{\gamma}{\nu}(\zeta)$ &
		$\ovrlp{\cdot}{pre}{\gamma}{\nu}(\zeta)$ &
		$\novrlp{\gamma}{\nu}(\zeta) T_\mathrm{post,pre} T_\mathrm{\cdot,pre} \left(\rho_\zeta + \left(\novrlp{\gamma}{\nu}(\zeta) - 1\right) \psi_\zeta\right) \sigma^2_\zeta$ 
		\\
		$\ovrlp{post}{pre}{\gamma}{\nu}(\zeta)$ &
		$\ovrlp{pre}{pre}{\gamma}{\nu}(\zeta)$ &
		$\novrlp{\gamma}{\nu}(\zeta) T_\mathrm{post,pre} T_\mathrm{pre,pre} \left(\rho_\zeta + \left(\novrlp{\gamma}{\nu}(\zeta) - 1\right) \psi_\zeta\right)$
		\\
		$\ovrlp{post}{pre}{\gamma}{\nu}(\zeta)$ &
		$\ovrlp{post}{pre}{\gamma}{\nu}(\zeta)$ &
		$\begin{aligned}
			\novrlp{\gamma}{\nu}(\zeta) T_\mathrm{post,pre} 		
			&\left( 1 + \left(T_\mathrm{post,post} - 1\right)\rho_\zeta \right. \\
			&\quad  + \left(\novrlp{\gamma}{\nu}(\zeta) - 1\right)  \left(\phi_\zeta \right. \left.\left. + \left(T_\mathrm{post,pre} - 1\right) \psi_\zeta\right)\right) \sigma^2_\zeta 
		\end{aligned}$ 
		\\
		$\ovrlp{post}{post}{\gamma}{\nu}(\zeta)$ &
		$\ovrlp{pre}{pre}{\gamma}{\nu}(\zeta)$ &
		$\novrlp{\gamma}{\nu}(\zeta) T_\mathrm{post,post} T_\mathrm{pre,pre} \left(\rho_\zeta + \left(\novrlp{\gamma}{\nu}(\zeta) - 1\right) \psi_\zeta\right) \sigma^2_\zeta$
		\\ 
		$\ovrlp{post}{post}{\gamma}{\nu}(\zeta)$ &
		$\ovrlp{\cdot}{pre}{\gamma}{\nu}(\zeta)$ &
		$0$ 
		\\
		$\ovrlp{post}{post}{\gamma}{\nu}(\zeta)$ &
		$\ovrlp{post}{pre}{\gamma}{\nu}(\zeta)$ &
		$\novrlp{\gamma}{\nu}(\zeta) T_\mathrm{post,post} T_\mathrm{post,pre} \left(\rho_\zeta + \left(\novrlp{\gamma}{\nu}(\zeta) - 1\right) \psi_\zeta\right) \sigma^2_\zeta$ 			
		\\
		\bottomrule
	\end{tabular}
\end{table}
\endgroup

\begingroup
\renewcommand{\arraystretch}{2.5}
\begin{table}
	\centering
	\caption{Closed-form expressions of component covariances of the covariance between the pre-treatment mean over control individuals for cohort $\mathcal{C}_\gamma$ and the post-treatment mean over control individuals for cohort $\mathcal{C}_\nu$. Note that some covariances are identically zero because the two windows cannot both have non-zero duration; see \cref{tab:windows}.}
	\label{tab:pre-post-blowup}
	\begin{tabular}{ccl}
		\toprule
		$X_1$ & $X_2$ & $\displaystyle \cov{\sum_{(i,t)\in X_1} Y_{\zeta it}}{\sum_{(i,t)\in X_2} Y_{\zeta it}}$ \\
		\midrule
		$\disj{pre}{\gamma}{\nu}(\zeta)$ &
		$\disj{post}{\nu}{\gamma}(\zeta)$ &
		$\ndisj{\gamma}{\nu}(\zeta) \ndisj{\nu}{\gamma}(\zeta) \Tpre\Tpost \psi_\zeta \sigma^2_\zeta$
		\\
		$\disj{pre}{\gamma}{\nu}(\zeta)$ &
		$\ovrlp{post}{post}{\gamma}{\nu}(\zeta)$ &
		$\ndisj{\gamma}{\nu}(\zeta) \novrlp{\gamma}{\nu}(\zeta) \Tpre T_\mathrm{post,post} \psi_\zeta \sigma^2_\zeta$
		\\
		$\disj{pre}{\gamma}{\nu}(\zeta)$ &
		$\ovrlp{\cdot}{post}{\gamma}{\nu}(\zeta)$ &
		$\ndisj{\gamma}{\nu}(\zeta) \novrlp{\gamma}{\nu}(\zeta) \Tpre T_\mathrm{\cdot,pre} \psi_\zeta \sigma^2_\zeta$ 
		\\
		$\ovrlp{pre}{\cdot}{\gamma}{\nu}(\zeta)$ & $\disj{post}{\nu}{\gamma}(\zeta)$ &
		$\ndisj{\nu}{\gamma}(\zeta) \novrlp{\gamma}{\nu}(\zeta) \Tpost T_\mathrm{pre,\cdot}\psi_\zeta \sigma^2_\zeta$ 
		\\
		$\ovrlp{pre}{pre}{\gamma}{\nu}(\zeta)$ &
		$\disj{post}{\nu}{\gamma}(\zeta)$ &
		$\ndisj{\nu}{\gamma}(\zeta) \novrlp{\gamma}{\nu}(\zeta) \Tpost T_\mathrm{pre,pre}\psi_\zeta \sigma^2_\zeta$
		\\
		$\ovrlp{pre}{\cdot}{\gamma}{\nu}(\zeta)$ &
		$\ovrlp{post}{post}{\gamma}{\nu}(\zeta)$ &
		$\novrlp{\gamma}{\nu}(\zeta) T_\mathrm{pre, \cdot} T_\mathrm{post,post} \left(\rho_\zeta + \left(\novrlp{\gamma}{\nu}(\zeta) - 1\right) \psi_\zeta\right) \sigma^2_\zeta$ 
		\\
		$\ovrlp{pre}{pre}{\gamma}{\nu}(\zeta)$ &
		$\ovrlp{post}{post}{\gamma}{\nu}(\zeta)$ &
		$\novrlp{\gamma}{\nu}(\zeta) \novrlp{\gamma}{\nu}(\zeta) T_\mathrm{pre,pre} T_\mathrm{post,post} \left(\rho_\zeta + \left(\novrlp{\gamma}{\nu}(\zeta) - 1\right) \psi_\zeta \right) \sigma^2_\zeta$
		\\
		$\ovrlp{pre}{\cdot}{\gamma}{\nu}(\zeta)$ &
		$\ovrlp{\cdot}{post}{\gamma}{\nu}(\zeta)$ &
		$\novrlp{\gamma}{\nu}(\zeta) T_\mathrm{pre,\cdot} T_\mathrm{\cdot, post} \left(\rho_\zeta + \left(\novrlp{\gamma}{\nu}(\zeta) - 1\right) \psi_\zeta\right) \sigma^2_\zeta \equiv 0$  
		\\
		$\ovrlp{pre}{pre}{\gamma}{\nu}(\zeta)$ &
		$\ovrlp{\cdot}{post}{\gamma}{\nu}(\zeta)$ &
		$\novrlp{\gamma}{\nu}(\zeta) T_\mathrm{pre,pre} T_\mathrm{\cdot,post} \left(\rho_\zeta + \left(\novrlp{\gamma}{\nu}(\zeta) - 1\right) \psi_\zeta\right) \sigma^2_\zeta$ 
		\\
		\bottomrule
	\end{tabular}
\end{table}
\endgroup

A closed-form expression of $\cov{\widehat{\mathrm{ATT}}_\gamma}{\widehat{\mathrm{ATT}}_\nu}$ can be derived as follows:
\begin{align*}
	\cov{\widehat{\mathrm{ATT}}_\gamma}{\widehat{\mathrm{ATT}}_\nu} =& \left(N_\gamma^\text{ctrl} N_\nu^\text{ctrl} \Tpost^2\right)^{-1} \times \left(\text{sum of expressions in \cref{tab:post-post-blowup}}\right) \\
	&{}\quad + \left(N_\gamma^\text{ctrl} N_\nu^\text{ctrl} \Tpre^2\right)^{-1} \times \left(\text{sum of expressions in \cref{tab:pre-pre-blowup}}\right) \\
	&{}\quad - \left(N_\gamma^\text{ctrl} N_\nu^\text{ctrl} \Tpost\Tpre\right)^{-1} \times \left(\text{sum of expressions in \cref{tab:post-pre-blowup}}\right) \\
	&{}\quad - \left(N_\gamma^\text{ctrl} N_\nu^\text{ctrl} \Tpre \Tpost\right)^{-1} \times \left(\text{sum of expressions in \cref{tab:pre-post-blowup}}\right).
\end{align*}
Combining terms, we have
\begin{align*}
	\cov{\widehat{\mathrm{ATT}}_\gamma}{\widehat{\mathrm{ATT}}_\nu} =& \left(\Tpre\Tpost\right)^{-2}\left(N_\gamma^\text{ctrl} N_\nu^\text{ctrl}\right)^{-1} \sum_{\zeta\in\Xi_\text{ctrl}} \sigma^2_\zeta \left[\ndisj{\gamma}{\nu}(\zeta) \ndisj{\nu}{\gamma}(\zeta) h_1(\Tpre,\Tpost, \Delta, \rho_\zeta, \phi_\zeta, \psi_\zeta)\right.
	\\
	&\phantom{\sigma^2_\zeta \left(\Tpre\Tpost\right)^{-2}\left(N_\gamma^\text{ctrl} N_\nu^\text{ctrl}\right)^{-1} \sum_{\zeta\in\Xi_\text{ctrl}}} \quad
	+ \novrlp{\gamma}{\nu}(\zeta) \ndisj{\gamma}{\nu}(\zeta) h_2(\Tpre,\Tpost, \Delta, \rho_\zeta, \phi_\zeta, \psi_\zeta) 
	\\
	&\phantom{\sigma^2_\zeta \left(\Tpre\Tpost\right)^{-2}\left(N_\gamma^\text{ctrl} N_\nu^\text{ctrl}\right)^{-1} \sum_{\zeta\in\Xi_\text{ctrl}}} \quad 
	+ \novrlp{\gamma}{\nu}(\zeta) \ndisj{\nu}{\gamma}(\zeta) h_3(\Tpre,\Tpost, \Delta, \rho_\zeta, \phi_\zeta, \psi_\zeta) 
	\\
	&\phantom{\sigma^2_\zeta \left(\Tpre\Tpost\right)^{-2}\left(N_\gamma^\text{ctrl} N_\nu^\text{ctrl}\right)^{-1} \sum_{\zeta\in\Xi_\text{ctrl}}} \quad 
	+ \novrlp{\gamma}{\nu}(\zeta) h_4(\Tpre,\Tpost, \Delta, \rho_\zeta, \phi_\zeta, \psi_\zeta) 
	\\
	&\phantom{\sigma^2_\zeta \left(\Tpre\Tpost\right)^{-2}\left(N_\gamma^\text{ctrl} N_\nu^\text{ctrl}\right)^{-1} \sum_{\zeta\in\Xi_\text{ctrl}}} \quad \left.
	+ \novrlp{\gamma}{\nu}(\zeta) \left(\novrlp{\gamma}{\nu}(\zeta) - 1\right)  h_5(\Tpre,\Tpost, \Delta, \rho_\zeta, \phi_\zeta, \psi_\zeta)  \right]. \numbereqn \label{eq:combined-covar-init-hfuns}
\end{align*}
We turn now to simplifying the $h(\cdot)$ functions above. Let $(x)_+ := \max\left(x, 0\right)$ and define 
\[
\min\left(\cdots\right) := \min\left(\Tpre, \Tpost, \Delta, \left(\Tpre + \Tpost - \Delta\right)_+\right).
\]
Collecting terms and substituting time durations from \cref{tab:windows} -- such that, e.g., $T_\mathrm{post,post} = \left(\Tpost - \Delta\right)_+$) -- we have
\begin{align*}
	h_1(\Tpre,\Tpost, \Delta, \rho_\zeta, \phi_\zeta, \psi_\zeta) &=
	\Tpre^2 (\Tpost - \Delta)_+ \phi_\zeta + \Tpre^2 \left(\Tpost^2 - (\Tpost - \Delta)_+\right) \psi_\zeta + \Tpost^2 (\Tpre-\Delta)_+ \phi_\zeta \\
	&{}\quad + \Tpost^2\left(\Tpre^2 - (\Tpre - \Delta)_+\right) \psi_\zeta - \Tpre\Tpost\min\left(\cdots\right) \phi_\zeta \\
	&{}\quad - \Tpre\Tpost \left(\Tpre\Tpost - \min\left(\cdots\right)\right) \psi_\zeta - \Tpre^2\Tpost^2\psi_\zeta
	\\
	&= \left(\Tpre^2 \left(\Tpost - \Delta\right)_+ + \Tpost^2 \left(\Tpre - \Delta\right)_+ - \Tpre\Tpost \min\left(\cdots\right)\right) \left(\phi_\zeta - \psi_\zeta\right) \numbereqn \label{eq:h1} \\
	&=: g_1(\Tpre, \Tpost, \Delta) \left(\phi_\zeta - \psi_\zeta\right).
\end{align*}

Notice that $\min\left(\Tpost, \Delta\right) + \left(\Tpost - \Delta\right)_+ = \Tpost$; similarly for other such combinations. Then, for $h_2$ and $h_3$, 
\begin{align*}
	h_2\left(\Tpre, \Tpost, \Delta, \rho_\zeta, \phi_\zeta, \psi_\zeta\right) &=
	\Tpost^2 \Tpre \left(\Delta - \Tpost\right)_+ \psi_\zeta + \Tpost^2 \left(\Tpre - \Delta\right)_+ \phi_\zeta + \Tpost\left(\Tpre - \Delta\right)_+ \left(\Tpre - 1\right) \psi_\zeta \\
	&{}\quad + \Tpost^2 \min\left(\cdots\right) \Tpre \psi_\zeta + \Tpre^2 \Tpost \min\left(\Tpost, \Delta\right) \psi_\zeta + \Tpre^2 \left(\Tpost - \Delta\right)_+ \phi_t \\
	&{}\quad + \Tpre^2 \left(\Tpost - \Delta\right)_+ \left(\Tpost - 1\right)\psi_\zeta - \Tpre \Tpost^2 \left(\Delta - \Tpost\right)_+ \psi_\zeta - \Tpre\Tpost\min\left(\cdots\right) \phi_\zeta \\
	&{}\quad - \Tpre \Tpost \min\left(\cdots\right) \left(\Tpost - 1\right) \psi_\zeta - \Tpre \Tpost^2 \left(\Tpre - \Delta\right)_+ \psi_\zeta \\
	&{}\quad - \Tpre^2 \Tpost \min\left(\Tpost, \Delta_+\right) \psi_\zeta - \Tpre^2 \Tpost \left(\Tpost - \Delta\right)_+ \psi_\zeta \\
	&= \left(\Tpre \left(\Tpost - \Delta\right)_+ + \Tpost^2 \left(\Tpre - \Delta\right)_+ - \Tpre\Tpost \min\left(\cdots\right)\right) \left(\phi_\zeta - \psi_\zeta\right) \numbereqn \label{eq:h2} \\
	&=: g_1(\Tpre, \Tpost, \Delta) \left(\phi_\zeta - \psi_\zeta\right).
\end{align*}
\begin{align*}
	h_3\left(\Tpre, \Tpost, \Delta, \rho_\zeta, \phi_\zeta, \psi_\zeta\right) 
	&= \Tpost^2 \left(\Tpre - \Delta\right)\phi_\zeta + \Tpost^2 \left(\Tpre - \Delta\right)_+ \left(\Tpre - 1\right)\psi_\zeta + \Tpre \Tpost^2 \min\left(\Tpre, \Delta\right) \psi_\zeta \\
	&{}\quad + \Tpre^2 \Tpost \min\left(\left(\Delta - \Tpre\right)_+, \Tpost\right) \psi_\zeta + \Tpre^2 \Tpost \min\left(\cdots\right) \psi_\zeta + \Tpre^2 \left(\Tpost - \Delta\right)_+ \phi_\zeta \\
	&{}\quad + \Tpre^2 \left(\Tpost - \Delta\right)_+ \left(\Tpost - 1\right) \psi_\zeta - \Tpre^2 \Tpost \min\left(\left(\Delta - \Tpre\right)_+, \Tpost\right) \psi_\zeta \\
	&{}\quad - \Tpre\Tpost\min\left(\cdots\right)\phi_\zeta - \Tpre\Tpost \min\left(\cdots\right) \left(\Tpre - 1\right) \psi_\zeta - \Tpre\Tpost^2 \left(\Tpre - \Delta\right)_+ \psi_\zeta \\
	&{}\quad - \Tpre \Tpost^2 \min\left(\Tpre, \Delta\right)\psi_\zeta \\
	&= \left(\Tpre \left(\Tpost - \Delta\right)_+ + \Tpost^2 \left(\Tpre - \Delta\right)_+ - \Tpre \Tpost \min\left(\cdots\right)\right) \left(\phi_\zeta - \psi_\zeta\right) \numbereqn \label{eq:h3} \\
	&=: g_1(\Tpre, \Tpost, \Delta) \left(\phi_\zeta - \psi_\zeta\right).
\end{align*}

Expressions for $h_1$, $h_2$, and $h_3$ have all been easily reached simply by cancelling terms. For $h_4$, we have
\begin{align*}
	h_4\left(\Tpre, \Tpost, \Delta, \rho_\zeta, \phi_\zeta, \psi_\zeta\right) 
	&= \Tpost^2 \min\left(\Tpre, \Delta\right) \left(\Tpre - \Delta\right)_+ \rho_\zeta + \Tpost^2 \min\left(\Tpre, \Delta\right) \min\left(\cdots\right) \rho_\zeta + \Tpost^2 \left(\Tpre - \Delta\right)_+ \\
	&{}\quad + \Tpost^2 \left(\Tpre - \Delta\right)_+ \left(\left(\Tpre - \Delta\right)_+ -1\right) \rho_\zeta + \Tpost^2\left(\Tpre-\Delta\right)_+ \left(\Delta - \Tpost\right)_+\rho_\zeta \\
	&{}\quad + \Tpost^2 \left(\Tpre - \Delta\right)_+ \min\left(\cdots\right) \rho_\zeta + \Tpost^2 \min\left(\Tpre, \Delta\right)\left(\Delta - \Tpost\right)_+ \rho_\zeta \\
	&{}\quad + \Tpre^2 \left(\Tpost - \Delta\right)_+ \min\left(\cdots\right) \rho_\zeta + \Tpre^2 \left(\Tpost - \Delta\right)_+ \\
	&{}\quad + \Tpre^2 \left(\Tpost-\Delta\right)_+ \left(\left(\Tpost - \Delta\right)_+ - 1\right) \rho_\zeta + \Tpre^2 \min\left(\left(\Delta - \Tpre\right)_+, \Tpost\right) \left(\Tpost - \Delta\right)_+ \rho_\zeta \\
	&{}\quad + \Tpre^2 \min\left(\Tpost, \Delta\right) \min\left(\cdots\right) \rho_\zeta + \Tpre^2\left(\Tpost - \Delta\right)_+ \min\left(\Tpost, \Delta\right) \rho_\zeta \\
	&{}\quad + \Tpre^2 \min\left(\Tpost, \Delta\right) \min\left(\left(\Delta - \Tpre\right)_+, \Tpost\right) \rho_\zeta - \Tpre \Tpost \left(\Tpre - \Delta\right)_+ \min\left(\cdots\right)\rho_\zeta \\
	&{}\quad - \Tpre\Tpost\left(\Tpost - \Delta\right)_+ \left(\Tpre - \Delta\right)_+ \rho_\zeta - \Tpre\Tpost \min\left(\left(\Delta - \Tpre\right)_+, \Tpost\right) \min\left(\cdots\right)\rho_\zeta \\
	&{}\quad - \Tpre\Tpost\min\left(\cdots\right) - \Tpre \Tpost \min\left(\cdots\right) \left(\min\left(\cdots\right) - 1\right) \rho_\zeta \\
	&{}\quad - \Tpre \Tpost \left(\Tpost - \Delta\right)_+ \min\left(\cdots\right)\rho_\zeta - \Tpre \Tpost  \min\left(\left(\Delta - \Tpre\right)_+, \Tpost\right) \left(\Delta - \Tpost\right)_+ \rho_\zeta \\
	&{}\quad -\Tpre \Tpost \min\left(\cdots\right) \left(\Delta - \Tpost\right)_+ \rho_\zeta - \Tpre\Tpost\min\left(\Tpre, \Delta\right)\left(\Tpost - \Delta\right) \rho_\zeta \\
	&{}\quad -\Tpre\Tpost\left(\Tpost - \Delta\right)_+ \left(\Tpre - \Delta\right)_+ \rho_\zeta - \Tpre\Tpost \min\left(\Tpre, \Delta\right) \min\left(\Tpost, \Delta\right) \rho_\zeta \\
	&{}\quad - \Tpre\Tpost \left(\Tpre - \Delta\right)_+ \min\left(\Tpost, \Delta\right) \rho_\zeta.
\end{align*}
We simplify this by again recognizing that $\min\left(\Tpost, \Delta\right) + \left(\Tpost - \Delta\right)_+ = \Tpost$, etc. After some algebra, we have
\begin{align*}
	h_4\left(\sim\right) &= 
	\left(1-\rho_\zeta\right) \left[\Tpre^2 \left(\Tpost - \Delta\right)_+ + \Tpost^2 \left(\Tpre - \Delta\right)_+ - \Tpre\Tpost \min\left(\cdots\right)\right]
	\\
	&\quad + \rho_\zeta \Tpre\Tpost \left\{\Tpost \left(\Delta - \Tpost\right)_+ 
	+ \Tpre \min\left(\left(\Delta - \Tpre\right)_+, \Tpost\right) \right.
	\\
	& \phantom{\quad + \rho_\zeta \Tpre\Tpost \big[}\quad - \min\left(\left(\Delta - \Tpre\right)_+, \Tpost\right) \left(\Delta - \Tpost\right)_+ - \min(\Tpre, \Delta) \min(\Tpost, \Delta) \\
	& \phantom{\quad + \rho_\zeta \Tpre\Tpost \big[}\quad + 
	\min(\cdots)  \left[\Tpre + \Tpost - \left(\Tpre - \Delta\right)_+ - \min\left(\left(\Delta - \Tpre\right)_+ ,\Tpost \right) \right.
	\\
	& \phantom{\quad + \rho_\zeta \Tpre\Tpost \big[ + \min(\cdots) \big[}\quad
	\left.\left. - \left(\Tpost - \Delta\right)_+ - \left(\Delta - \Tpost\right)_+ 
	- \min(\cdots) \right] \right\}. \\
	&=: (1-\rho_\zeta)g_{1}\left(\Tpre, \Tpost, \Delta\right) + \rho_\zeta g_{2}\left(\Tpre, \Tpost, \Delta\right) \numbereqn \label{eq:h4-with-extras}
\end{align*}

We now attempt to understand $g_{2}\left(\Tpre, \Tpost, \Delta\right)$ in \cref{eq:h4-with-extras}. We consider various values of $\Delta$ relative to $\Tpre$ and $\Tpost$. Note that if $\Delta < \Tpre + \Tpost$ then $\min\left(\left(\Delta - \Tpre\right)_+, \Tpost\right) = \left(\Delta - \Tpre\right)_+$.
\begin{itemize}
	\item If $\Delta < \min(\Tpre, \Tpost)$, then
	$\min\left(\Tpre, \Tpost, \Delta, \left(\Tpre + \Tpost - \Delta\right)_+\right) = \Delta$ 
	and therefore 
	\begin{align*}
		g_2\left(\sim\right) &= \Tpre \Tpost \left\{\Tpost \cdot 0 + \Tpre \cdot 0 - 0\cdot 0 - \Delta^2 \right. \\
		&\quad\phantom{\Tpre \Tpost \big\{} \left.
		+ \Delta \left[\Tpre + \Tpost - (\Tpre - \Delta) - 0 - (\Tpost - \Delta) - 0 - \Delta\right] \right\} \\
		&= 0.
	\end{align*}
	
	\item If $\Tpre < \Delta < \Tpost$, then
	$\min\left(\Tpre, \Tpost, \Delta, \left(\Tpre + \Tpost - \Delta\right)_+\right) = \Tpre$ 
	and therefore 
	\begin{align*}
		g_2\left(\sim\right) &= \Tpre \Tpost \left\{\Tpost \cdot 0 + \Tpre \left(\Delta - \Tpre\right) - \left(\Delta - \Tpre\right)\cdot 0 - \Tpre\Delta \right. \\
		&\quad\phantom{\Tpre \Tpost \big\{} \left.
		+ \Tpre \left[\Tpre + \Tpost - 0 - (\Delta - \Tpre) - (\Tpost - \Delta) - 0 - \Tpre\right] \right\} \\
		&= 0.
	\end{align*}
	
	\item If $\Tpost < \Delta < \Tpre$, then
	$\min\left(\Tpre, \Tpost, \Delta, \left(\Tpre + \Tpost - \Delta\right)_+\right) = \Tpre$ 
	and therefore 
	\begin{align*}
		g_2\left(\sim\right) &= \Tpre \Tpost \left\{\Tpost \cdot 0 + \Tpre \left(\Delta - \Tpre\right) - \left(\Delta - \Tpre\right)\cdot 0 - \Tpre\Delta \right. \\
		&\quad\phantom{\Tpre \Tpost \big\{} \left.
		+ \Tpre \left[\Tpre + \Tpost - 0 - (\Delta - \Tpre) - (\Tpost - \Delta) - 0 - \Tpre\right] \right\} \\
		&= 0.
	\end{align*}
	
	\item For $\max(\Tpre, \Tpost) < \Delta < \Tpre + \Tpost$, we have $\min\left(\Tpre, \Tpost, \Delta, \left(\Tpre + \Tpost - \Delta\right)_+\right) = \Tpre + \Tpost - \Delta$ and therefore
	\begin{align*}
		g_2\left(\sim\right) &= \Tpre \Tpost \left\{
		\Tpost (\Delta - \Tpost) + \Tpre (\Delta - \Tpre) - (\Delta - \Tpre) (\Delta - \Tpost) - \Tpre\Tpost 
		\right. \\
		&\quad\phantom{\Tpre \Tpost \big\{} \left.
		+ \left(\Tpre + \Tpost - \Delta\right) \left[ \Tpre + \Tpost - 0 - (\Delta - \Tpre) - 0 - (\Delta - \Tpost) - (\Tpre + \Tpost - \Delta) \right] \right\} \\
		&= \Tpre \Tpost \left\{-\left(\Tpre + \Tpost\right)^2 + 2\Delta(\Tpre + \Tpost) - \Delta^2 + (\Tpre + \Tpost - \Delta)^2\right\} \\
		&= 0.
	\end{align*}
	
	\item For $\max(\Tpre, \Tpost) < \Tpre + \Tpost < \Delta$, we have $\min\left(\Tpre, \Tpost, \Delta, \left(\Tpre + \Tpost - \Delta\right)_+\right) = 0$ and, since $\Delta - \Tpre > \Tpost$, $\min\left(\left(\Delta - \Tpre\right)_+, \Tpost\right) = \Tpost$; therefore
	\begin{align*}
		g_2\left(\sim\right) &= \Tpre \Tpost 
		\left\{ \Tpost\left(\Delta - \Tpost\right) + \Tpre\Tpost - \Tpost\left(\Delta - \Tpost\right) - \Tpre\Tpost + 0 \right\} \\
		&= 0
	\end{align*}
\end{itemize}

Since we have partitioned the positive real line with the above, $g_2(\Tpre,\Tpost,\Delta) = 0$ for all positive $\Tpre$, $\Tpost$, and $\Delta$, so that 
\begin{align*}
	h_4\left(\Tpre, \Tpost, \Delta, \rho_\zeta, \phi_\zeta, \psi_\zeta\right) &=  \left(\Tpre^2 \left(\Tpost - \Delta\right)_+ + \Tpost^2 \left(\Tpre - \Delta\right)_+ - \Tpre\Tpost \min\left(\cdots\right)\right) \left(1-\rho_\zeta\right) \numbereqn \label{eq:h4} \\
	&= g_1(\Tpre, \Tpost, \Delta) (1-\rho_\zeta).
\end{align*}

Finally, we turn to $h_5$:
\begin{align*}
	h_5\left(\Tpre, \Tpost, \Delta, \rho_\zeta, \phi_\zeta, \psi_\zeta\right) 
	&= \Tpost^2\min(\Tpre, \Delta) \left(\Tpre-\Delta\right)_+ \psi_\zeta 
	+ \Tpost^2 \min(\Tpre, \Delta) \min(\cdots) \psi_\zeta \\
	&\quad + \Tpost^2 \left(\Tpre - \Delta\right)_+ \phi_\zeta +
	\Tpost^2\left(\Tpre - \Delta\right)_+ \left(\left(\Tpre - \Delta\right)_+ - 1\right) \psi_\zeta \\
	&\quad + \Tpost^2 \left(\Tpre - \Delta\right)_+ \left(\Delta - \Tpost\right)_+ \psi_\zeta +
	\Tpost^2 \left(\Tpre - \Delta\right)_+ \min(\cdots) \psi_\zeta \\
	&\quad + \Tpost^2 \min(\Tpre, \Delta) \left(\Delta - \Tpost\right)_+ \psi_\zeta + \Tpre^2 \min(\cdots) \left(\Tpost - \Delta\right)_+ \psi_\zeta \\
	&\quad + \Tpre^2 \left(\Tpost - \Delta\right)_+ \phi_\zeta + \Tpre^2 \left(\left(\Tpost - \Delta\right) - 1 \right) \psi_\zeta \\
	&\quad + \Tpre^2 \min\left( \left(\Delta - \Tpre\right)_+, \Tpost\right) \left(\Tpost - \Delta\right)_+ \psi_\zeta + \Tpre^2 \min(\cdots) \min(\Tpost, \Delta)\psi_\zeta \\
	&\quad + \Tpre^2 \left(\Tpost - \Delta\right)_+ \min(\Tpost, \Delta) \psi_\zeta + \Tpre^2 \min\left( \left(\Delta - \Tpre\right)_+, \Tpost\right) \min(\Tpost, \Delta) \psi_\zeta \\
	&\quad - \Tpre\Tpost\min(\cdots) \left(\Tpre - \Delta\right)_+ \psi_\zeta - \Tpre\Tpost \left(\Tpost - \Delta\right)_+ \left(\Tpre - \Delta\right)_+ \psi_\zeta \\
	&\quad - \Tpre\Tpost \min\left( \left(\Delta - \Tpre\right)_+, \Tpost\right) \min(\cdots) \psi_\zeta - \Tpre\Tpost\min(\cdots) \phi_\zeta \\
	&\quad - \Tpre\Tpost \min(\cdots) \left(\min(\cdots) - 1 \right) \psi_\zeta - \Tpre\Tpost\left(\Tpost - \Delta\right)_+ \min(\cdots) \psi_\zeta \\
	&\quad - \Tpre\Tpost \min\left( \left(\Delta - \Tpre\right)_+, \Tpost\right) \left(\Delta - \Tpost\right)_+ \psi_\zeta - \Tpre\Tpost \min(\cdots) \left(\Delta - \Tpost\right)_+ \psi_\zeta \\
	&\quad - \Tpre\Tpost \min(\Tpre, \Delta) \left(\Tpost - \Delta\right)_+ \psi_\zeta - \Tpre\Tpost\left(\Tpost - \Delta\right)_+ \left(\Tpre - \Delta \right)_+ \psi_\zeta \\
	&\quad - \Tpre\Tpost\min(\Tpre,\Delta)\min(\Tpost, \Delta)\psi_\zeta - \Tpre\Tpost (\Tpre - \Delta)_+ \min(\Tpost,\Delta)\psi_\zeta.
\end{align*}

Gathering terms as above, we have
\begin{align*}
	h_5(\sim) &= \left(\Tpost^2\left(\Tpre - \Delta\right)_+ + \Tpre^2 \left(\Tpost - \Delta\right)_+ - \Tpre\Tpost\min(\cdots)\right)\left(\phi_\zeta - \psi_\zeta\right) \\
	&\qquad + \psi_\zeta \left\{\Tpre\Tpost \min(\cdots) \left[
	\Tpost + \Tpre - \left(\Tpre - \Delta\right)_+ - \min\left(\left( \Delta - \Tpre\right)_+, \Tpost\right) - \left(\Tpost - \Delta\right)_+ \right.\right.\\
	&\qquad\phantom{+ \psi_\zeta \Tpre\Tpost \big[\min(\cdots) \big[} \left.\quad 
	- \left(\Delta - \Tpost\right)_+ - \min(\cdots) \right] \\
	&\qquad \phantom{+ \psi_\zeta \big[} \quad
	+ \Tpost^2\left(\Tpre - \Delta\right)_+ \left(\Delta - \Tpost\right)_+ + \Tpre^2 \left(\Tpost - \Delta\right)_+ \min\left(\left(\Delta - \Tpre\right)_+, \Tpost\right) \\
	&\qquad \phantom{+ \psi_\zeta \big[} \quad
	+ \Tpost^2 \min(\Tpre, \Delta)\left(\Delta - \Tpost\right)_+ + \Tpre^2 \min(\Tpost, \Delta) \min\left(\left(\Delta - \Tpre\right)_+, \Tpost\right) \\
	&\qquad \phantom{+ \psi_\zeta \big[} \quad
	\left.- \Tpre\Tpost \min\left(\left(\Delta - \Tpre\right)_+, \Tpost\right) \left(\Delta - \Tpost\right)_+ - \Tpre\Tpost \min(\Tpre, \Delta) \min(\Tpost, \Delta)\right\} \\
	&= \left(\Tpost^2\left(\Tpre - \Delta\right)_+ + \Tpre^2 \left(\Tpost - \Delta\right)_+ - \Tpre\Tpost\min(\cdots)\right)\left(\phi_\zeta - \psi_\zeta\right) \\
	&\qquad + \psi_\zeta \Tpre\Tpost \left\{ \min(\cdots) \left[
	\Tpost + \Tpre - \left(\Tpre - \Delta\right)_+ - \min\left(\left( \Delta - \Tpre\right)_+, \Tpost\right) - \left(\Tpost - \Delta\right)_+ \right.\right.\\
	&\qquad\phantom{+ \psi_\zeta \Tpre\Tpost \big[\min(\cdots) \big[} \left.\quad 
	- \left(\Delta - \Tpost\right)_+ - \min(\cdots) \right] \\
	&\qquad \phantom{+ \psi_\zeta \Tpre\Tpost \big[} \quad 
	+ \Tpost \left(\Delta - \Tpost\right)_+ + \Tpre \min\left(\left(\Delta - \Tpre\right)_+, \Tpost\right)  \\
	&\qquad \phantom{+ \psi_\zeta \Tpre\Tpost \big[} \quad \left.
	- \min\left(\left(\Delta - \Tpre\right)_+, \Tpost\right) \left(\Delta - \Tpost\right) - \min(\Tpre,\Delta) \min(\Tpost, \Delta)\right\} \\
	&= \left(\Tpost^2\left(\Tpre - \Delta\right)_+ + \Tpre^2 \left(\Tpost - \Delta\right)_+ - \Tpre\Tpost\min(\cdots)\right)\left(\phi_\zeta - \psi_\zeta\right) + \psi_\zeta g_2(\Tpre, \Tpost, \Delta) \\
	&= \left(\Tpost^2\left(\Tpre - \Delta\right)_+ + \Tpre^2 \left(\Tpost - \Delta\right)_+ - \Tpre\Tpost\min(\cdots)\right)\left(\phi_\zeta - \psi_\zeta\right) \numbereqn \label{eq:h5} \\
	&= g_1(\Tpre, \Tpost, \Delta) (\phi_\zeta - \psi_\zeta),
\end{align*}
since we saw above that $g_2(\Tpre,\Tpost,\Delta)$ is identically zero for non-negative $\Tpre$, $\Tpost$, and $\Delta$.

Now, we substitute \cref{eq:h1,eq:h2,eq:h3,eq:h4,eq:h5} back into \cref{eq:combined-covar-init-hfuns}:
\begin{align*}
	\cov{\widehat{\mathrm{ATT}}_\gamma}{\widehat{\mathrm{ATT}}_\nu} =& \left(\Tpre\Tpost\right)^{-2}\left(N_\gamma^\text{ctrl} N_\nu^\text{ctrl}\right)^{-1} g_1(\Tpre, \Tpost, \Delta) \\
	& \times \sum_{\zeta\in\Xi_\text{ctrl}}
	\left[ (\phi_\zeta - \psi_\zeta) \left( \ndisj{\gamma}{\nu}(\zeta) \left(\ndisj{\nu}{\gamma}(\zeta) + \novrlp{\gamma}{\nu}(\zeta)\right) + \novrlp{\gamma}{\nu}(\zeta) \left(\ndisj{\nu}{\gamma}(\zeta) + \novrlp{\gamma}{\nu}(\zeta) - 1\right) \right) \right. \\
	& \phantom{\times \sum_{\zeta\in\Xi_\text{ctrl}}\big[ \times \big[}
	\left. + (1-\rho_\zeta) \novrlp{\gamma}{\nu}(\zeta) \right] \\
	=& \left(\Tpre\Tpost\right)^{-2}\left(N_\gamma^\text{ctrl} N_\nu^\text{ctrl}\right)^{-1} g_1(\Tpre, \Tpost, \Delta) \\
	& \times \sum_{\zeta\in\Xi_\text{ctrl}} \left[ (\phi_\zeta - \psi_\zeta) \left( \left(\ndisj{\gamma}{\nu}(\zeta) + \novrlp{\gamma}{\nu}(\zeta)\right) N_\nu(\zeta) - \novrlp{\gamma}{\nu}(\zeta) \right) + (1-\rho_\zeta) \novrlp{\gamma}{\nu}(\zeta) \right] \\
	=& \left(\Tpre\Tpost\right)^{-2}\left(N_\gamma^\text{ctrl} N_\nu^\text{ctrl}\right)^{-1} g_1(\Tpre, \Tpost, \Delta) \\
	& \times \sum_{\zeta\in\Xi_\text{ctrl}} \left[ N_\gamma(\zeta) N_\nu(\zeta) (\phi_\zeta - \psi_\zeta)  - \novrlp{\gamma}{\nu}(\zeta) \left((1-\rho_\zeta) - (\phi_\zeta - \psi_\zeta)\right)\right].
\end{align*}

Substituting the definition of $g_1(\Tpre, \Tpost, \Delta)$, we have 
\begin{equation}
	\boxed{
		\begin{split}
			\cov{\widehat{\mathrm{ATT}}_\gamma}{\widehat{\mathrm{ATT}}_\nu} = \left(\Tpre\Tpost\right)^{-2}\left(N_\gamma^\text{ctrl} N_\nu^\text{ctrl}\right)^{-1} \left(\Tpost^2\left(\Tpre - \Delta\right)_+ + \Tpre^2 \left(\Tpost - \Delta\right)_+ - \Tpre\Tpost\min(\cdots)\right) \\
			\times \sum_{\zeta\in\Xi_\text{ctrl}} \left[ N_\gamma(\zeta) N_\nu(\zeta) (\phi_\zeta - \psi_\zeta)  - \novrlp{\gamma}{\nu}(\zeta) \left((1-\rho_\zeta) - (\phi_\zeta - \psi_\zeta)\right)\right].
		\end{split}
	}
\end{equation}


\section{Additional Simulation Results} \label{appendix:sims}

We reproduce Table 1 in the main manuscript, this time with 10 control states (vs. 3). Interpretation of results remains unchanged, though because the expected correlations are smaller, the improvements in coverage and standard error from $\widehat{\ATT}_\text{gls}$ over $\widehat{\ATT}_\text{ivw}$ are minimal.

\begin{table}
\centering
\begingroup\scriptsize
\begin{tabular}{lllllll|llllllll}
  \toprule & & & & & & & \multicolumn{2}{c}{$\cor{\widehat{\ATT}_\gamma}{\widehat{\ATT}_\nu}$} & \multicolumn{3}{c}{$\widehat{\mathrm{ATT}}_\text{ivw}$} & \multicolumn{3}{c}{$\widehat{\mathrm{ATT}}_\text{gls}$} \\ \cmidrule{8-15} 
$T_\text{pre}$ & $T_\text{post}$ & $\Delta$ & \% Shared & $\rho$ & $\phi$ & $\psi$ & True Cor. & Est. Bias & Bias & SE & 95\% Covg. & Bias & SE & 95\% Covg. \\ 
  \midrule
1 & 1 & 1 & 0.25 & 0.10 & 0.06 & 0.02 & -0.039 & 0.00 & 0.00 & 0.25 & 0.955 & 0.00 & 0.24 & 0.951 \\ 
   &  &  &  & 0.60 & 0.40 & 0.20 & -0.045 & 0.00 & -0.01 & 1.05 & 0.955 & -0.01 & 1.03 & 0.951 \\ 
   &  &  & 0.75 & 0.10 & 0.06 & 0.02 & -0.043 & 0.01 & 0.00 & 0.25 & 0.954 & 0.00 & 0.24 & 0.948 \\ 
   &  &  &  & 0.60 & 0.40 & 0.20 & -0.045 & -0.01 & 0.01 & 1.05 & 0.954 & 0.01 & 1.03 & 0.949 \\ 
   &  & 2 & 0.25 & 0.10 & 0.06 & 0.02 & 0.000 & 0.00 & 0.00 & 0.25 & 0.951 & 0.00 & 0.25 & 0.951 \\ 
   &  &  &  & 0.60 & 0.40 & 0.20 & 0.000 & -0.00 & 0.01 & 1.05 & 0.950 & 0.01 & 1.05 & 0.950 \\ 
   &  &  & 0.75 & 0.10 & 0.06 & 0.02 & 0.000 & -0.00 & -0.00 & 0.25 & 0.952 & -0.00 & 0.25 & 0.952 \\ 
   &  &  &  & 0.60 & 0.40 & 0.20 & 0.000 & 0.01 & -0.02 & 1.05 & 0.950 & -0.02 & 1.05 & 0.950 \\ 
  5 & 5 & 3 & 0.25 & 0.10 & 0.06 & 0.02 & 0.008 & 0.01 & 0.00 & 0.11 & 0.949 & 0.00 & 0.11 & 0.949 \\ 
   &  &  &  & 0.60 & 0.40 & 0.20 & 0.009 & 0.00 & 0.00 & 0.47 & 0.950 & 0.00 & 0.47 & 0.951 \\ 
   &  &  & 0.75 & 0.10 & 0.06 & 0.02 & 0.009 & 0.01 & 0.00 & 0.11 & 0.948 & 0.00 & 0.11 & 0.949 \\ 
   &  &  &  & 0.60 & 0.40 & 0.20 & 0.009 & -0.01 & -0.01 & 0.47 & 0.950 & -0.01 & 0.47 & 0.952 \\ 
   &  & 6 & 0.25 & 0.10 & 0.06 & 0.02 & -0.032 & -0.00 & 0.00 & 0.11 & 0.951 & 0.00 & 0.11 & 0.949 \\ 
   &  &  &  & 0.60 & 0.40 & 0.20 & -0.036 & -0.00 & -0.00 & 0.47 & 0.953 & -0.00 & 0.46 & 0.949 \\ 
   &  &  & 0.75 & 0.10 & 0.06 & 0.02 & -0.035 & -0.01 & 0.00 & 0.11 & 0.956 & 0.00 & 0.11 & 0.952 \\ 
   &  &  &  & 0.60 & 0.40 & 0.20 & -0.036 & -0.00 & -0.00 & 0.47 & 0.958 & -0.00 & 0.46 & 0.952 \\ 
  7 & 3 & 3 & 0.25 & 0.10 & 0.06 & 0.02 & -0.010 & 0.01 & 0.00 & 0.12 & 0.949 & 0.00 & 0.12 & 0.948 \\ 
   &  &  &  & 0.60 & 0.40 & 0.20 & -0.012 & 0.01 & 0.00 & 0.51 & 0.953 & 0.00 & 0.51 & 0.951 \\ 
   &  &  & 0.75 & 0.10 & 0.06 & 0.02 & -0.011 & -0.02 & 0.00 & 0.12 & 0.950 & 0.00 & 0.12 & 0.949 \\ 
   &  &  &  & 0.60 & 0.40 & 0.20 & -0.012 & -0.01 & 0.00 & 0.51 & 0.952 & 0.00 & 0.51 & 0.951 \\ 
   &  & 6 & 0.25 & 0.10 & 0.06 & 0.02 & -0.020 & -0.00 & 0.00 & 0.12 & 0.955 & 0.00 & 0.12 & 0.953 \\ 
   &  &  &  & 0.60 & 0.40 & 0.20 & -0.023 & -0.02 & -0.00 & 0.51 & 0.957 & -0.00 & 0.51 & 0.955 \\ 
   &  &  & 0.75 & 0.10 & 0.06 & 0.02 & -0.022 & 0.01 & 0.00 & 0.12 & 0.953 & 0.00 & 0.12 & 0.952 \\ 
   &  &  &  & 0.60 & 0.40 & 0.20 & -0.023 & 0.00 & -0.01 & 0.51 & 0.953 & -0.01 & 0.51 & 0.950 \\ 
   \bottomrule
\end{tabular}
\endgroup
\caption{Simulated between-estimate correlations along with standard error and 95\% confidence interval coverage for aggregated estimates of $\widehat{\ATT}$ for a variety of generative model parameters, 100 individuals per state, and 10 control states. Reported correlations are averages of 100 estimates generated from 100 simulations.} 
\label{tab:sims10}
\end{table}

\section{Medical Cannabis Law Study}\label{appendix:medical-cannabis-law-study}
\singlespacing

Here, we present sample sizes and other information needed to reproduce
the correlation results presented in the main manuscript.

\begin{itemize}
	\item \textbf{Treated States:} AR, CT, FL, LA, MD, MN, ND, NH, NY, OH, OK, PA
	\item \textbf{Control States:} AL, GA, ID, IN, IA, KS, KY, MS, NE, NC, SC, SD, TN, TX, VA, WI, WY
\end{itemize}

\subsection{Total Counts Per Analysis}\label{total-counts-per-analysis}

\begin{table}[H]
	\centering
	\begingroup\footnotesize
	\begin{tabular}{llll}
		\toprule
		Cohort & N Treated & N Control & Total \\
		\midrule
		AR & 2,788 & 202,282 & 205,070 \\
		CT & 1,919 & 111,339 & 113,258 \\ 
		FL & 45,816 & 139,519 & 185,335 \\ 
		LA & 6,253 & 207,053 & 213,306 \\ 
		MD & 12,047 & 159,088 & 171,135 \\ 
		MN & 23,216 & 121,426 & 144,642 \\ 
		ND & 5,005 & 235,505 & 240,510 \\ 
		NH & 1,241 & 134,254 & 135,495 \\ 
		NY & 26,010 & 140,642 & 166,652 \\ 
		OH & 19,907 & 235,544 & 255,451 \\ 
		OK & 3,028 & 176,845 & 179,873 \\
		PA & 6,144 & 171,175 & 177,319 \\
		\bottomrule
	\end{tabular}
	\endgroup
	\caption{Sample sizes for each treated state's cohort, including the numbers of treated and control individuals. Using the notation from the manuscript, the number of treated individuals is $N^\text{tx}_\gamma := N_{\gamma}(\gamma)$, the number of control individuals is $N^\text{ctrl}_\gamma$, and the total sample size is $N_\gamma$, for $\gamma = \text{AR, CT, ...}$.} 
\end{table}

\begin{table}[H]
	\centering
	\begingroup\footnotesize
	\begin{tabular}{rrrrrrrrrrrrr}
		\toprule
		& AR & CT & FL & LA & MD & MN & ND & NH & NY & OH & OK & PA \\ 
		\midrule
		AL & 7518 & 5665 & 7006 & 7564 & 7438 & 6019 & 8565 & 6868 & 6882 & 8563 & 8022 & 7942 \\ 
		GA & 48054 & 10180 & 12120 & 49047 & 13792 & 10502 & 49190 & 11896 & 12091 & 49102 & 14529 & 14062 \\ 
		ID & 2555 & 1439 & 2068 & 2583 & 2301 & 1702 & 2954 & 2000 & 2006 & 2962 & 2680 & 2609 \\ 
		IN & 13489 & 6179 & 9225 & 13650 & 11666 & 7629 & 15207 & 9069 & 9277 & 15193 & 14089 & 13434 \\ 
		IA & 9595 & 6357 & 8112 & 9642 & 9334 & 7107 & 10464 & 7959 & 8018 & 10464 & 10146 & 9855 \\ 
		KS & 3574 & 1910 & 2638 & 3614 & 2850 & 2112 & 3613 & 2541 & 2514 & 3641 & 3137 & 3030 \\ 
		KY & 9295 & 2011 & 2588 & 12366 & 2805 & 2119 & 13503 & 2534 & 2615 & 13407 & 3049 & 2966 \\ 
		MS & 2251 & 1284 & 1830 & 2250 & 2035 & 1389 & 2293 & 1814 & 1819 & 2289 & 2506 & 2404 \\ 
		NE & 4255 & 3589 & 7087 & 4286 & 7515 & 3814 & 4600 & 4637 & 4816 & 4564 & 4727 & 4637 \\ 
		NC & 20656 & 14726 & 19416 & 20608 & 21391 & 16994 & 25402 & 19035 & 19434 & 25419 & 23413 & 23033 \\ 
		SC & 2634 & 1521 & 1825 & 2687 & 1951 & 1628 & 2773 & 1773 & 1903 & 2793 & 2378 & 2246 \\ 
		SD & 1551 & 490 & 2087 & 1601 & 3291 & 1019 & 5368 & 1916 & 1732 & 5332 & 4399 & 3885 \\ 
		TN & 8933 & 6557 & 7737 & 8976 & 9105 & 6826 & 10388 & 7635 & 7675 & 10407 & 10052 & 9646 \\ 
		TX & 27937 & 23093 & 21978 & 27924 & 23071 & 23150 & 29395 & 21647 & 26043 & 29553 & 28273 & 27661 \\ 
		VA & 9919 & 6189 & 8622 & 9962 & 9764 & 7394 & 11126 & 8398 & 8729 & 11162 & 10149 & 9935 \\ 
		WI & 29471 & 19847 & 24755 & 29696 & 30241 & 21656 & 40049 & 24106 & 24676 & 40067 & 34674 & 33196 \\ 
		WY & 595 & 302 & 425 & 597 & 538 & 366 & 615 & 426 & 412 & 626 & 622 & 634 \\ 
		\bottomrule
	\end{tabular}
	\endgroup
	\caption{Number of individuals in each control state, by cohort} 
\end{table}

\subsection{Counts of Disjoint Control
	Individuals}\label{counts-of-disjoint-control-individuals}

\begin{table}[H]
	\centering
	\begingroup\footnotesize
	\begin{tabular}{llllllllllll}
		\toprule
		Ctrl. State & CT & FL & LA & MD & MN & ND & NH & NY & OH & OK & PA \\ 
		\midrule
		AL & 5,737 & 4,286 & 497 & 3,272 & 5,355 & 342 & 4,492 & 4,759 & 495 & 1,594 & 2,331 \\ 
		GA & 45,578 & 43,057 & 2,606 & 40,749 & 45,027 & 2,273 & 43,427 & 43,959 & 3,170 & 36,927 & 38,951 \\ 
		IA & 7,798 & 5,884 & 619 & 4,177 & 7,253 & 461 & 6,144 & 6,579 & 626 & 1,626 & 2,908 \\ 
		ID & 2,152 & 1,652 & 165 & 1,279 & 1,983 & 123 & 1,723 & 1,804 & 163 & 583 & 863 \\ 
		IN & 11,927 & 9,711 & 907 & 7,169 & 11,191 & 653 & 9,964 & 10,304 & 949 & 3,001 & 4,858 \\ 
		KS & 3,085 & 2,448 & 255 & 2,002 & 2,938 & 204 & 2,552 & 2,687 & 280 & 1,141 & 1,553 \\ 
		KY & 8,900 & 8,389 & 551 & 8,051 & 8,809 & 350 & 8,457 & 8,550 & 544 & 7,307 & 7,671 \\ 
		MS & 2,002 & 1,568 & 171 & 1,211 & 1,947 & 162 & 1,616 & 1,698 & 225 & 510 & 888 \\ 
		NC & 16,873 & 12,957 & 1,664 & 10,108 & 15,630 & 1,059 & 13,468 & 14,198 & 1,477 & 4,950 & 7,309 \\ 
		NE & 3,567 & 2,663 & 282 & 2,082 & 3,382 & 245 & 2,769 & 2,930 & 383 & 1,125 & 1,620 \\ 
		SC & 2,381 & 1,989 & 181 & 1,738 & 2,274 & 158 & 2,035 & 2,098 & 209 & 1,007 & 1,338 \\ 
		SD & 1,439 & 1,208 & 324 & 940 & 1,370 & 316 & 1,244 & 1,278 & 338 & 540 & 719 \\ 
		TN & 7,258 & 5,683 & 655 & 4,265 & 6,782 & 472 & 5,848 & 6,141 & 649 & 1,957 & 3,142 \\ 
		TX & 24,646 & 20,712 & 2,414 & 17,138 & 23,569 & 1,635 & 21,241 & 22,010 & 2,227 & 6,721 & 10,655 \\ 
		VA & 8,197 & 6,172 & 698 & 4,717 & 7,584 & 529 & 6,451 & 6,814 & 714 & 2,339 & 3,506 \\ 
		WI & 24,902 & 19,775 & 2,303 & 14,551 & 23,279 & 1,761 & 20,466 & 21,445 & 2,300 & 6,659 & 10,329 \\ 
		WY & 526 & 426 & 41 & 298 & 503 & 32 & 434 & 466 & 42 & 105 & 178 \\ 
		\bottomrule
	\end{tabular}
	\endgroup
	\caption{Counts of disjoint control individuals in Arkansas cohort, by paired analysis. Each entry is the number of individuals in a control state (row) who contribute to only the analysis for Arkansas and not the other, paired treated state (column).} 
\end{table}
\begin{table}[H]
	\centering
	\begingroup\footnotesize
	\begin{tabular}{llllllllllll}
		\toprule
		Ctrl. State & AR & FL & LA & MD & MN & ND & NH & NY & OH & OK & PA \\ 
		\midrule
		AL & 3,884 & 1,949 & 3,946 & 2,622 & 1,071 & 3,520 & 1,816 & 1,493 & 3,485 & 3,326 & 2,986 \\ 
		GA & 7,704 & 4,382 & 7,784 & 5,680 & 2,477 & 7,556 & 4,120 & 3,379 & 7,517 & 7,079 & 6,527 \\ 
		IA & 4,560 & 2,696 & 4,619 & 3,449 & 1,381 & 4,358 & 2,502 & 1,965 & 4,335 & 4,176 & 3,874 \\ 
		ID & 1,036 & 582 & 1,052 & 757 & 319 & 966 & 541 & 451 & 964 & 929 & 856 \\ 
		IN & 4,617 & 2,656 & 4,672 & 3,355 & 1,429 & 4,357 & 2,493 & 1,943 & 4,322 & 4,099 & 3,762 \\ 
		KS & 1,421 & 832 & 1,441 & 1,065 & 489 & 1,414 & 765 & 656 & 1,398 & 1,278 & 1,197 \\ 
		KY & 1,616 & 867 & 1,628 & 1,066 & 517 & 1,592 & 809 & 670 & 1,579 & 1,300 & 1,207 \\ 
		MS & 1,035 & 585 & 1,039 & 765 & 269 & 1,027 & 538 & 429 & 1,013 & 919 & 861 \\ 
		NC & 10,943 & 5,721 & 11,118 & 7,498 & 3,107 & 9,831 & 5,373 & 4,256 & 9,769 & 9,214 & 8,454 \\ 
		NE & 2,901 & 1,583 & 2,923 & 2,055 & 808 & 2,794 & 1,480 & 1,123 & 2,783 & 2,611 & 2,434 \\ 
		SC & 1,268 & 758 & 1,279 & 944 & 419 & 1,254 & 730 & 562 & 1,251 & 1,098 & 1,029 \\ 
		SD & 378 & 203 & 385 & 252 & 121 & 308 & 191 & 152 & 307 & 302 & 282 \\ 
		TN & 4,882 & 2,712 & 4,963 & 3,380 & 1,559 & 4,525 & 2,538 & 2,123 & 4,493 & 4,163 & 3,848 \\ 
		TX & 19,802 & 13,691 & 19,956 & 16,491 & 7,070 & 19,512 & 13,144 & 9,048 & 19,451 & 18,882 & 18,037 \\ 
		VA & 4,467 & 2,677 & 4,529 & 3,335 & 1,409 & 4,232 & 2,550 & 1,933 & 4,222 & 4,073 & 3,752 \\ 
		WI & 15,278 & 8,769 & 15,435 & 10,935 & 5,176 & 13,926 & 8,331 & 6,761 & 13,832 & 13,161 & 12,206 \\ 
		WY & 233 & 124 & 234 & 169 & 63 & 230 & 109 & 91 & 225 & 209 & 194 \\ 
		\bottomrule
	\end{tabular}
	\endgroup
	\caption{Counts of disjoint control individuals in Connecticut cohort, by paired analysis. Each entry is the number of individuals in a control state (row) who contribute to only the analysis for Connecticut and not the other, paired treated state (column).} 
\end{table}
\begin{table}[H]
	\centering
	\begingroup\footnotesize
	\begin{tabular}{llllllllllll}
		\toprule
		Ctrl. State & AR & CT & LA & MD & MN & ND & NH & NY & OH & OK & PA \\ 
		\midrule
		AL & 3,774 & 3,290 & 3,904 & 1,384 & 2,319 & 3,151 & 579 & 1,183 & 3,085 & 2,731 & 2,143 \\ 
		GA & 7,123 & 6,322 & 7,344 & 2,860 & 4,540 & 6,708 & 1,180 & 2,412 & 6,587 & 5,731 & 4,637 \\ 
		IA & 4,401 & 4,451 & 4,573 & 1,726 & 2,967 & 3,880 & 787 & 1,650 & 3,822 & 3,413 & 2,710 \\ 
		ID & 1,165 & 1,211 & 1,210 & 426 & 791 & 974 & 203 & 396 & 955 & 851 & 672 \\ 
		IN & 5,447 & 5,702 & 5,620 & 2,240 & 3,716 & 4,813 & 852 & 1,801 & 4,729 & 4,190 & 3,402 \\ 
		KS & 1,512 & 1,560 & 1,566 & 610 & 1,086 & 1,453 & 305 & 596 & 1,416 & 1,186 & 946 \\ 
		KY & 1,682 & 1,444 & 1,722 & 576 & 1,064 & 1,617 & 260 & 530 & 1,595 & 1,155 & 916 \\ 
		MS & 1,147 & 1,131 & 1,168 & 485 & 939 & 1,116 & 210 & 440 & 1,090 & 923 & 728 \\ 
		NC & 11,717 & 10,411 & 12,140 & 4,118 & 6,678 & 9,404 & 1,753 & 3,516 & 9,256 & 8,000 & 6,401 \\ 
		NE & 5,495 & 5,081 & 5,552 & 1,510 & 4,416 & 5,264 & 2,816 & 3,282 & 5,224 & 4,756 & 4,353 \\ 
		SC & 1,180 & 1,062 & 1,203 & 457 & 743 & 1,115 & 186 & 363 & 1,095 & 845 & 669 \\ 
		SD & 1,744 & 1,800 & 1,758 & 364 & 1,324 & 828 & 283 & 644 & 818 & 745 & 605 \\ 
		TN & 4,487 & 3,892 & 4,658 & 1,489 & 2,644 & 3,762 & 674 & 1,370 & 3,695 & 3,071 & 2,449 \\ 
		TX & 14,753 & 12,576 & 15,154 & 6,796 & 8,541 & 13,985 & 2,171 & 4,492 & 13,784 & 12,460 & 10,507 \\ 
		VA & 4,875 & 5,110 & 5,040 & 1,842 & 3,270 & 4,236 & 787 & 1,654 & 4,166 & 3,733 & 3,013 \\ 
		WI & 15,059 & 13,677 & 15,449 & 5,076 & 8,832 & 11,810 & 2,358 & 4,702 & 11,538 & 10,025 & 7,980 \\ 
		WY & 256 & 247 & 264 & 104 & 156 & 250 & 35 & 93 & 241 & 207 & 150 \\ 
		\bottomrule
	\end{tabular}
	\endgroup
	\caption{Counts of disjoint control individuals in Florida cohort, by paired analysis. Each entry is the number of individuals in a control state (row) who contribute to only the analysis for Florida and not the other, paired treated state (column).} 
\end{table}
\begin{table}[H]
	\centering
	\begingroup\footnotesize
	\begin{tabular}{llllllllllll}
		\toprule
		Ctrl. State & AR & CT & FL & MD & MN & ND & NH & NY & OH & OK & PA \\ 
		\midrule
		AL & 543 & 5,845 & 4,462 & 3,476 & 5,502 & 781 & 4,678 & 4,927 & 900 & 1,872 & 2,624 \\ 
		GA & 3,599 & 46,651 & 44,271 & 42,090 & 46,174 & 5,291 & 44,685 & 45,199 & 6,067 & 38,459 & 40,475 \\ 
		IA & 666 & 7,904 & 6,103 & 4,462 & 7,402 & 1,014 & 6,367 & 6,784 & 1,152 & 2,040 & 3,313 \\ 
		ID & 193 & 2,196 & 1,725 & 1,359 & 2,038 & 285 & 1,798 & 1,873 & 314 & 698 & 976 \\ 
		IN & 1,068 & 12,143 & 10,045 & 7,598 & 11,458 & 1,494 & 10,303 & 10,631 & 1,733 & 3,624 & 5,457 \\ 
		KS & 295 & 3,145 & 2,542 & 2,128 & 3,015 & 464 & 2,653 & 2,785 & 522 & 1,311 & 1,714 \\ 
		KY & 3,622 & 11,983 & 11,500 & 11,177 & 11,904 & 1,165 & 11,567 & 11,654 & 1,359 & 10,476 & 10,828 \\ 
		MS & 170 & 2,005 & 1,588 & 1,249 & 1,957 & 276 & 1,647 & 1,725 & 326 & 568 & 951 \\ 
		NC & 1,616 & 17,000 & 13,332 & 10,616 & 15,887 & 2,323 & 13,894 & 14,561 & 2,648 & 5,775 & 8,083 \\ 
		NE & 313 & 3,620 & 2,751 & 2,197 & 3,463 & 486 & 2,876 & 3,029 & 599 & 1,280 & 1,783 \\ 
		SC & 234 & 2,445 & 2,065 & 1,825 & 2,350 & 333 & 2,126 & 2,184 & 379 & 1,141 & 1,474 \\ 
		SD & 374 & 1,496 & 1,272 & 1,012 & 1,433 & 425 & 1,312 & 1,344 & 445 & 632 & 804 \\ 
		TN & 698 & 7,382 & 5,897 & 4,551 & 6,951 & 996 & 6,089 & 6,366 & 1,137 & 2,365 & 3,541 \\ 
		TX & 2,401 & 24,787 & 21,100 & 17,699 & 23,836 & 3,583 & 21,670 & 22,388 & 4,058 & 8,031 & 11,885 \\ 
		VA & 741 & 8,302 & 6,380 & 4,985 & 7,730 & 1,142 & 6,683 & 7,031 & 1,281 & 2,742 & 3,889 \\ 
		WI & 2,528 & 25,284 & 20,390 & 15,378 & 23,781 & 3,587 & 21,106 & 22,031 & 4,030 & 7,852 & 11,532 \\ 
		WY & 43 & 529 & 436 & 313 & 507 & 60 & 442 & 474 & 69 & 124 & 202 \\ 
		\bottomrule
	\end{tabular}
	\endgroup
	\caption{Counts of disjoint control individuals in Louisiana cohort, by paired analysis. Each entry is the number of individuals in a control state (row) who contribute to only the analysis for Louisiana and not the other, paired treated state (column).} 
\end{table}
\begin{table}[H]
	\centering
	\begingroup\footnotesize
	\begin{tabular}{llllllllllll}
		\toprule
		Ctrl. State & AR & CT & FL & LA & MN & ND & NH & NY & OH & OK & PA \\ 
		\midrule
		AL & 3,192 & 4,395 & 1,816 & 3,350 & 3,608 & 2,423 & 2,192 & 2,668 & 2,336 & 1,900 & 1,119 \\ 
		GA & 6,487 & 9,292 & 4,532 & 6,835 & 7,932 & 5,920 & 5,273 & 6,238 & 5,777 & 4,538 & 2,856 \\ 
		IA & 3,916 & 6,426 & 2,948 & 4,154 & 5,227 & 3,202 & 3,438 & 4,161 & 3,117 & 2,556 & 1,546 \\ 
		ID & 1,025 & 1,619 & 659 & 1,077 & 1,271 & 768 & 800 & 951 & 745 & 613 & 366 \\ 
		IN & 5,346 & 8,842 & 4,681 & 5,614 & 7,312 & 4,277 & 5,164 & 5,837 & 4,131 & 3,320 & 2,053 \\ 
		KS & 1,278 & 2,005 & 822 & 1,364 & 1,644 & 1,198 & 1,008 & 1,248 & 1,155 & 869 & 515 \\ 
		KY & 1,561 & 1,860 & 793 & 1,616 & 1,571 & 1,482 & 948 & 1,143 & 1,455 & 881 & 555 \\ 
		MS & 995 & 1,516 & 690 & 1,034 & 1,369 & 953 & 809 & 978 & 914 & 700 & 398 \\ 
		NC & 10,843 & 14,163 & 6,093 & 11,399 & 11,211 & 7,754 & 7,212 & 8,636 & 7,542 & 5,863 & 3,592 \\ 
		NE & 5,342 & 5,981 & 1,938 & 5,426 & 5,439 & 5,058 & 4,138 & 4,525 & 5,011 & 4,352 & 3,819 \\ 
		SC & 1,055 & 1,374 & 583 & 1,089 & 1,129 & 961 & 704 & 841 & 935 & 620 & 365 \\ 
		SD & 2,680 & 3,053 & 1,568 & 2,702 & 2,662 & 919 & 1,781 & 2,082 & 896 & 774 & 499 \\ 
		TN & 4,437 & 5,928 & 2,857 & 4,680 & 4,886 & 3,446 & 3,265 & 3,857 & 3,355 & 2,522 & 1,548 \\ 
		TX & 12,272 & 16,469 & 7,889 & 12,846 & 13,705 & 11,138 & 9,087 & 10,774 & 10,884 & 8,904 & 5,985 \\ 
		VA & 4,562 & 6,910 & 2,984 & 4,787 & 5,440 & 3,684 & 3,486 & 4,155 & 3,588 & 3,002 & 1,974 \\ 
		WI & 15,321 & 21,329 & 10,562 & 15,923 & 17,401 & 10,429 & 12,076 & 13,969 & 10,081 & 7,917 & 4,774 \\ 
		WY & 241 & 405 & 217 & 254 & 335 & 228 & 237 & 286 & 219 & 174 & 84 \\ 
		\bottomrule
	\end{tabular}
	\endgroup
	\caption{Counts of disjoint control individuals in Maryland cohort, by paired analysis. Each entry is the number of individuals in a control state (row) who contribute to only the analysis for Maryland and not the other, paired treated state (column).} 
\end{table}
\begin{table}[H]
	\centering
	\begingroup\footnotesize
	\begin{tabular}{llllllllllll}
		\toprule
		Ctrl. State & AR & CT & FL & LA & MD & ND & NH & NY & OH & OK & PA \\ 
		\midrule
		AL & 3,856 & 1,425 & 1,332 & 3,957 & 2,189 & 3,396 & 1,144 & 746 & 3,345 & 3,125 & 2,695 \\ 
		GA & 7,475 & 2,799 & 2,922 & 7,629 & 4,642 & 7,229 & 2,514 & 1,508 & 7,135 & 6,527 & 5,797 \\ 
		IA & 4,765 & 2,131 & 1,962 & 4,867 & 3,000 & 4,411 & 1,693 & 997 & 4,371 & 4,116 & 3,646 \\ 
		ID & 1,130 & 582 & 425 & 1,157 & 672 & 1,021 & 368 & 236 & 1,010 & 952 & 834 \\ 
		IN & 5,331 & 2,879 & 2,120 & 5,437 & 3,275 & 4,924 & 1,873 & 1,065 & 4,868 & 4,542 & 4,014 \\ 
		KS & 1,476 & 691 & 560 & 1,513 & 906 & 1,445 & 472 & 324 & 1,414 & 1,249 & 1,132 \\ 
		KY & 1,633 & 625 & 595 & 1,657 & 885 & 1,597 & 497 & 283 & 1,576 & 1,239 & 1,097 \\ 
		MS & 1,085 & 374 & 498 & 1,096 & 723 & 1,069 & 415 & 276 & 1,052 & 947 & 857 \\ 
		NC & 11,968 & 5,375 & 4,256 & 12,273 & 6,814 & 10,346 & 3,713 & 2,205 & 10,233 & 9,414 & 8,283 \\ 
		NE & 2,941 & 1,033 & 1,143 & 2,991 & 1,738 & 2,797 & 999 & 567 & 2,775 & 2,542 & 2,283 \\ 
		SC & 1,268 & 526 & 546 & 1,291 & 806 & 1,240 & 500 & 266 & 1,229 & 1,048 & 948 \\ 
		SD & 838 & 650 & 256 & 851 & 390 & 557 & 220 & 122 & 555 & 531 & 477 \\ 
		TN & 4,675 & 1,828 & 1,733 & 4,801 & 2,607 & 4,176 & 1,490 & 949 & 4,128 & 3,696 & 3,266 \\ 
		TX & 18,782 & 7,127 & 9,713 & 19,062 & 13,784 & 18,273 & 8,882 & 3,581 & 18,149 & 17,298 & 16,058 \\ 
		VA & 5,059 & 2,614 & 2,042 & 5,162 & 3,070 & 4,619 & 1,836 & 1,051 & 4,582 & 4,305 & 3,821 \\ 
		WI & 15,464 & 6,985 & 5,733 & 15,741 & 8,816 & 13,335 & 5,064 & 2,959 & 13,146 & 12,155 & 10,784 \\ 
		WY & 274 & 127 & 97 & 276 & 163 & 268 & 80 & 48 & 261 & 237 & 199 \\ 
		\bottomrule
	\end{tabular}
	\endgroup
	\caption{Counts of disjoint control individuals in Minnesota cohort, by paired analysis. Each entry is the number of individuals in a control state (row) who contribute to only the analysis for Minnesota and not the other, paired treated state (column).} 
\end{table}
\begin{table}[H]
	\centering
	\begingroup\footnotesize
	\begin{tabular}{llllllllllll}
		\toprule
		Ctrl. State & AR & CT & FL & LA & MD & MN & NH & NY & OH & OK & PA \\ 
		\midrule
		AL & 1,389 & 6,420 & 4,710 & 1,782 & 3,550 & 5,942 & 4,957 & 5,239 & 209 & 1,505 & 2,395 \\ 
		GA & 3,409 & 46,566 & 43,778 & 5,434 & 41,318 & 45,917 & 44,211 & 44,739 & 1,153 & 37,138 & 39,262 \\ 
		IA & 1,330 & 8,465 & 6,232 & 1,836 & 4,332 & 7,768 & 6,528 & 6,991 & 237 & 1,436 & 2,827 \\ 
		ID & 522 & 2,481 & 1,860 & 656 & 1,421 & 2,273 & 1,949 & 2,039 & 57 & 579 & 910 \\ 
		IN & 2,371 & 13,385 & 10,795 & 3,051 & 7,818 & 12,502 & 11,099 & 11,462 & 394 & 2,921 & 5,002 \\ 
		KS & 243 & 3,117 & 2,428 & 463 & 1,961 & 2,946 & 2,539 & 2,674 & 101 & 1,038 & 1,480 \\ 
		KY & 4,558 & 13,084 & 12,532 & 2,302 & 12,180 & 12,981 & 12,612 & 12,703 & 313 & 11,356 & 11,737 \\ 
		MS & 204 & 2,036 & 1,579 & 319 & 1,211 & 1,973 & 1,638 & 1,711 & 93 & 458 & 832 \\ 
		NC & 5,805 & 20,507 & 15,390 & 7,117 & 11,765 & 18,754 & 16,073 & 16,946 & 655 & 5,313 & 8,141 \\ 
		NE & 590 & 3,805 & 2,777 & 800 & 2,143 & 3,583 & 2,905 & 3,068 & 179 & 1,050 & 1,584 \\ 
		SC & 297 & 2,506 & 2,063 & 419 & 1,783 & 2,385 & 2,121 & 2,189 & 62 & 971 & 1,331 \\ 
		SD & 4,133 & 5,186 & 4,109 & 4,192 & 2,996 & 4,906 & 4,263 & 4,489 & 148 & 1,304 & 2,110 \\ 
		TN & 1,927 & 8,356 & 6,413 & 2,408 & 4,729 & 7,738 & 6,620 & 6,971 & 248 & 1,932 & 3,309 \\ 
		TX & 3,093 & 25,814 & 21,402 & 5,054 & 17,462 & 24,518 & 22,006 & 22,787 & 815 & 5,972 & 10,185 \\ 
		VA & 1,736 & 9,169 & 6,740 & 2,306 & 5,046 & 8,351 & 7,058 & 7,450 & 275 & 2,269 & 3,582 \\ 
		WI & 12,339 & 34,128 & 27,104 & 13,940 & 20,237 & 31,728 & 28,013 & 29,219 & 957 & 9,764 & 14,559 \\ 
		WY & 52 & 543 & 440 & 78 & 305 & 517 & 450 & 480 & 15 & 95 & 179 \\ 
		\bottomrule
	\end{tabular}
	\endgroup
	\caption{Counts of disjoint control individuals in North Dakota cohort, by paired analysis. Each entry is the number of individuals in a control state (row) who contribute to only the analysis for North Dakota and not the other, paired treated state (column).} 
\end{table}
\begin{table}[H]
	\centering
	\begingroup\footnotesize
	\begin{tabular}{llllllllllll}
		\toprule
		Ctrl. State & AR & CT & FL & LA & MD & MN & ND & NY & OH & OK & PA \\ 
		\midrule
		AL & 3,842 & 3,019 & 441 & 3,982 & 1,622 & 1,993 & 3,260 & 768 & 3,200 & 2,881 & 2,303 \\ 
		GA & 7,269 & 5,836 & 956 & 7,534 & 3,377 & 3,908 & 6,917 & 1,556 & 6,807 & 6,023 & 4,976 \\ 
		IA & 4,508 & 4,104 & 634 & 4,684 & 2,063 & 2,545 & 4,023 & 1,059 & 3,970 & 3,625 & 2,942 \\ 
		ID & 1,168 & 1,102 & 135 & 1,215 & 499 & 666 & 995 & 236 & 979 & 889 & 717 \\ 
		IN & 5,544 & 5,383 & 696 & 5,722 & 2,567 & 3,313 & 4,961 & 1,192 & 4,887 & 4,399 & 3,630 \\ 
		KS & 1,519 & 1,396 & 208 & 1,580 & 699 & 901 & 1,467 & 345 & 1,429 & 1,211 & 1,003 \\ 
		KY & 1,696 & 1,332 & 206 & 1,735 & 677 & 912 & 1,643 & 322 & 1,617 & 1,212 & 990 \\ 
		MS & 1,179 & 1,068 & 194 & 1,211 & 588 & 840 & 1,159 & 285 & 1,137 & 993 & 801 \\ 
		NC & 11,847 & 9,682 & 1,372 & 12,321 & 4,856 & 5,754 & 9,706 & 2,242 & 9,577 & 8,452 & 6,902 \\ 
		NE & 3,151 & 2,528 & 366 & 3,227 & 1,260 & 1,822 & 2,942 & 591 & 2,910 & 2,476 & 2,088 \\ 
		SC & 1,174 & 982 & 134 & 1,212 & 526 & 645 & 1,121 & 227 & 1,103 & 874 & 718 \\ 
		SD & 1,609 & 1,617 & 112 & 1,627 & 406 & 1,117 & 811 & 403 & 802 & 745 & 607 \\ 
		TN & 4,550 & 3,616 & 572 & 4,748 & 1,795 & 2,299 & 3,867 & 891 & 3,815 & 3,247 & 2,656 \\ 
		TX & 14,951 & 11,698 & 1,840 & 15,393 & 7,663 & 7,379 & 14,258 & 2,924 & 14,087 & 12,912 & 11,024 \\ 
		VA & 4,930 & 4,759 & 563 & 5,119 & 2,120 & 2,840 & 4,330 & 1,057 & 4,269 & 3,873 & 3,175 \\ 
		WI & 15,101 & 12,590 & 1,709 & 15,516 & 5,941 & 7,514 & 12,070 & 2,912 & 11,828 & 10,494 & 8,538 \\ 
		WY & 265 & 233 & 36 & 271 & 125 & 140 & 261 & 66 & 252 & 222 & 169 \\ 
		\bottomrule
	\end{tabular}
	\endgroup
	\caption{Counts of disjoint control individuals in New Hampshire cohort, by paired analysis. Each entry is the number of individuals in a control state (row) who contribute to only the analysis for New Hampshire and not the other, paired treated state (column).} 
\end{table}
\begin{table}[H]
	\centering
	\begingroup\footnotesize
	\begin{tabular}{llllllllllll}
		\toprule
		Ctrl. State & AR & CT & FL & LA & MD & MN & ND & NH & OH & OK & PA \\ 
		\midrule
		AL & 4,123 & 2,710 & 1,059 & 4,245 & 2,112 & 1,609 & 3,556 & 782 & 3,501 & 3,220 & 2,674 \\ 
		GA & 7,996 & 5,290 & 2,383 & 8,243 & 4,537 & 3,097 & 7,640 & 1,751 & 7,533 & 6,831 & 5,845 \\ 
		IA & 5,002 & 3,626 & 1,556 & 5,160 & 2,845 & 1,908 & 4,545 & 1,118 & 4,486 & 4,205 & 3,541 \\ 
		ID & 1,255 & 1,018 & 334 & 1,296 & 656 & 540 & 1,091 & 242 & 1,075 & 994 & 835 \\ 
		IN & 6,092 & 5,041 & 1,853 & 6,258 & 3,448 & 2,713 & 5,532 & 1,400 & 5,457 & 5,028 & 4,305 \\ 
		KS & 1,627 & 1,260 & 472 & 1,685 & 912 & 726 & 1,575 & 318 & 1,537 & 1,339 & 1,147 \\ 
		KY & 1,870 & 1,274 & 557 & 1,903 & 953 & 779 & 1,815 & 403 & 1,793 & 1,406 & 1,204 \\ 
		MS & 1,266 & 964 & 429 & 1,294 & 762 & 706 & 1,237 & 290 & 1,215 & 1,083 & 913 \\ 
		NC & 12,976 & 8,964 & 3,534 & 13,387 & 6,679 & 4,645 & 10,978 & 2,641 & 10,833 & 9,833 & 8,339 \\ 
		NE & 3,491 & 2,350 & 1,011 & 3,559 & 1,826 & 1,569 & 3,284 & 770 & 3,252 & 2,870 & 2,513 \\ 
		SC & 1,367 & 944 & 441 & 1,400 & 793 & 541 & 1,319 & 357 & 1,306 & 1,092 & 953 \\ 
		SD & 1,459 & 1,394 & 289 & 1,475 & 523 & 835 & 853 & 219 & 845 & 800 & 685 \\ 
		TN & 4,883 & 3,241 & 1,308 & 5,065 & 2,427 & 1,798 & 4,258 & 931 & 4,204 & 3,697 & 3,144 \\ 
		TX & 20,116 & 11,998 & 8,557 & 20,507 & 13,746 & 6,474 & 19,435 & 7,320 & 19,273 & 18,230 & 16,506 \\ 
		VA & 5,624 & 4,473 & 1,761 & 5,798 & 3,120 & 2,386 & 5,053 & 1,388 & 4,993 & 4,655 & 3,984 \\ 
		WI & 16,650 & 11,590 & 4,623 & 17,011 & 8,404 & 5,979 & 13,846 & 3,482 & 13,610 & 12,449 & 10,547 \\ 
		WY & 283 & 201 & 80 & 289 & 160 & 94 & 277 & 52 & 267 & 241 & 194 \\ 
		\bottomrule
	\end{tabular}
	\endgroup
	\caption{Counts of disjoint control individuals in New York cohort, by paired analysis. Each entry is the number of individuals in a control state (row) who contribute to only the analysis for New York and not the other, paired treated state (column).} 
\end{table}
\begin{table}[H]
	\centering
	\begingroup\footnotesize
	\begin{tabular}{llllllllllll}
		\toprule
		Ctrl. State & AR & CT & FL & LA & MD & MN & ND & NH & NY & OK & PA \\ 
		\midrule
		AL & 1,540 & 6,383 & 4,642 & 1,899 & 3,461 & 5,889 & 207 & 4,895 & 5,182 & 1,362 & 2,246 \\ 
		GA & 4,218 & 46,439 & 43,569 & 6,122 & 41,087 & 45,735 & 1,065 & 44,013 & 44,544 & 36,774 & 38,921 \\ 
		IA & 1,495 & 8,442 & 6,174 & 1,974 & 4,247 & 7,728 & 237 & 6,475 & 6,932 & 1,282 & 2,678 \\ 
		ID & 570 & 2,487 & 1,849 & 693 & 1,406 & 2,270 & 65 & 1,941 & 2,031 & 538 & 875 \\ 
		IN & 2,653 & 13,336 & 10,697 & 3,276 & 7,658 & 12,432 & 380 & 11,011 & 11,373 & 2,656 & 4,747 \\ 
		KS & 347 & 3,129 & 2,419 & 549 & 1,946 & 2,943 & 129 & 2,529 & 2,664 & 996 & 1,436 \\ 
		KY & 4,656 & 12,975 & 12,414 & 2,400 & 12,057 & 12,864 & 217 & 12,490 & 12,585 & 11,216 & 11,595 \\ 
		MS & 263 & 2,018 & 1,549 & 365 & 1,168 & 1,952 & 89 & 1,612 & 1,685 & 400 & 774 \\ 
		NC & 6,240 & 20,462 & 15,259 & 7,459 & 11,570 & 18,658 & 672 & 15,961 & 16,818 & 4,917 & 7,741 \\ 
		NE & 692 & 3,758 & 2,701 & 877 & 2,060 & 3,525 & 143 & 2,837 & 3,000 & 933 & 1,475 \\ 
		SC & 368 & 2,523 & 2,063 & 485 & 1,777 & 2,394 & 82 & 2,123 & 2,196 & 941 & 1,301 \\ 
		SD & 4,119 & 5,149 & 4,063 & 4,176 & 2,937 & 4,868 & 112 & 4,218 & 4,445 & 1,194 & 2,021 \\ 
		TN & 2,123 & 8,343 & 6,365 & 2,568 & 4,657 & 7,709 & 267 & 6,587 & 6,936 & 1,766 & 3,168 \\ 
		TX & 3,843 & 25,911 & 21,359 & 5,687 & 17,366 & 24,552 & 973 & 21,993 & 22,783 & 5,505 & 9,764 \\ 
		VA & 1,957 & 9,195 & 6,706 & 2,481 & 4,986 & 8,350 & 311 & 7,033 & 7,426 & 2,113 & 3,432 \\ 
		WI & 12,896 & 34,052 & 26,850 & 14,401 & 19,907 & 31,557 & 975 & 27,789 & 29,001 & 9,166 & 14,003 \\ 
		WY & 73 & 549 & 442 & 98 & 307 & 521 & 26 & 452 & 481 & 86 & 169 \\ 
		\bottomrule
	\end{tabular}
	\endgroup
	\caption{Counts of disjoint control individuals in Ohio cohort, by paired analysis. Each entry is the number of individuals in a control state (row) who contribute to only the analysis for Ohio and not the other, paired treated state (column).} 
\end{table}
\begin{table}[H]
	\centering
	\begingroup\footnotesize
	\begin{tabular}{llllllllllll}
		\toprule
		Ctrl. State & AR & CT & FL & LA & MD & MN & ND & NH & NY & OH & PA \\ 
		\midrule
		AL & 2,098 & 5,683 & 3,747 & 2,330 & 2,484 & 5,128 & 962 & 4,035 & 4,360 & 821 & 1,244 \\ 
		GA & 3,402 & 11,428 & 8,140 & 3,941 & 5,275 & 10,554 & 2,477 & 8,656 & 9,269 & 2,201 & 3,000 \\ 
		IA & 2,177 & 7,965 & 5,447 & 2,544 & 3,368 & 7,155 & 1,118 & 5,812 & 6,333 & 964 & 1,801 \\ 
		ID & 708 & 2,170 & 1,463 & 795 & 992 & 1,930 & 305 & 1,569 & 1,668 & 256 & 453 \\ 
		IN & 3,601 & 12,009 & 9,054 & 4,063 & 5,743 & 11,002 & 1,803 & 9,419 & 9,840 & 1,552 & 2,731 \\ 
		KS & 704 & 2,505 & 1,685 & 834 & 1,156 & 2,274 & 562 & 1,807 & 1,962 & 492 & 603 \\ 
		KY & 1,061 & 2,338 & 1,616 & 1,159 & 1,125 & 2,169 & 902 & 1,727 & 1,840 & 858 & 596 \\ 
		MS & 765 & 2,141 & 1,599 & 824 & 1,171 & 2,064 & 671 & 1,685 & 1,770 & 617 & 587 \\ 
		NC & 7,707 & 17,901 & 11,997 & 8,580 & 7,885 & 15,833 & 3,324 & 12,830 & 13,812 & 2,911 & 3,976 \\ 
		NE & 1,597 & 3,749 & 2,396 & 1,721 & 1,564 & 3,455 & 1,177 & 2,566 & 2,781 & 1,096 & 852 \\ 
		SC & 751 & 1,955 & 1,398 & 832 & 1,047 & 1,798 & 576 & 1,479 & 1,567 & 526 & 535 \\ 
		SD & 3,388 & 4,211 & 3,057 & 3,430 & 1,882 & 3,911 & 335 & 3,228 & 3,467 & 261 & 975 \\ 
		TN & 3,076 & 7,658 & 5,386 & 3,441 & 3,469 & 6,922 & 1,596 & 5,664 & 6,074 & 1,411 & 1,918 \\ 
		TX & 7,057 & 24,062 & 18,755 & 8,380 & 14,106 & 22,421 & 4,850 & 19,538 & 20,460 & 4,225 & 5,914 \\ 
		VA & 2,569 & 8,033 & 5,260 & 2,929 & 3,387 & 7,060 & 1,292 & 5,624 & 6,075 & 1,100 & 1,826 \\ 
		WI & 11,862 & 27,988 & 19,944 & 12,830 & 12,350 & 25,173 & 4,389 & 21,062 & 22,447 & 3,773 & 6,343 \\ 
		WY & 132 & 529 & 404 & 149 & 258 & 493 & 102 & 418 & 451 & 82 & 120 \\ 
		\bottomrule
	\end{tabular}
	\endgroup
	\caption{Counts of disjoint control individuals in Oklahoma cohort, by paired analysis. Each entry is the number of individuals in a control state (row) who contribute to only the analysis for Oklahoma and not the other, paired treated state (column).} 
\end{table}
\begin{table}[H]
	\centering
	\begingroup\footnotesize
	\begin{tabular}{llllllllllll}
		\toprule
		Ctrl. State & AR & CT & FL & LA & MD & MN & ND & NH & NY & OH & OK \\ 
		\midrule
		AL & 2,755 & 5,263 & 3,079 & 3,002 & 1,623 & 4,618 & 1,772 & 3,377 & 3,734 & 1,625 & 1,164 \\ 
		GA & 4,959 & 10,409 & 6,579 & 5,490 & 3,126 & 9,357 & 4,134 & 7,142 & 7,816 & 3,881 & 2,533 \\ 
		IA & 3,168 & 7,372 & 4,453 & 3,526 & 2,067 & 6,394 & 2,218 & 4,838 & 5,378 & 2,069 & 1,510 \\ 
		ID & 917 & 2,026 & 1,213 & 1,002 & 674 & 1,741 & 565 & 1,326 & 1,438 & 522 & 382 \\ 
		IN & 4,803 & 11,017 & 7,611 & 5,241 & 3,821 & 9,819 & 3,229 & 7,995 & 8,462 & 2,988 & 2,076 \\ 
		KS & 1,009 & 2,317 & 1,338 & 1,130 & 695 & 2,050 & 897 & 1,492 & 1,663 & 825 & 496 \\ 
		KY & 1,342 & 2,162 & 1,294 & 1,428 & 716 & 1,944 & 1,200 & 1,422 & 1,555 & 1,154 & 513 \\ 
		MS & 1,041 & 1,981 & 1,302 & 1,105 & 767 & 1,872 & 943 & 1,391 & 1,498 & 889 & 485 \\ 
		NC & 9,686 & 16,761 & 10,018 & 10,508 & 5,234 & 14,322 & 5,772 & 10,900 & 11,938 & 5,355 & 3,596 \\ 
		NE & 2,002 & 3,482 & 1,903 & 2,134 & 941 & 3,106 & 1,621 & 2,088 & 2,334 & 1,548 & 762 \\ 
		SC & 950 & 1,754 & 1,090 & 1,033 & 660 & 1,566 & 804 & 1,191 & 1,296 & 754 & 403 \\ 
		SD & 3,053 & 3,677 & 2,403 & 3,088 & 1,093 & 3,343 & 627 & 2,576 & 2,838 & 574 & 461 \\ 
		TN & 3,855 & 6,937 & 4,358 & 4,211 & 2,089 & 6,086 & 2,567 & 4,667 & 5,115 & 2,407 & 1,512 \\ 
		TX & 10,379 & 22,605 & 16,190 & 11,622 & 10,575 & 20,569 & 8,451 & 17,038 & 18,124 & 7,872 & 5,302 \\ 
		VA & 3,522 & 7,498 & 4,326 & 3,862 & 2,145 & 6,362 & 2,391 & 4,712 & 5,190 & 2,205 & 1,612 \\ 
		WI & 14,054 & 25,555 & 16,421 & 15,032 & 7,729 & 22,324 & 7,706 & 17,628 & 19,067 & 7,132 & 4,865 \\ 
		WY & 217 & 526 & 359 & 239 & 180 & 467 & 198 & 377 & 416 & 177 & 132 \\ 
		\bottomrule
	\end{tabular}
	\endgroup
	\caption{Counts of disjoint control individuals in Pennsylvania cohort, by paired analysis. Each entry is the number of individuals in a control state (row) who contribute to only the analysis for Pennsylvania and not the other, paired treated state (column).} 
\end{table}

\subsection{Counts of Shared Control Individuals}\label{counts-of-shared-control-individuals}

\begin{table}[H]
	\centering
	\begingroup\footnotesize
	\begin{tabular}{llllllllllll}
		\toprule
		Ctrl. State & CT & FL & LA & MD & MN & ND & NH & NY & OH & OK & PA \\ 
		\midrule
		AL & 1,781 & 3,232 & 7,021 & 4,246 & 2,163 & 7,176 & 3,026 & 2,759 & 7,023 & 5,924 & 5,187 \\ 
		GA & 2,476 & 4,997 & 45,448 & 7,305 & 3,027 & 45,781 & 4,627 & 4,095 & 44,884 & 11,127 & 9,103 \\ 
		IA & 1,797 & 3,711 & 8,976 & 5,418 & 2,342 & 9,134 & 3,451 & 3,016 & 8,969 & 7,969 & 6,687 \\ 
		ID & 403 & 903 & 2,390 & 1,276 & 572 & 2,432 & 832 & 751 & 2,392 & 1,972 & 1,692 \\ 
		IN & 1,562 & 3,778 & 12,582 & 6,320 & 2,298 & 12,836 & 3,525 & 3,185 & 12,540 & 10,488 & 8,631 \\ 
		KS & 489 & 1,126 & 3,319 & 1,572 & 636 & 3,370 & 1,022 & 887 & 3,294 & 2,433 & 2,021 \\ 
		KY & 395 & 906 & 8,744 & 1,244 & 486 & 8,945 & 838 & 745 & 8,751 & 1,988 & 1,624 \\ 
		MS & 249 & 683 & 2,080 & 1,040 & 304 & 2,089 & 635 & 553 & 2,026 & 1,741 & 1,363 \\ 
		NC & 3,783 & 7,699 & 18,992 & 10,548 & 5,026 & 19,597 & 7,188 & 6,458 & 19,179 & 15,706 & 13,347 \\ 
		NE & 688 & 1,592 & 3,973 & 2,173 & 873 & 4,010 & 1,486 & 1,325 & 3,872 & 3,130 & 2,635 \\ 
		SC & 253 & 645 & 2,453 & 896 & 360 & 2,476 & 599 & 536 & 2,425 & 1,627 & 1,296 \\ 
		SD & 112 & 343 & 1,227 & 611 & 181 & 1,235 & 307 & 273 & 1,213 & 1,011 & 832 \\ 
		TN & 1,675 & 3,250 & 8,278 & 4,668 & 2,151 & 8,461 & 3,085 & 2,792 & 8,284 & 6,976 & 5,791 \\ 
		TX & 3,291 & 7,225 & 25,523 & 10,799 & 4,368 & 26,302 & 6,696 & 5,927 & 25,710 & 21,216 & 17,282 \\ 
		VA & 1,722 & 3,747 & 9,221 & 5,202 & 2,335 & 9,390 & 3,468 & 3,105 & 9,205 & 7,580 & 6,413 \\ 
		WI & 4,569 & 9,696 & 27,168 & 14,920 & 6,192 & 27,710 & 9,005 & 8,026 & 27,171 & 22,812 & 19,142 \\ 
		WY & 69 & 169 & 554 & 297 & 92 & 563 & 161 & 129 & 553 & 490 & 417 \\ 
		\bottomrule
	\end{tabular}
	\endgroup
	\caption{Counts of control individuals shared between Arkansas and each other cohort. Each entry is the number of individuals in a control state (row) who contribute to the analysis for Arkansas and the analysis for the other, paired treated state (column).} 
\end{table}
\begin{table}[H]
	\centering
	\begingroup\footnotesize
	\begin{tabular}{lllllllllll}
		\toprule
		Ctrl. State & FL & LA & MD & MN & ND & NH & NY & OH & OK & PA \\ 
		\midrule
		AL & 3,716 & 1,719 & 3,043 & 4,594 & 2,145 & 3,849 & 4,172 & 2,180 & 2,339 & 2,679 \\ 
		GA & 5,798 & 2,396 & 4,500 & 7,703 & 2,624 & 6,060 & 6,801 & 2,663 & 3,101 & 3,653 \\ 
		IA & 3,661 & 1,738 & 2,908 & 4,976 & 1,999 & 3,855 & 4,392 & 2,022 & 2,181 & 2,483 \\ 
		ID & 857 & 387 & 682 & 1,120 & 473 & 898 & 988 & 475 & 510 & 583 \\ 
		IN & 3,523 & 1,507 & 2,824 & 4,750 & 1,822 & 3,686 & 4,236 & 1,857 & 2,080 & 2,417 \\ 
		KS & 1,078 & 469 & 845 & 1,421 & 496 & 1,145 & 1,254 & 512 & 632 & 713 \\ 
		KY & 1,144 & 383 & 945 & 1,494 & 419 & 1,202 & 1,341 & 432 & 711 & 804 \\ 
		MS & 699 & 245 & 519 & 1,015 & 257 & 746 & 855 & 271 & 365 & 423 \\ 
		NC & 9,005 & 3,608 & 7,228 & 11,619 & 4,895 & 9,353 & 10,470 & 4,957 & 5,512 & 6,272 \\ 
		NE & 2,006 & 666 & 1,534 & 2,781 & 795 & 2,109 & 2,466 & 806 & 978 & 1,155 \\ 
		SC & 763 & 242 & 577 & 1,102 & 267 & 791 & 959 & 270 & 423 & 492 \\ 
		SD & 287 & 105 & 238 & 369 & 182 & 299 & 338 & 183 & 188 & 208 \\ 
		TN & 3,845 & 1,594 & 3,177 & 4,998 & 2,032 & 4,019 & 4,434 & 2,064 & 2,394 & 2,709 \\ 
		TX & 9,402 & 3,137 & 6,602 & 16,023 & 3,581 & 9,949 & 14,045 & 3,642 & 4,211 & 5,056 \\ 
		VA & 3,512 & 1,660 & 2,854 & 4,780 & 1,957 & 3,639 & 4,256 & 1,967 & 2,116 & 2,437 \\ 
		WI & 11,078 & 4,412 & 8,912 & 14,671 & 5,921 & 11,516 & 13,086 & 6,015 & 6,686 & 7,641 \\ 
		WY & 178 & 68 & 133 & 239 & 72 & 193 & 211 & 77 & 93 & 108 \\ 
		\bottomrule
	\end{tabular}
	\endgroup
	\caption{Counts of control individuals shared between Connecticut and each other cohort. Each entry is the number of individuals in a control state (row) who contribute to the analysis for Connecticut and the analysis for the other, paired treated state (column).} 
\end{table}
\begin{table}[H]
	\centering
	\begingroup\footnotesize
	\begin{tabular}{llllllllll}
		\toprule
		Ctrl. State & LA & MD & MN & ND & NH & NY & OH & OK & PA \\ 
		\midrule
		AL & 3,102 & 5,622 & 4,687 & 3,855 & 6,427 & 5,823 & 3,921 & 4,275 & 4,863 \\ 
		GA & 4,776 & 9,260 & 7,580 & 5,412 & 10,940 & 9,708 & 5,533 & 6,389 & 7,483 \\ 
		IA & 3,539 & 6,386 & 5,145 & 4,232 & 7,325 & 6,462 & 4,290 & 4,699 & 5,402 \\ 
		ID & 858 & 1,642 & 1,277 & 1,094 & 1,865 & 1,672 & 1,113 & 1,217 & 1,396 \\ 
		IN & 3,605 & 6,985 & 5,509 & 4,412 & 8,373 & 7,424 & 4,496 & 5,035 & 5,823 \\ 
		KS & 1,072 & 2,028 & 1,552 & 1,185 & 2,333 & 2,042 & 1,222 & 1,452 & 1,692 \\ 
		KY & 866 & 2,012 & 1,524 & 971 & 2,328 & 2,058 & 993 & 1,433 & 1,672 \\ 
		MS & 662 & 1,345 & 891 & 714 & 1,620 & 1,390 & 740 & 907 & 1,102 \\ 
		NC & 7,276 & 15,298 & 12,738 & 10,012 & 17,663 & 15,900 & 10,160 & 11,416 & 13,015 \\ 
		NE & 1,535 & 5,577 & 2,671 & 1,823 & 4,271 & 3,805 & 1,863 & 2,331 & 2,734 \\ 
		SC & 622 & 1,368 & 1,082 & 710 & 1,639 & 1,462 & 730 & 980 & 1,156 \\ 
		SD & 329 & 1,723 & 763 & 1,259 & 1,804 & 1,443 & 1,269 & 1,342 & 1,482 \\ 
		TN & 3,079 & 6,248 & 5,093 & 3,975 & 7,063 & 6,367 & 4,042 & 4,666 & 5,288 \\ 
		TX & 6,824 & 15,182 & 13,437 & 7,993 & 19,807 & 17,486 & 8,194 & 9,518 & 11,471 \\ 
		VA & 3,582 & 6,780 & 5,352 & 4,386 & 7,835 & 6,968 & 4,456 & 4,889 & 5,609 \\ 
		WI & 9,306 & 19,679 & 15,923 & 12,945 & 22,397 & 20,053 & 13,217 & 14,730 & 16,775 \\ 
		WY & 161 & 321 & 269 & 175 & 390 & 332 & 184 & 218 & 275 \\ 
		\bottomrule
	\end{tabular}
	\endgroup
	\caption{Counts of control individuals shared between Florida and each other cohort. Each entry is the number of individuals in a control state (row) who contribute to the analysis for Florida and the analysis for the other, paired treated state (column).} 
\end{table}
\begin{table}[H]
	\centering
	\begingroup\footnotesize
	\begin{tabular}{lllllllll}
		\toprule
		Ctrl. State & MD & MN & ND & NH & NY & OH & OK & PA \\ 
		\midrule
		AL & 4,088 & 2,062 & 6,783 & 2,886 & 2,637 & 6,664 & 5,692 & 4,940 \\ 
		GA & 6,957 & 2,873 & 43,756 & 4,362 & 3,848 & 42,980 & 10,588 & 8,572 \\ 
		IA & 5,180 & 2,240 & 8,628 & 3,275 & 2,858 & 8,490 & 7,602 & 6,329 \\ 
		ID & 1,224 & 545 & 2,298 & 785 & 710 & 2,269 & 1,885 & 1,607 \\ 
		IN & 6,052 & 2,192 & 12,156 & 3,347 & 3,019 & 11,917 & 10,026 & 8,193 \\ 
		KS & 1,486 & 599 & 3,150 & 961 & 829 & 3,092 & 2,303 & 1,900 \\ 
		KY & 1,189 & 462 & 11,201 & 799 & 712 & 11,007 & 1,890 & 1,538 \\ 
		MS & 1,001 & 293 & 1,974 & 603 & 525 & 1,924 & 1,682 & 1,299 \\ 
		NC & 9,992 & 4,721 & 18,285 & 6,714 & 6,047 & 17,960 & 14,833 & 12,525 \\ 
		NE & 2,089 & 823 & 3,800 & 1,410 & 1,257 & 3,687 & 3,006 & 2,503 \\ 
		SC & 862 & 337 & 2,354 & 561 & 503 & 2,308 & 1,546 & 1,213 \\ 
		SD & 589 & 168 & 1,176 & 289 & 257 & 1,156 & 969 & 797 \\ 
		TN & 4,425 & 2,025 & 7,980 & 2,887 & 2,610 & 7,839 & 6,611 & 5,435 \\ 
		TX & 10,225 & 4,088 & 24,341 & 6,254 & 5,536 & 23,866 & 19,893 & 16,039 \\ 
		VA & 4,977 & 2,232 & 8,820 & 3,279 & 2,931 & 8,681 & 7,220 & 6,073 \\ 
		WI & 14,318 & 5,915 & 26,109 & 8,590 & 7,665 & 25,666 & 21,844 & 18,164 \\ 
		WY & 284 & 90 & 537 & 155 & 123 & 528 & 473 & 395 \\ 
		\bottomrule
	\end{tabular}
	\endgroup
	\caption{Counts of control individuals shared between Louisiana and each other cohort. Each entry is the number of individuals in a control state (row) who contribute to the analysis for Louisiana and the analysis for the other, paired treated state (column).} 
\end{table}
\begin{table}[H]
	\centering
	\begingroup\footnotesize
	\begin{tabular}{llllllll}
		\toprule
		Ctrl. State & MN & ND & NH & NY & OH & OK & PA \\ 
		\midrule
		AL & 3,830 & 5,015 & 5,246 & 4,770 & 5,102 & 5,538 & 6,319 \\ 
		GA & 5,860 & 7,872 & 8,519 & 7,554 & 8,015 & 9,254 & 10,936 \\ 
		IA & 4,107 & 6,132 & 5,896 & 5,173 & 6,217 & 6,778 & 7,788 \\ 
		ID & 1,030 & 1,533 & 1,501 & 1,350 & 1,556 & 1,688 & 1,935 \\ 
		IN & 4,354 & 7,389 & 6,502 & 5,829 & 7,535 & 8,346 & 9,613 \\ 
		KS & 1,206 & 1,652 & 1,842 & 1,602 & 1,695 & 1,981 & 2,335 \\ 
		KY & 1,234 & 1,323 & 1,857 & 1,662 & 1,350 & 1,924 & 2,250 \\ 
		MS & 666 & 1,082 & 1,226 & 1,057 & 1,121 & 1,335 & 1,637 \\ 
		NC & 10,180 & 13,637 & 14,179 & 12,755 & 13,849 & 15,528 & 17,799 \\ 
		NE & 2,076 & 2,457 & 3,377 & 2,990 & 2,504 & 3,163 & 3,696 \\ 
		SC & 822 & 990 & 1,247 & 1,110 & 1,016 & 1,331 & 1,586 \\ 
		SD & 629 & 2,372 & 1,510 & 1,209 & 2,395 & 2,517 & 2,792 \\ 
		TN & 4,219 & 5,659 & 5,840 & 5,248 & 5,750 & 6,583 & 7,557 \\ 
		TX & 9,366 & 11,933 & 13,984 & 12,297 & 12,187 & 14,167 & 17,086 \\ 
		VA & 4,324 & 6,080 & 6,278 & 5,609 & 6,176 & 6,762 & 7,790 \\ 
		WI & 12,840 & 19,812 & 18,165 & 16,272 & 20,160 & 22,324 & 25,467 \\ 
		WY & 203 & 310 & 301 & 252 & 319 & 364 & 454 \\ 
		\bottomrule
	\end{tabular}
	\endgroup
	\caption{Counts of control individuals shared between Maryland and each other cohort. Each entry is the number of individuals in a control state (row) who contribute to the analysis for Maryland and the analysis for the other, paired treated state (column).} 
\end{table}
\begin{table}[H]
	\centering
	\begingroup\footnotesize
	\begin{tabular}{lllllll}
		\toprule
		Ctrl. State & ND & NH & NY & OH & OK & PA \\ 
		\midrule
		AL & 2,623 & 4,875 & 5,273 & 2,674 & 2,894 & 3,324 \\ 
		GA & 3,273 & 7,988 & 8,994 & 3,367 & 3,975 & 4,705 \\ 
		IA & 2,696 & 5,414 & 6,110 & 2,736 & 2,991 & 3,461 \\ 
		ID & 681 & 1,334 & 1,466 & 692 & 750 & 868 \\ 
		IN & 2,705 & 5,756 & 6,564 & 2,761 & 3,087 & 3,615 \\ 
		KS & 667 & 1,640 & 1,788 & 698 & 863 & 980 \\ 
		KY & 522 & 1,622 & 1,836 & 543 & 880 & 1,022 \\ 
		MS & 320 & 974 & 1,113 & 337 & 442 & 532 \\ 
		NC & 6,648 & 13,281 & 14,789 & 6,761 & 7,580 & 8,711 \\ 
		NE & 1,017 & 2,815 & 3,247 & 1,039 & 1,272 & 1,531 \\ 
		SC & 388 & 1,128 & 1,362 & 399 & 580 & 680 \\ 
		SD & 462 & 799 & 897 & 464 & 488 & 542 \\ 
		TN & 2,650 & 5,336 & 5,877 & 2,698 & 3,130 & 3,560 \\ 
		TX & 4,877 & 14,268 & 19,569 & 5,001 & 5,852 & 7,092 \\ 
		VA & 2,775 & 5,558 & 6,343 & 2,812 & 3,089 & 3,573 \\ 
		WI & 8,321 & 16,592 & 18,697 & 8,510 & 9,501 & 10,872 \\ 
		WY & 98 & 286 & 318 & 105 & 129 & 167 \\ 
		\bottomrule
	\end{tabular}
	\endgroup
	\caption{Counts of control individuals shared between Minnesota and each other cohort. Each entry is the number of individuals in a control state (row) who contribute to the analysis for Minnesota and the analysis for the other, paired treated state (column).} 
\end{table}
\begin{table}[H]
	\centering
	\begingroup\footnotesize
	\begin{tabular}{llllll}
		\toprule
		Ctrl. State & NH & NY & OH & OK & PA \\ 
		\midrule
		AL & 3,608 & 3,326 & 8,356 & 7,060 & 6,170 \\ 
		GA & 4,979 & 4,451 & 48,037 & 12,052 & 9,928 \\ 
		IA & 3,936 & 3,473 & 10,227 & 9,028 & 7,637 \\ 
		ID & 1,005 & 915 & 2,897 & 2,375 & 2,044 \\ 
		IN & 4,108 & 3,745 & 14,813 & 12,286 & 10,205 \\ 
		KS & 1,074 & 939 & 3,512 & 2,575 & 2,133 \\ 
		KY & 891 & 800 & 13,190 & 2,147 & 1,766 \\ 
		MS & 655 & 582 & 2,200 & 1,835 & 1,461 \\ 
		NC & 9,329 & 8,456 & 24,747 & 20,089 & 17,261 \\ 
		NE & 1,695 & 1,532 & 4,421 & 3,550 & 3,016 \\ 
		SC & 652 & 584 & 2,711 & 1,802 & 1,442 \\ 
		SD & 1,105 & 879 & 5,220 & 4,064 & 3,258 \\ 
		TN & 3,768 & 3,417 & 10,140 & 8,456 & 7,079 \\ 
		TX & 7,389 & 6,608 & 28,580 & 23,423 & 19,210 \\ 
		VA & 4,068 & 3,676 & 10,851 & 8,857 & 7,544 \\ 
		WI & 12,036 & 10,830 & 39,092 & 30,285 & 25,490 \\ 
		WY & 165 & 135 & 600 & 520 & 436 \\ 
		\bottomrule
	\end{tabular}
	\endgroup
	\caption{Counts of control individuals shared between North Dakota and each other cohort. Each entry is the number of individuals in a control state (row) who contribute to the analysis for North Dakota and the analysis for the other, paired treated state (column).} 
\end{table}
\begin{table}[H]
	\centering
	\begingroup\footnotesize
	\begin{tabular}{lllll}
		\toprule
		Ctrl. State & NY & OH & OK & PA \\ 
		\midrule
		AL & 6,100 & 3,668 & 3,987 & 4,565 \\ 
		GA & 10,340 & 5,089 & 5,873 & 6,920 \\ 
		IA & 6,900 & 3,989 & 4,334 & 5,017 \\ 
		ID & 1,764 & 1,021 & 1,111 & 1,283 \\ 
		IN & 7,877 & 4,182 & 4,670 & 5,439 \\ 
		KS & 2,196 & 1,112 & 1,330 & 1,538 \\ 
		KY & 2,212 & 917 & 1,322 & 1,544 \\ 
		MS & 1,529 & 677 & 821 & 1,013 \\ 
		NC & 16,793 & 9,458 & 10,583 & 12,133 \\ 
		NE & 4,046 & 1,727 & 2,161 & 2,549 \\ 
		SC & 1,546 & 670 & 899 & 1,055 \\ 
		SD & 1,513 & 1,114 & 1,171 & 1,309 \\ 
		TN & 6,744 & 3,820 & 4,388 & 4,979 \\ 
		TX & 18,723 & 7,560 & 8,735 & 10,623 \\ 
		VA & 7,341 & 4,129 & 4,525 & 5,223 \\ 
		WI & 21,194 & 12,278 & 13,612 & 15,568 \\ 
		WY & 360 & 174 & 204 & 257 \\ 
		\bottomrule
	\end{tabular}
	\endgroup
	\caption{Counts of control individuals shared between New Hampshire and each other cohort. Each entry is the number of individuals in a control state (row) who contribute to the analysis for New Hampshire and the analysis for the other, paired treated state (column).} 
\end{table}
\begin{table}[H]
	\centering
	\begingroup\footnotesize
	\begin{tabular}{llll}
		\toprule
		Ctrl. State & OH & OK & PA \\ 
		\midrule
		AL & 3,381 & 3,662 & 4,208 \\ 
		GA & 4,558 & 5,260 & 6,246 \\ 
		IA & 3,532 & 3,813 & 4,477 \\ 
		ID & 931 & 1,012 & 1,171 \\ 
		IN & 3,820 & 4,249 & 4,972 \\ 
		KS & 977 & 1,175 & 1,367 \\ 
		KY & 822 & 1,209 & 1,411 \\ 
		MS & 604 & 736 & 906 \\ 
		NC & 8,601 & 9,601 & 11,095 \\ 
		NE & 1,564 & 1,946 & 2,303 \\ 
		SC & 597 & 811 & 950 \\ 
		SD & 887 & 932 & 1,047 \\ 
		TN & 3,471 & 3,978 & 4,531 \\ 
		TX & 6,770 & 7,813 & 9,537 \\ 
		VA & 3,736 & 4,074 & 4,745 \\ 
		WI & 11,066 & 12,227 & 14,129 \\ 
		WY & 145 & 171 & 218 \\ 
		\bottomrule
	\end{tabular}
	\endgroup
	\caption{Counts of control individuals shared between New York and each other cohort. Each entry is the number of individuals in a control state (row) who contribute to the analysis for New York and the analysis for the other, paired treated state (column).} 
\end{table}
\begin{table}[H]
	\centering
	\begingroup\footnotesize
	\begin{tabular}{lll}
		\toprule
		Ctrl. State & OK & PA \\ 
		\midrule
		AL & 7,201 & 6,317 \\ 
		GA & 12,328 & 10,181 \\ 
		IA & 9,182 & 7,786 \\ 
		ID & 2,424 & 2,087 \\ 
		IN & 12,537 & 10,446 \\ 
		KS & 2,645 & 2,205 \\ 
		KY & 2,191 & 1,812 \\ 
		MS & 1,889 & 1,515 \\ 
		NC & 20,502 & 17,678 \\ 
		NE & 3,631 & 3,089 \\ 
		SC & 1,852 & 1,492 \\ 
		SD & 4,138 & 3,311 \\ 
		TN & 8,641 & 7,239 \\ 
		TX & 24,048 & 19,789 \\ 
		VA & 9,049 & 7,730 \\ 
		WI & 30,901 & 26,064 \\ 
		WY & 540 & 457 \\ 
		\bottomrule
	\end{tabular}
	\endgroup
	\caption{Counts of control individuals shared between Ohio and each other cohort. Each entry is the number of individuals in a control state (row) who contribute to the analysis for Ohio and the analysis for the other, paired treated state (column).} 
\end{table}
\begin{table}[H]
	\centering
	\begingroup\footnotesize
	\begin{tabular}{ll}
		\toprule
		Ctrl. State & PA \\ 
		\midrule
		AL & 6,778 \\ 
		GA & 11,529 \\ 
		IA & 8,345 \\ 
		ID & 2,227 \\ 
		IN & 11,358 \\ 
		KS & 2,534 \\ 
		KY & 2,453 \\ 
		MS & 1,919 \\ 
		NC & 19,437 \\ 
		NE & 3,875 \\ 
		SC & 1,843 \\ 
		SD & 3,424 \\ 
		TN & 8,134 \\ 
		TX & 22,359 \\ 
		VA & 8,323 \\ 
		WI & 28,331 \\ 
		WY & 502 \\ 
		\bottomrule
	\end{tabular}
	\endgroup
	\caption{Counts of control individuals shared between Oklahoma and each other cohort. Each entry is the number of individuals in a control state (row) who contribute to the analysis for Oklahoma and the analysis for the other, paired treated state (column).} 
\end{table}

\end{document}